%% file: cfie_notes.tex
\documentclass[preprint,10pt]{elsarticle}
\pdfoutput=1


\makeatletter
\def\ps@pprintTitle{%
   \let\@oddhead\@empty
   \let\@evenhead\@empty
   \def\@oddfoot{\reset@font\hfil\thepage\hfil}
   \let\@evenfoot\@oddfoot
}
\makeatother

\input{packages.tex}
\usepackage[margin=1in]{geometry}
\biboptions{sort&compress}
\begin{document}

\begin{frontmatter}
\title{Code-Verification Techniques for the Method-of-Moments Implementation of the Combined-Field Integral Equation}
\author[freno]{Brian A.\ Freno}
\ead{bafreno@sandia.gov}
\author[freno]{Neil R.\ Matula}
\address[freno]{Sandia National Laboratories, Albuquerque, NM 87185}
\begin{abstract}
\input{abstract.tex}
\end{abstract}


\end{frontmatter}

\input{introduction.tex}
\input{cfie.tex}

\input{mms.tex}

\input{g_ms.tex}

\input{results.tex}

\appendix
\renewcommand{\thesection}{Appendix~\Alph{section}}
\input{appendix.tex}

%
\phantomsection\addcontentsline{toc}{section}{\refname}
\bibliographystyle{elsarticle-num}
\bibliography{../quadrature_manuscript/quadrature.bib}

\end{document}

%% file: packages.tex
\usepackage{titlesec}
\usepackage{xcolor}
\usepackage{enumitem}
\usepackage{hyperref}
\hypersetup{colorlinks=true,urlcolor=black}
\usepackage[T1]{fontenc}\usepackage{lmodern}

\usepackage{mathtools,amsfonts,amssymb}
\usepackage{caption}
\usepackage{subcaption}
\usepackage{booktabs}
\usepackage{pifont}
\usepackage{placeins}

\definecolor{green}{rgb}{0,0.5,0}

\usepackage{tikz}
\usetikzlibrary{calc}
\usepackage{tkz-euclide}
\usepackage{calc}
\usepackage{tikz,pgfplots}
\pgfplotsset{compat=1.12}
\usetikzlibrary{arrows.meta,3d,matrix,positioning,decorations.markings}

\newcommand{\pz}{\phantom{0}}
\newcommand{\pn}{\phantom{-}}
\newcommand{\vpad}{\vphantom{\bigg(}}
\newcommand{\ntriangles}{n_t}

\newcommand{\nbasis}{n_b}
\newcommand{\srcidx}{j}
\newcommand{\testidx}{i}

\makeatletter
\tikzoption{canvas is xy plane at z}[]{%
  \def\tikz@plane@origin{\pgfpointxyz{0}{0}{#1}}%
  \def\tikz@plane@x{\pgfpointxyz{1}{0}{#1}}%
  \def\tikz@plane@y{\pgfpointxyz{0}{1}{#1}}%
  \tikz@canvas@is@plane
}
\makeatother

\makeatletter
\tikzoption{canvas is plane}[]{\@setOxy#1}
\def\@setOxy O(#1,#2,#3)x(#4,#5,#6)y(#7,#8,#9)%
  {\def\tikz@plane@origin{\pgfpointxyz{#1}{#2}{#3}}%
   \def\tikz@plane@x{\pgfpointxyz{#4}{#5}{#6}}%
   \def\tikz@plane@y{\pgfpointxyz{#7}{#8}{#9}}%
   \tikz@canvas@is@plane
  }
\makeatother 

\definecolor{orange}{rgb}{1,0.5,0}
\definecolor{green}{rgb}{0,0.5,0}
\definecolor{purple}{rgb}{0.5,0,0.5}
\newcommand{\reviewerOne}[1]{#1}

\newcommand{\reviewerTwo}[1]{#1}

%% file: abstract.tex
Code verification plays an important role in establishing the credibility of computational simulations by assessing the correctness of the implementation of the underlying numerical methods.  In computational electromagnetics, the numerical solution to integral equations incurs multiple interacting sources of numerical error, as well as other challenges, which render traditional code-verification approaches ineffective.  In this paper, we provide approaches to separately measure the numerical errors arising from these different error sources for the method-of-moments implementation of the combined-field integral equation.  We demonstrate the effectiveness of these approaches for cases with and without coding errors.

%% file: introduction.tex
\section{Introduction}

For electromagnetic scatterers, Maxwell's equations, together with appropriate boundary conditions, may be formulated as surface integral equations (SIEs).  
The most common SIEs for modeling time-harmonic electromagnetic phenomena are the electric-field integral equation (EFIE), which relates the surface current to the scattered electric field, and the magnetic-field integral equation (MFIE), which relates the surface current to the scattered magnetic field.  
The EFIE arises from the condition that the total tangential electric field on the surface of a perfect electric conductor is zero, whereas the MFIE arises from the condition that the component of the total magnetic field tangent to the surface of a perfect electric conductor is equal to the surface current density.  At certain frequencies, the accuracy of the solutions to the EFIE and MFIE deteriorates due to the internal resonances of the scatterer.  Therefore, the combined-field integral equation (CFIE), which is a linear combination of the EFIE and MFIE, is employed to overcome this problem.

These SIEs are typically solved through the method of moments (MoM), wherein the surface of the electromagnetic scatterer is discretized using planar or curvilinear mesh elements, and four-dimensional integrals are evaluated over two-dimensional source and test elements.  
These integrals contain a Green's function, which yields singularities when the test and source elements share one or more edges or vertices, and near-singularities when they are otherwise close.  The accurate evaluation of these integrals is an active research topic, with many approaches being developed to address the (near-)singularity for the inner, source-element integral~\cite{graglia_1993,wilton_1984,rao_1982,khayat_2005,fink_2008,khayat_2008,vipiana_2011,vipiana_2012,botha_2013,rivero_2019}, as well as for the outer, test-element integral~\cite{vipiana_2013,polimeridis_2013,wilton_2017,rivero_2019b,freno_em}.

Code verification plays an important role in establishing the credibility of results from computational physics simulations~\cite{roache_1998,knupp_2022,oberkampf_2010} by assessing the correctness of the implementation of the underlying numerical methods.  The discretization of differential, integral, or integro-differential equations incurs some truncation error, and thus the approximate solutions produced from the discretized equations will incur an associated discretization error.  If the discretization error tends to zero as the discretization is refined, the consistency of the code is verified~\cite{roache_1998}.  This may be taken a step further by examining not only consistency, but the rate at which the error decreases as the discretization is refined, thereby verifying the order of accuracy of the discretization scheme.  The correctness of the numerical-method implementation may then be verified by comparing the expected and observed orders of accuracy obtained from numerous test cases with known solutions.

To measure the discretization error, a known solution is required to compare with the discretized solution.  Exact solutions are generally limited and may not sufficiently exercise the capabilities of the code.  Therefore, manufactured solutions~\cite{roache_2001} are a popular alternative, permitting the construction of arbitrarily complex problems with known solutions.  Through the method of manufactured solutions (MMS), a solution is manufactured and substituted directly into the governing equations to yield a residual term, which is added as a source term to coerce the solution to the manufactured solution.

For code verification, integral equations yield an additional challenge.  While analytical differentiation is straightforward, analytical integration is not always possible.  Therefore, the residual source term arising from the manufactured solution may not be representable in closed form, and its implementation may be accompanied by numerical techniques that carry their own numerical errors.  Furthermore, for the EFIE, MFIE, and CFIE, the aforementioned (nearly) singular integrals can further complicate the numerical evaluation of the source term.  Therefore, many of the benefits associated with MMS are lost when applied straightforwardly to these integral equations.

There are many examples of code verification in the literature for different computational physics disciplines.  These disciplines include aerodynamics~\cite{nishikawa_2022}, fluid dynamics~\cite{roy_2004,bond_2007,veluri_2010,oliver_2012,eca_2016,hennink_2021,freno_2021}, solid mechanics~\cite{chamberland_2010}, fluid--structure interaction~\cite{etienne_2012,bukac_2023}, heat transfer in fluid--solid interaction~\cite{veeraragavan_2016}, multiphase flows~\cite{brady_2012,lovato_2021}, radiation hydrodynamics~\cite{mcclarren_2008}, electrostatics~\cite{tranquilli_2022}, electrodynamics~\cite{ellis_2009,amormartin_2021}, magneto-hydrodynamics~\cite{rueda_2023}, and ablation~\cite{amar_2008,amar_2009,amar_2011,freno_ablation,freno_ablation_2022}.  
For electromagnetic SIEs, code-verification activities that employ manufactured solutions have been limited to the EFIE~\cite{marchand_2013,marchand_2014,freno_em_mms_2020,freno_em_mms_quad_2021} and, more recently, the MFIE~\cite{freno_mfie_2022}.

As described in~\cite{freno_em_mms_2020,freno_mfie_2022}, the EFIE and MFIE, and consequently the CFIE, incur numerical error due to curved surfaces being approximated by planar elements (domain-discretization error), the solution being approximated as a linear combination of a finite number of basis functions (solution-discretization error), and the approximate evaluation of integrals using quadrature rules (numerical-integration error).

For the EFIE, Marchand et al.~\cite{marchand_2013,marchand_2014} compute the MMS source term using additional quadrature points.  Freno et al.~\cite{freno_em_mms_2020} manufacture the Green's function, permitting the numerical-integration error to be eliminated and the solution-discretization error to be isolated.  Freno et al.~\cite{freno_em_mms_quad_2021} also provide approaches to isolate the numerical-integration error.
For the MFIE, Freno and Matula~\cite{freno_mfie_2022} isolate and measure the solution-discretization error and numerical-integration error.

In this paper, we present code-verification techniques for the MoM implementation of the CFIE that isolate and measure the solution-discretization error and numerical-integration error.  
For curved surfaces, the domain-discretization error cannot be completely isolated or eliminated, but methods are presented in~\cite{freno_mfie_2022} to account for it in the MFIE.  These methods can be applied to the CFIE straightforwardly.  In this work, we avoid the domain-discretization error by considering only planar surfaces.  
We isolate the solution-discretization error by approximating the Green's function in terms of even powers of the distance between the test and source points.  Through this approximation, we can evaluate the integrals exactly, thereby avoiding numerical-integration error.  The approximated Green's function differs from the previously manufactured Green's function~\cite{freno_em_mms_2020,freno_mfie_2022} by more closely resembling the actual Green's function with higher-degree polynomials of the distance.  We isolate the numerical-integration error by canceling the influence of the basis functions.  We perform convergence studies for different wavenumbers and combination parameters to vary the relative weights between the terms in the CFIE within a physically realistic range.

\reviewerTwo{By approximating the Green's function in terms of even powers of the distance, we avoid the challenges associated with evaluating the aforementioned (nearly) singular integrals.  Given the computational expense of computing accurate reference solutions, assessing these integral evaluations is best accomplished through extensive unit testing as complementary code verification.  Examples for the EFIE are included in~\cite{freno_quad,freno_em}.}

This paper is organized as follows.  In Section~\ref{sec:cfie}, we describe the MoM implementation of the CFIE.  In Section~\ref{sec:mms}, we describe the challenges of using MMS with the MoM implementation of the CFIE, as well as our approach to mitigating them.  In Section~\ref{sec:g_ms}, we describe our approach to approximating the Green's function and evaluating integrals containing it.  In Section~\ref{sec:results}, we demonstrate the effectiveness of our approaches for several different configurations with and without coding errors.  In Section~\ref{sec:conclusions}, we summarize this work.

%% file: cfie.tex
\section{The Method-of-Moments Implementation of the Combined-Field Integral Equation} 
\label{sec:cfie}

In time-harmonic form, the scattered electric field $\mathbf{E}^\mathcal{S}$ and magnetic field $\mathbf{H}^\mathcal{S}$ due to induced surface currents on a scatterer can be computed by~\cite{harrington_2001}
\begin{align}
\mathbf{E}^\mathcal{S}(\mathbf{x}) &{}= -(j\omega\mathbf{A}(\mathbf{x})+\nabla\Phi(\mathbf{x})), 
\label{eq:Es} \\
\mathbf{H}^\mathcal{S}(\mathbf{x}) &{}= \frac{1}{\mu}\nabla\times\mathbf{A}(\mathbf{x}), 
\label{eq:Hs}
\end{align}
where the magnetic vector potential $\mathbf{A}$ is defined by
\begin{align}
\mathbf{A}(\mathbf{x})= \mu \int_{S'} \mathbf{J} (\mathbf{x}')G(\mathbf{x},\mathbf{x}')dS',
\label{eq:A}
\end{align}
and, by employing the Lorenz gauge condition and the continuity equation, the electric scalar potential $\Phi$ is defined by 
\begin{align}
\Phi(\mathbf{x})=  \frac{j}{\epsilon\omega} \int_{S'} \nabla'\cdot\mathbf{J}(\mathbf{x}')G(\mathbf{x},\mathbf{x}')dS'.
\label{eq:Phi}
\end{align}
In~\eqref{eq:A} and~\eqref{eq:Phi}, the integration domain $S'=S$ is the closed surface of a perfect electric conductor, where the prime notation is introduced here to distinguish the inner and outer integration domains later in this section.  Additionally, $\mathbf{J}$ is the electric surface current density, $\mu$ and $\epsilon$ are the permeability and permittivity of the surrounding medium, and $G$ is the Green's function 
\begin{align}
G(\mathbf{x},\mathbf{x}') = \frac{e^{-jkR}}{4\pi R},
\label{eq:G}
\end{align}
where $R=\|\mathbf{R}\|_2$, $\mathbf{R}=\mathbf{x}-\mathbf{x}'$, and $k=\omega\sqrt{\mu\epsilon}$ is the wavenumber.  

\subsection{The Electric-Field Integral Equation} 
The total electric field $\mathbf{E}$ is the sum of the incident electric field $\mathbf{E}^\mathcal{I}$, which induces $\mathbf{J}$, and $\mathbf{E}^\mathcal{S}$.  On $S$, the tangential component of $\mathbf{E}$ is zero, such that
\begin{align}
\mathbf{E}_t^\mathcal{S}=-\mathbf{E}_t^\mathcal{I},
\label{eq:tan_BC}
\end{align}
where $(\cdot)_t$ denotes the tangential component.
Substituting~\eqref{eq:Es} into~\eqref{eq:tan_BC} yields the EFIE at a point on the surface of the scatterer, from which we can compute $\mathbf{J}$ from $\mathbf{E}^\mathcal{I}$:
\begin{align}
(j\omega\mathbf{A} + \nabla\Phi)_t = \mathbf{E}_t^\mathcal{I}.
\label{eq:tan}
\end{align}
%

\subsection{The Magnetic-Field Integral Equation} 

The total magnetic field $\mathbf{H}$ is the sum of the incident magnetic field $\mathbf{H}^\mathcal{I}$ and $\mathbf{H}^\mathcal{S}$.  On $S$,
\begin{align}
\mathbf{n}\times \mathbf{H} = \mathbf{J},
\label{eq:mfie_bc}
\end{align}
where $\mathbf{n}$ is the unit vector normal to $S$.  
Noting that $\nabla\times\big[\mathbf{J}(\mathbf{x}')G(\mathbf{x},\mathbf{x}')\big]=\nabla G(\mathbf{x},\mathbf{x}')\times\mathbf{J}(\mathbf{x}')$ and 
\begin{align}
\nabla G(\mathbf{x},\mathbf{x}') = -\nabla' G(\mathbf{x},\mathbf{x}'),
\label{eq:grad_G}
\end{align}
from~\eqref{eq:Hs} and~\eqref{eq:A}, 
\begin{align*}
\mathbf{H}^\mathcal{S}(\mathbf{x}) = \int_{S'} \mathbf{J}(\mathbf{x}')\times\nabla' G(\mathbf{x},\mathbf{x}')dS'
\end{align*}
when $\mathbf{x}$ is just outside of $S$.  Therefore, at $S$,
\begin{align}
\mathbf{n}\times\mathbf{H}^\mathcal{S} = \lim_{\mathbf{x}\to S}\mathbf{n}\times \int_{S'} \mathbf{J}(\mathbf{x}')\times\nabla' G(\mathbf{x},\mathbf{x}')dS' = \frac{1}{2}\mathbf{J} + \mathbf{n}\times \int_{S'} \mathbf{J}(\mathbf{x}')\times\nabla' G(\mathbf{x},\mathbf{x}')dS',
\label{eq:limit}
\end{align}
where the final term is evaluated through principal value integration.
From~\eqref{eq:mfie_bc} and~\eqref{eq:limit} the MFIE at a point on the surface of the scatterer is~\cite{chew_1995, balanis_2012}
\begin{align}
\frac{1}{2}\mathbf{J} - \mathbf{n}\times \int_{S'} \mathbf{J}(\mathbf{x}')\times\nabla' G(\mathbf{x},\mathbf{x}')dS' = \mathbf{n}\times \mathbf{H}^\mathcal{I}.
\label{eq:mfie}
\end{align}

\subsection{The Method of Moments} 

Inserting~\eqref{eq:A} and~\eqref{eq:Phi} into~\eqref{eq:tan}, projecting~\eqref{eq:tan} onto an appropriate space $\mathbb{V}$ containing vector fields that are tangent to $S$, and integrating by parts yields the variational form of the EFIE: find $\mathbf{J}\in\mathbb{V}$, such that
\begin{align}
\int_S \mathbf{E}^\mathcal{I}\cdot \bar{\mathbf{v}} dS =
j\omega\mu\int_S \bar{\mathbf{v}}(\mathbf{x})\cdot \int_{S'} \mathbf{J}(\mathbf{x}')G(\mathbf{x},\mathbf{x}')dS' dS - \frac{j}{\epsilon\omega}\int_S \nabla\cdot\bar{\mathbf{v}}(\mathbf{x}) \int_{S'} \nabla'\cdot \mathbf{J}(\mathbf{x}')G(\mathbf{x},\mathbf{x}')dS' dS 
\label{eq:efie_variational}
\end{align} 
for all $\mathbf{v}\in\mathbb{V}$, where the overbar denotes complex conjugation.

We can write~\eqref{eq:efie_variational} more succinctly in terms of a sesquilinear form and inner product: 
\begin{align}
a^\mathcal{E}(\mathbf{J},\mathbf{v}) = b^\mathcal{E}\big(\mathbf{E}^\mathcal{I}, \mathbf{v}\big),
\label{eq:efie_sesquilinear}
\end{align}
where the sesquilinear form and inner product are defined by
\begin{align*}
a^\mathcal{E}(\mathbf{u},\mathbf{v}) &{}= j\omega\mu \int_S \bar{\mathbf{v}}(\mathbf{x})\cdot\int_{S'} \mathbf{u}(\mathbf{x}')G(\mathbf{x},\mathbf{x}')dS'dS -\frac{j}{\epsilon\omega} \int_S \nabla\cdot\bar{\mathbf{v}}(\mathbf{x})\int_{S'} \nabla'\cdot\mathbf{u}(\mathbf{x}')G(\mathbf{x},\mathbf{x}')dS' dS, 
\\
b^\mathcal{E}(\mathbf{u},\mathbf{v})  &{}= \int_S \mathbf{u}(\mathbf{x})\cdot \bar{\mathbf{v}}(\mathbf{x}) dS. 
\end{align*}

Projecting~\eqref{eq:mfie} onto $\mathbb{V}$ yields the variational form of the MFIE: find $\mathbf{J}\in\mathbb{V}$, such that
\begin{align}
\frac{1}{2}\int_S \bar{\mathbf{v}}\cdot\mathbf{J} dS - \int_S \bar{\mathbf{v}}(\mathbf{x})\cdot \bigg(\mathbf{n}(\mathbf{x})\times \int_{S'} \mathbf{J}(\mathbf{x}')\times\nabla' G(\mathbf{x},\mathbf{x}')dS'\bigg) dS = \int_S \bar{\mathbf{v}}\cdot \big(\mathbf{n}\times \mathbf{H}^\mathcal{I}\big) dS
\label{eq:mfie_variational}
\end{align} 
for all $\mathbf{v}\in\mathbb{V}$.
We can write~\eqref{eq:mfie_variational} more succinctly as
\begin{align}
a^\mathcal{M}(\mathbf{J},\mathbf{v}) = b^\mathcal{M}\big(\mathbf{H}^\mathcal{I}, \mathbf{v}\big),
\label{eq:mfie_sesquilinear}
\end{align}
where the sesquilinear forms are defined by
\begin{align*}
a^\mathcal{M}(\mathbf{u}, \mathbf{v}) &{}= \frac{1}{2}\int_S \bar{\mathbf{v}}(\mathbf{x})\cdot\mathbf{u}(\mathbf{x}) dS - \int_S \bar{\mathbf{v}}(\mathbf{x})\cdot \bigg(\mathbf{n}(\mathbf{x})\times \int_{S'} \mathbf{u}(\mathbf{x}')\times\nabla' G(\mathbf{x},\mathbf{x}')dS'\bigg) dS, 
\\
b^\mathcal{M}(\mathbf{u}, \mathbf{v}) &{}= \int_S \bar{\mathbf{v}}(\mathbf{x})\cdot [\mathbf{n}(\mathbf{x})\times \mathbf{u}(\mathbf{x})] dS. 
\end{align*}

The CFIE linearly combines the EFIE~\eqref{eq:efie_sesquilinear} and MFIE~\eqref{eq:mfie_sesquilinear} to yield the variational problem: find $\mathbf{J}\in\mathbb{V}$, such that
\begin{align}
a(\mathbf{J},\mathbf{v}) 
{}={} 
b\big(\mathbf{E}^\mathcal{I} , \mathbf{H}^\mathcal{I} , \mathbf{v}\big) \qquad \forall \mathbf{v}\in\mathbb{V},
\label{eq:sesquilinear}
\end{align}
where 
\begin{align}
a(\mathbf{J},\mathbf{v}) 
&{}={} 
 \frac{\alpha}{\eta}a^\mathcal{E}(\mathbf{J},\mathbf{v}) 
{}+{}
(1-\alpha) a^\mathcal{M}(\mathbf{J},\mathbf{v}), \label{eq:a}
\\
b\big(\mathbf{E}^\mathcal{I} , \mathbf{H}^\mathcal{I} , \mathbf{v}\big)
&{}={} 
\frac{\alpha}{\eta} b^\mathcal{E}\big(\mathbf{E}^\mathcal{I} , \mathbf{v}\big)
{}+{}
(1-\alpha) b^\mathcal{M}\big(\mathbf{H}^\mathcal{I}, \mathbf{v}\big). \label{eq:b}
\end{align}
%
In~\eqref{eq:a} and~\eqref{eq:b}, $\alpha\in[0,1]$ is the combination parameter, and $\eta=\sqrt{\mu/\epsilon}$ is the characteristic impedance of the surrounding medium.  It should be noted this is one of multiple choices for the CFIE~\cite{chew_2001,ylaoijala_2003}; however, the verification methods presented in this paper can be applied to the other CFIE choices.

To solve the variational problem~\eqref{eq:sesquilinear}, we discretize $S$ with a mesh composed of triangular elements and approximate $\mathbf{J}$ with $\mathbf{J}_h$ in terms of the Rao--Wilton--Glisson (RWG) basis functions $\boldsymbol{\Lambda}_{\srcidx}(\mathbf{x})$~\cite{rao_1982}:
\begin{align}
\mathbf{J}_h(\mathbf{x}) = \sum_{\srcidx=1}^{\nbasis} J_{\srcidx} \boldsymbol{\Lambda}_{\srcidx}(\mathbf{x}),
\label{eq:J_h}
\end{align}%
where $\nbasis$ is the total number of basis functions. 
The RWG basis functions are second-order accurate~\cite[pp.\ 155--156]{warnick_2008}, and are defined for a triangle pair by
\begin{align*}
\boldsymbol{\Lambda}_{\srcidx}(\mathbf{x}) = \left\{
\begin{matrix}
\displaystyle\frac{\ell_{\srcidx}}{2A_{\srcidx}^+}\boldsymbol{\rho}_{\srcidx}^+, & \text{for }\mathbf{x}\in T_{\srcidx}^+ \\[1em]
\displaystyle\frac{\ell_{\srcidx}}{2A_{\srcidx}^-}\boldsymbol{\rho}_{\srcidx}^-, & \text{for }\mathbf{x}\in T_{\srcidx}^- \\[1em]
\mathbf{0}, & \text{otherwise}
\end{matrix}
\right.,
\end{align*}
where $\ell_{\srcidx}$ is the length of the edge shared by the triangle pair, and $A_{\srcidx}^+$ and $A_{\srcidx}^-$ are the areas of the triangles $T_{\srcidx}^+$ and $T_{\srcidx}^-$ associated with basis function $\srcidx$.  $\boldsymbol{\rho}_{\srcidx}^+$ denotes the vector from the vertex of $T_{\srcidx}^+$ opposite the shared edge to $\mathbf{x}$, and $\boldsymbol{\rho}_{\srcidx}^-$ denotes the vector to the vertex of $T_{\srcidx}^-$ opposite the shared edge from $\mathbf{x}$.

These basis functions ensure that $\mathbf{J}_h$ is tangent to the mesh when using planar triangular elements.
Additionally, along the shared edge of the triangle pair, the component of $\boldsymbol{\Lambda}_{\srcidx}(\mathbf{x})$ normal to that edge is unity.  Therefore, for a triangle edge shared by only two triangles, the component of $\mathbf{J}_h$ normal to that edge is $J_\srcidx$.  The solution is considered most accurate at the midpoint of the edge~\cite[pp.\ 155--156]{warnick_2008}; therefore, we measure the solution at the midpoints.

Defining $\mathbb{V}_h$ to be the span of RWG basis functions associated with the mesh on $S$, the Galerkin approximation of~\eqref{eq:sesquilinear} is now: find $\mathbf{J}_h\in\mathbb{V}_h$, such that
\begin{align}
a(\mathbf{J}_h,\boldsymbol{\Lambda}_{\testidx}) = b\big(\mathbf{E}^\mathcal{I} , \mathbf{H}^\mathcal{I} ,  \boldsymbol{\Lambda}_{\testidx}\big)
\label{eq:proj_disc}
\end{align}
for $i=1,\hdots,\nbasis$.  
Letting $\mathbf{J}^h$ denote the vector of coefficients used to construct $\mathbf{J}_h$~\eqref{eq:J_h}, \eqref{eq:proj_disc} can be written in matrix form as $\mathbf{Z}\mathbf{J}^h = \mathbf{V}$,
%
%
where $Z_{\testidx,\srcidx} = a(\boldsymbol{\Lambda}_{\srcidx},\boldsymbol{\Lambda}_{\testidx})$ \reviewerTwo{is the impedance matrix}, 
$J_{\srcidx}^h = J_{\srcidx}$ \reviewerTwo{is the current vector}, and
$V_{\testidx} =b\big(\mathbf{E}^\mathcal{I} , \mathbf{H}^\mathcal{I} ,  \boldsymbol{\Lambda}_{\testidx}\big)$ \reviewerTwo{is the excitation vector}.
%

%
%

%% file: mms.tex
\section{Manufactured Solutions} 
\label{sec:mms}

We define the residual functional for each test basis function as
\begin{align}
r_{\testidx}(\mathbf{u}) = a(\mathbf{u},\boldsymbol{\Lambda}_{\testidx}) -b\big(\mathbf{E}^\mathcal{I} , \mathbf{H}^\mathcal{I}, \boldsymbol{\Lambda}_{\testidx}\big).
\label{eq:res_func}
\end{align}
%
We can write the variational form~\eqref{eq:sesquilinear} in terms of~\eqref{eq:res_func} as
\begin{align}
r_{\testidx}(\mathbf{J}) = a(\mathbf{J},\boldsymbol{\Lambda}_{\testidx}) -b\big(\mathbf{E}^\mathcal{I} , \mathbf{H}^\mathcal{I}, \boldsymbol{\Lambda}_{\testidx}\big) = 0.
\label{eq:res}
\end{align}
Similarly, we can write the discretized problem~\eqref{eq:proj_disc} in terms of~\eqref{eq:res_func} as
\begin{align}
r_{\testidx}(\mathbf{J}_h) = a(\mathbf{J}_h,\boldsymbol{\Lambda}_{\testidx}) -b\big(\mathbf{E}^\mathcal{I} , \mathbf{H}^\mathcal{I}, \boldsymbol{\Lambda}_{\testidx}\big) = 0.
\label{eq:res_disc}
\end{align}
%

The method of manufactured solutions modifies~\eqref{eq:res_disc} to be
\begin{align}
r_{\testidx}(\mathbf{J}_h) = r_{\testidx}(\mathbf{J}_\text{MS}),
\label{eq:mms}
\end{align}
where $\mathbf{J}_\text{MS}$ is the manufactured solution, and $\mathbf{r}(\mathbf{J}_\text{MS})$ is computed exactly.
%

%

%
Inserting~\eqref{eq:res} and~\eqref{eq:res_disc} into~\eqref{eq:mms} yields
\begin{align}
a(\mathbf{J}_h,\boldsymbol{\Lambda}_{\testidx}) = a(\mathbf{J}_\text{MS},\boldsymbol{\Lambda}_{\testidx}).
\label{eq:proj_disc_mms}
\end{align}
%
%
However, instead of solving~\eqref{eq:proj_disc_mms}, we can equivalently solve~\eqref{eq:proj_disc} by setting 
\begin{align}
b\big(\mathbf{E}^\mathcal{I} , \mathbf{H}^\mathcal{I}, \boldsymbol{\Lambda}_{\testidx}\big) = a(\mathbf{J}_\text{MS},\boldsymbol{\Lambda}_{\testidx}).
\label{eq:H_mms_1}
\end{align}
Equation~\eqref{eq:H_mms_1} is satisfied by~\cite{freno_em_mms_2020}
\begin{align*}
\mathbf{E}^\mathcal{I} &{}= \frac{j}{\epsilon\omega} \int_{S'}\big[ k^2\mathbf{J}_\text{MS} (\mathbf{x}')G(\mathbf{x},\mathbf{x}') +\nabla'\cdot\mathbf{J}_\text{MS}(\mathbf{x}')\nabla G(\mathbf{x},\mathbf{x}')\big]dS', 
\end{align*}
which, from~\eqref{eq:grad_G}, is equivalent to
\begin{align}
\mathbf{E}^\mathcal{I} &{}= \frac{j}{\epsilon\omega} \int_{S'}\big[ k^2\mathbf{J}_\text{MS} (\mathbf{x}')G(\mathbf{x},\mathbf{x}') -\nabla'\cdot\mathbf{J}_\text{MS}(\mathbf{x}')\nabla' G(\mathbf{x},\mathbf{x}')\big]dS', 
\label{eq:E_i}
\end{align}
and~\cite{freno_mfie_2022}
\begin{align}
\mathbf{H}^\mathcal{I} = \frac{1}{2}\mathbf{J}_\text{MS}\times\mathbf{n} - \int_{S'}\mathbf{J}_\text{MS}(\mathbf{x}')\times\nabla'G(\mathbf{x},\mathbf{x}') dS'.
\label{eq:H_i}
\end{align}
%

\subsection{Solution-Discretization Error} 
\label{sec:sde}

In~\eqref{eq:proj_disc}, if the integrals in $a(\cdot,\cdot)$~\eqref{eq:a} and $b(\cdot,\cdot)$~\eqref{eq:b} are evaluated exactly, the only contribution to the discretization error is the solution-discretization error.  Solving for $\mathbf{J}^h$ enables us to compute the discretization error 
\begin{align}
\mathbf{e}_\mathbf{J} = \mathbf{J}^h - \mathbf{J}_n,
\label{eq:solution_error}
\end{align}
where $J_{n_\srcidx}$ denotes the component of $\mathbf{J}_\text{MS}$ flowing from $T_\srcidx^+$ to $T_\srcidx^-$.  The norm of~\eqref{eq:solution_error} has the property $\|\mathbf{e}_\mathbf{J}\|\le C_\mathbf{J} h^{p_\mathbf{J}}$, where
$C_\mathbf{J}$ is a function of the solution derivatives, $h$ is representative of the mesh size, and $p_\mathbf{J}$ is the order of accuracy.  By performing a mesh-convergence study of the norm of the discretization error, we can ensure the expected order of accuracy is obtained.  \reviewerTwo{For the RWG basis functions, the expectation is second-order accuracy $(p_\mathbf{J}=2)$ when the error is evaluated at the edge centers~\cite{warnick_2008}}.

\subsection{Numerical-Integration Error} 
\label{sec:nie}

In practice, the integrals in $a(\cdot,\cdot)$~\eqref{eq:a} and $b(\cdot,\cdot)$~\eqref{eq:b} are evaluated numerically, yielding the approximations $a^q(\cdot,\cdot)$ and $b^q(\cdot,\cdot)$.  $a^q(\cdot,\cdot)$ and $b^q(\cdot,\cdot)$ are obtained by integrating over each triangular element using quadrature, and generally incur a numerical-integration error.  Therefore, it is important to measure the numerical-integration error without contamination from the solution-discretization error.

In~\cite{freno_mfie_2022}, approaches are presented to isolate the numerical-integration error by canceling or eliminating the solution-discretization error.  In this paper, we cancel the solution-discretization error and measure the numerical-integration error from
\begin{alignat}{19}
&e_a&(\mathbf{J}_{h_\text{MS}}) = {} &a^q(\mathbf{J}_{h_\text{MS}}        ,\mathbf{J}_{h_\text{MS}}    ) {}-{} a(\mathbf{J}_{h_\text{MS}}        ,\mathbf{J}_{h_\text{MS}}    ),  \label{eq:a_error_cancel}\\
&e_b&(\mathbf{J}_{h_\text{MS}}) = {} & b^q\big(\mathbf{E}^\mathcal{I}_\text{MS},\mathbf{H}^\mathcal{I}_\text{MS},\mathbf{J}_{h_\text{MS}}\big) {}-{} b\big(\mathbf{E}^\mathcal{I}_\text{MS},\mathbf{H}^\mathcal{I}_\text{MS},\mathbf{J}_{h_\text{MS}}\big), \label{eq:b_error_cancel}
\end{alignat}
where $\mathbf{J}_{h_\text{MS}}$ is the basis-function representation of $\mathbf{J}_\text{MS}$, obtained from~\eqref{eq:J_h} by setting the coefficients $J_\srcidx$ equal to the normal component of $\mathbf{J}_\text{MS}$ at the midpoint of each edge associated with $\boldsymbol{\Lambda}_\srcidx(\mathbf{x})$.  The presence of the basis functions in the minuend and subtrahend of~\eqref{eq:a_error_cancel} and~\eqref{eq:b_error_cancel} cancels the solution-discretization error.
Equations~\eqref{eq:a_error_cancel} and~\eqref{eq:b_error_cancel} have the properties $|e_a(\mathbf{J}_{h_\text{MS}})| \le C_a h^{p_a}$ and $|e_b(\mathbf{J}_{h_\text{MS}})| \le C_b h^{p_b}$, where $C_a$ and $C_b$ are functions of the integrand derivatives, and $p_a$ and $p_b$ depend on the quadrature accuracy.

Reference~\cite{freno_mfie_2022} shows that $e_a(\mathbf{J}_{h_\text{MS}})$~\eqref{eq:a_error_cancel}  and $e_b(\mathbf{J}_{h_\text{MS}})$~\eqref{eq:b_error_cancel} are proportional to their influence on the discretization error $\mathbf{e}_\mathbf{J}$~\eqref{eq:solution_error}.

%% file: g_ms.tex
\section{Electromagnetic Integral Evaluations} 
\label{sec:g_ms}

As described in the introduction, integrals containing the Green's function~\eqref{eq:G} or its derivatives, such as those appearing in the manufactured electric field $\mathbf{E}^\mathcal{I}$~\eqref{eq:E_i} and magnetic field $\mathbf{H}^\mathcal{I}$~\eqref{eq:H_i} in $b\big(\mathbf{E}^\mathcal{I},\mathbf{H}^\mathcal{I},\boldsymbol{\Lambda}_{\testidx}\big)$~\eqref{eq:proj_disc}, are unable to be computed analytically.  Additionally, the singularity when $R\to 0$ complicates their accurate approximation, potentially contaminating convergence studies.  

In~\cite{freno_em_mms_2020} and~\cite{freno_mfie_2022}, this problem is mitigated by manufacturing the Green's function in terms of low, even powers of $R$, permitting the integrals in $a(\mathbf{J}_h,\boldsymbol{\Lambda}_{\testidx})$ and $b\big(\mathbf{E}^\mathcal{I},\mathbf{H}^\mathcal{I},\boldsymbol{\Lambda}_{\testidx}\big)$~\eqref{eq:proj_disc} to be computed analytically for many choices of $\mathbf{J}_\text{MS}$.  The drawbacks of the manufactured Green's function are the lack of physical realism and the practically singular matrices arising from terms containing integrals of the Green's function or its derivatives.  To mitigate the latter concern, an effective optimization approach was presented to provide a unique solution and detect coding errors.

In this paper, we optimally approximate the Green's function in terms of even powers of $R$.  This enables us to compute the integrals in~\eqref{eq:proj_disc} exactly, while maintaining some of the physical realism of the actual Green's function.  Additionally, the condition number of the \reviewerTwo{impedance matrix} remains low enough to conduct meaningful mesh-convergence studies. 

\subsection{Green's Function Approximation} 

We can write the Green's function~\eqref{eq:G} alternatively as
%
$G(R) = G_r(R) + j G_i(R)$,
%
where
\begin{align}
G_r(R) = \frac{\cos(kR)}{4\pi R}, \qquad\qquad 
G_i(R) = -\frac{\sin(kR)}{4\pi R}. \label{eq:Gri}
\end{align}
The Taylor series expansions of $G_r(R)$ and $G_i(R)$~\eqref{eq:Gri} at $R=0$ are
\begin{align}
G_r(R) = \frac{1}{4\pi}\sum_{n=0}^\infty \frac{(-1)^n k^{2n}}{(2n)!}R^{2n-1}, \qquad\qquad 
G_i(R) = -\frac{1}{4\pi}\sum_{n=0}^\infty \frac{(-1)^n k^{2n+1}}{(2n+1)!}R^{2n}. \label{eq:Gri_taylor}
\end{align}
\reviewerTwo{As $R\to 0$, integrands with negative or odd powers of $R$, such as those appearing in $G_r(R)$~\eqref{eq:Gri_taylor}, are singular~\cite{freno_em}.}

To avoid contamination from integrating singular integrands, we approximate $G(R)$ using even, nonnegative powers of $R$ for $R\in[\delta,\,R_m]$, where $R_m=\max_{\mathbf{x},\mathbf{x}'\in S} R$ is the maximum possible distance between two points on $S$ and $\delta\to 0$.  We introduce the inner product
\begin{align}
\langle u,v\rangle 
= \int_{\delta}^{R_m} u(R) \bar{v}(R) R dR,
\label{eq:inner_product}
\end{align}
with the norm $\|\cdot\|_R=\langle\cdot,\cdot\rangle^{1/2}$.
%
Let $\phi_n(R)$ denote the orthonormal basis arising from the functions $R^{2n}$ for $n\in\mathbb{N}_0$, subject to~\eqref{eq:inner_product}.  $G(R)$ is approximated by
\begin{align}
G(R) \approx \tilde{G}(R) = \sum_{n=0}^{n_m} G_n\phi_n(R), \label{eq:G_approx} 
\end{align}
where $n_m$ denotes the truncation of an infinite series, and the coefficients $G_n$ are obtained \reviewerTwo{by} minimizing 
$e_G^2 = \|\tilde{G}-G\|_R^2 = \|\tilde{G}_r-G_r\|_R^2 + \|\tilde{G}_i-G_i\|_R^2$.  By setting $\partial e_G/\partial G_n=0$,
%
$G_n = \langle G,\phi_n\rangle$. 

For $G_i$, $\|\tilde{G}_i-G_i\|_R^2$ and $G_{i_n}$ are bounded for $\delta=0$ in~\eqref{eq:inner_product}.  For $G_r$, $\|\tilde{G}_r-G_r\|_R^2$ is bounded for $\delta>0$ but not for $\delta=0$.  However, $G_{r_n}$ is bounded for $\delta\ge 0$.  Therefore, we compute the approximation~\eqref{eq:G_approx} for $\delta=0$ in~\eqref{eq:inner_product}.

\begin{figure}
\centering
\begin{subfigure}[b]{.49\textwidth}
\includegraphics[scale=.64,clip=true,trim=2.3in 0in 2.8in 0in]{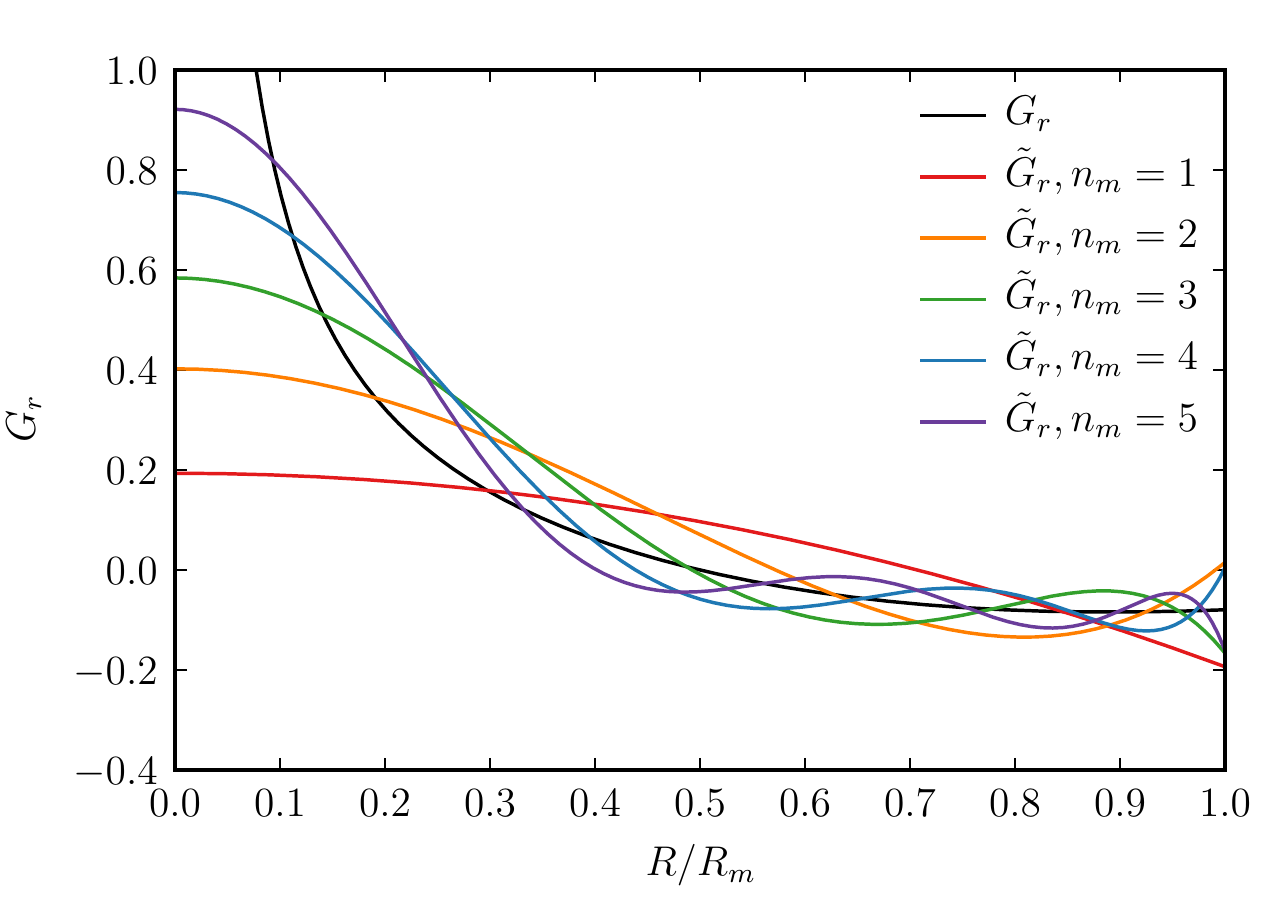}
\caption{$G_r$, \reviewerOne{$k=\pi/L$}\vpad}
\end{subfigure}
\hspace{0.25em}
\begin{subfigure}[b]{.49\textwidth}
\includegraphics[scale=.64,clip=true,trim=2.3in 0in 2.8in 0in]{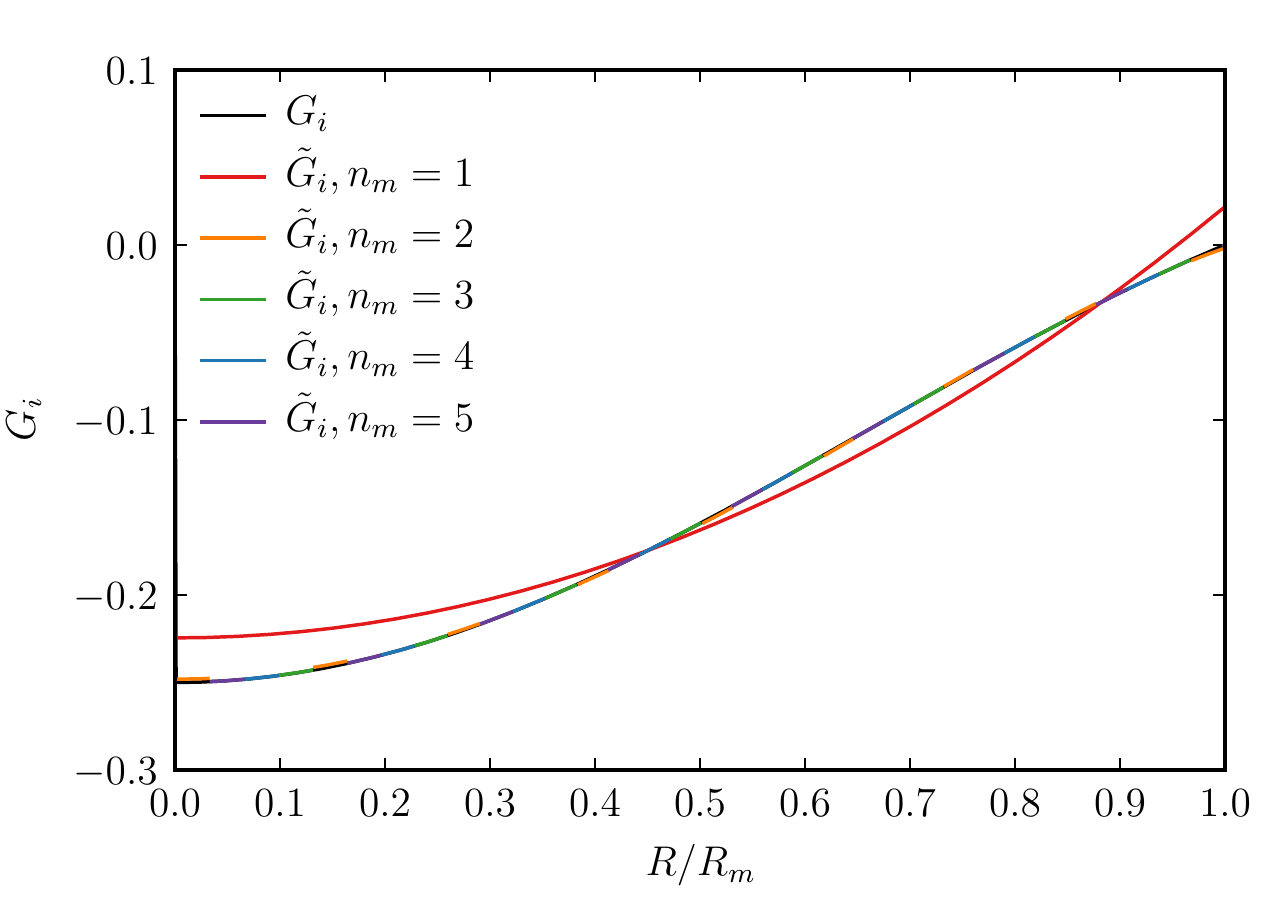}
\caption{$G_i$, \reviewerOne{$k=\pi/L$}\vpad}
\end{subfigure}
\\
\begin{subfigure}[b]{.49\textwidth}
\includegraphics[scale=.64,clip=true,trim=2.3in 0in 2.8in 0in]{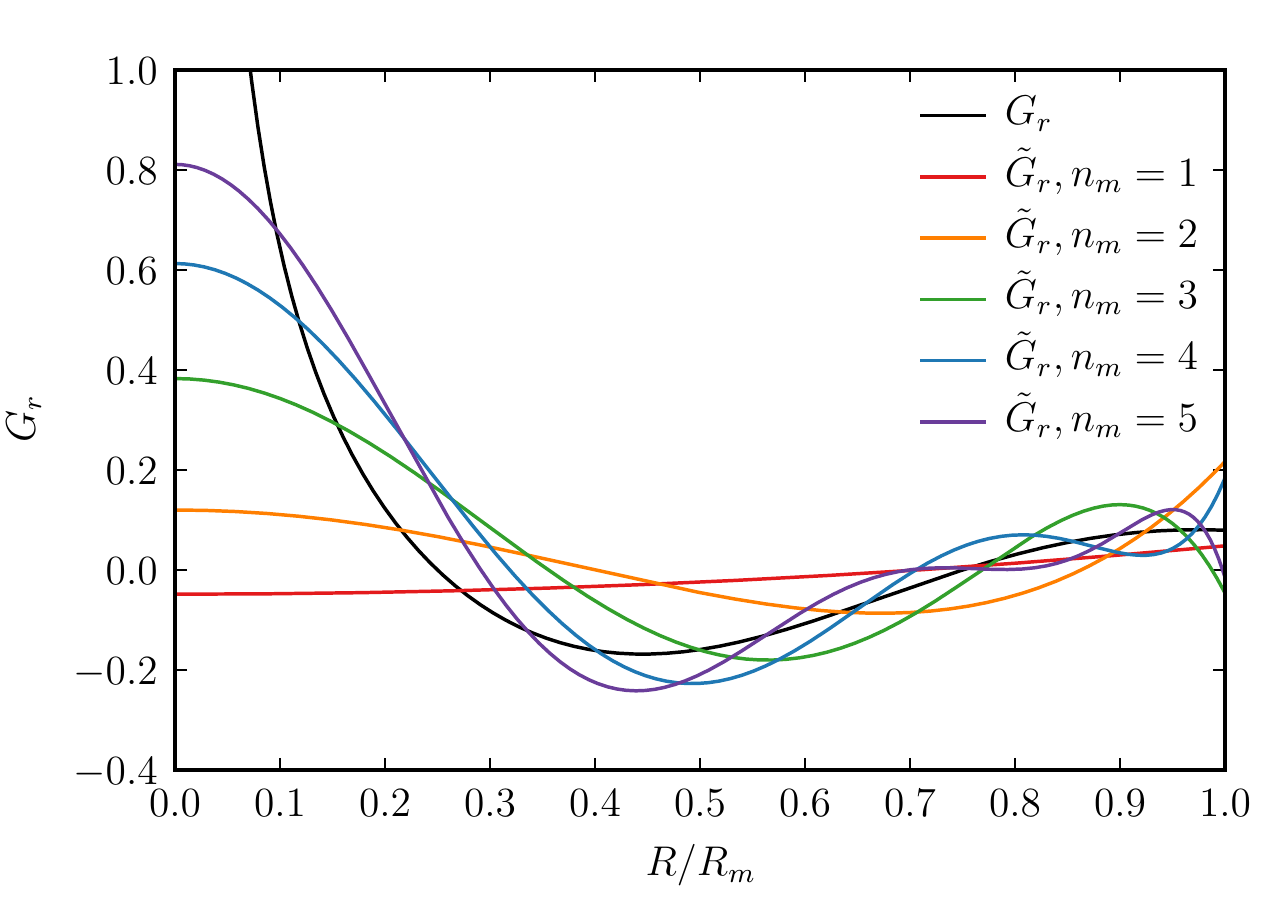}
\caption{$G_r$, \reviewerOne{$k=2\pi/L$}\vpad}
\end{subfigure}
\hspace{0.25em}
\begin{subfigure}[b]{.49\textwidth}
\includegraphics[scale=.64,clip=true,trim=2.3in 0in 2.8in 0in]{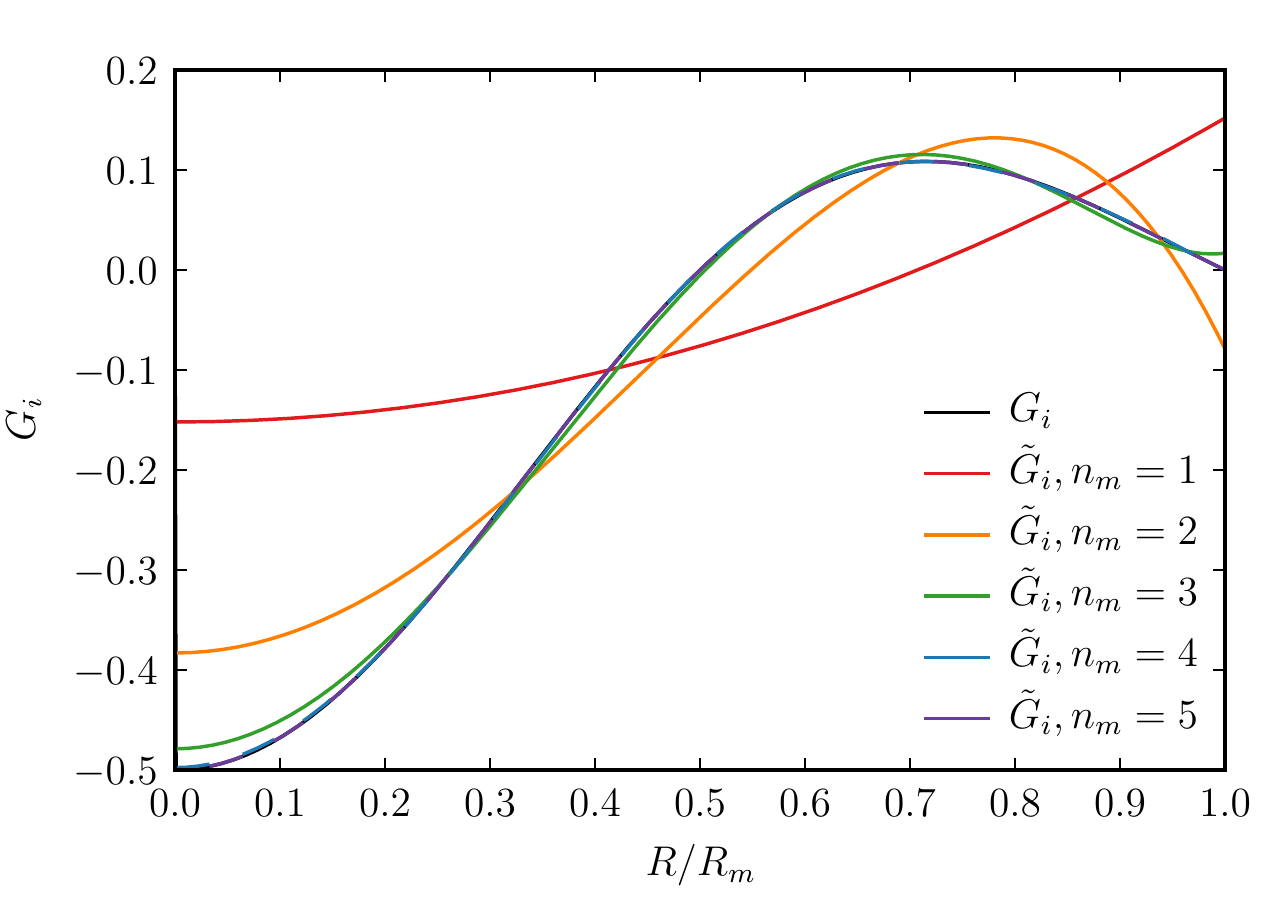}
\caption{$G_i$, \reviewerOne{$k=2\pi/L$}\vpad}
\end{subfigure}
\\
\begin{subfigure}[b]{.49\textwidth}
\includegraphics[scale=.64,clip=true,trim=2.3in 0in 2.8in 0in]{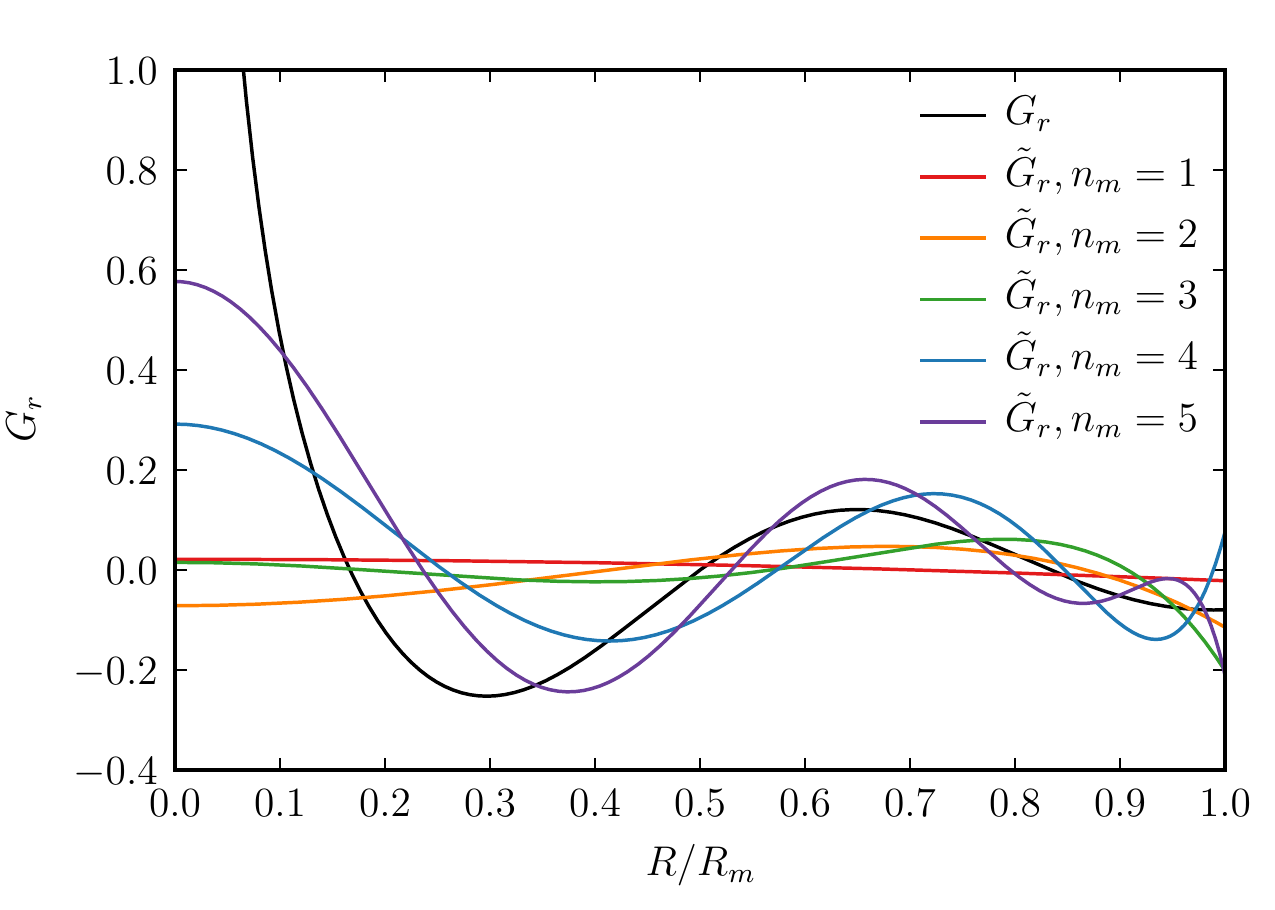}
\caption{$G_r$, \reviewerOne{$k=3\pi/L$}\vpad}
\end{subfigure}
\hspace{0.25em}
\begin{subfigure}[b]{.49\textwidth}
\includegraphics[scale=.64,clip=true,trim=2.3in 0in 2.8in 0in]{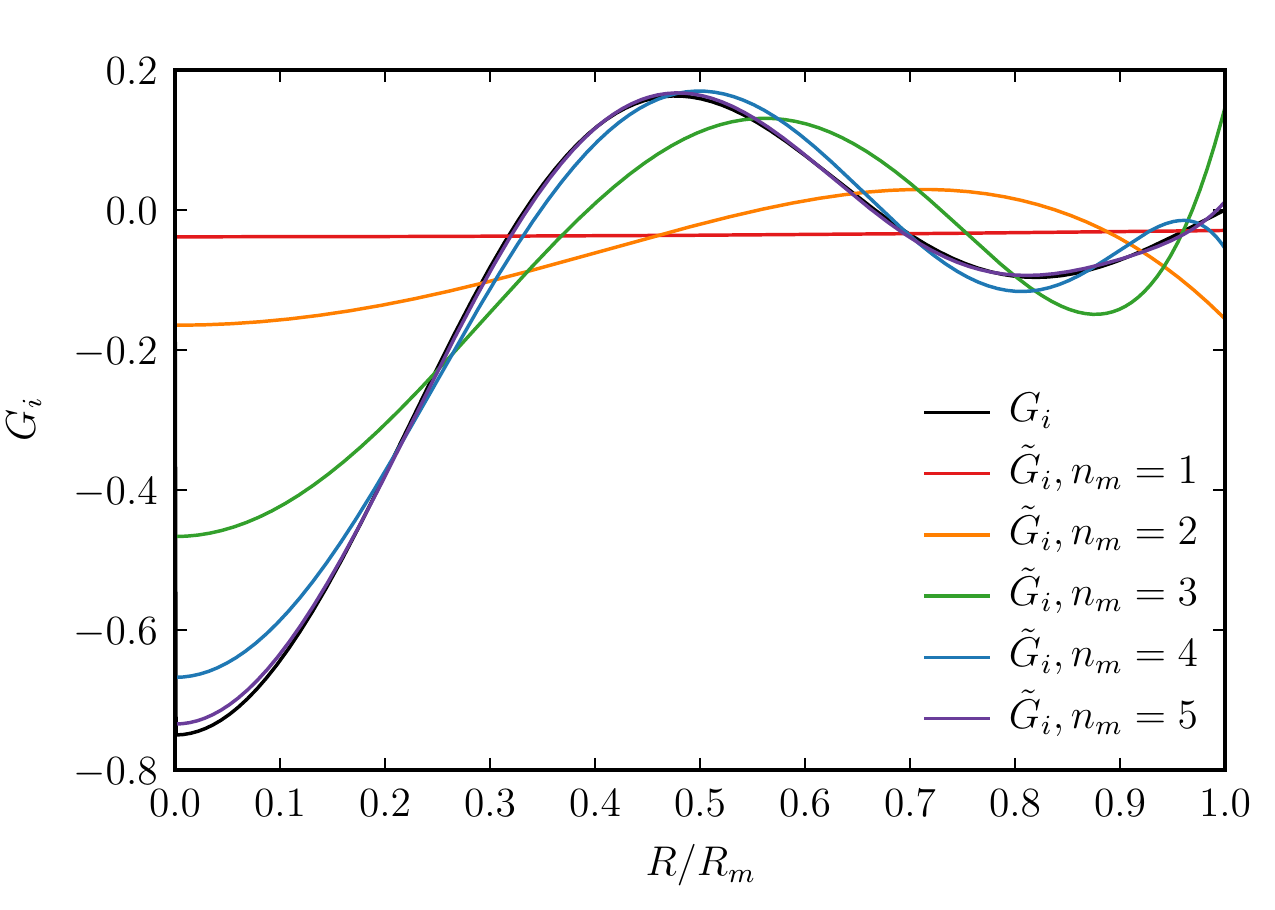}
\caption{$G_i$, \reviewerOne{$k=3\pi/L$}\vpad}
\end{subfigure}
\caption{Approximations $\tilde{G}=\tilde{G}_r+j\tilde{G}_i$ for $G=G_r+jG_i$ \reviewerOne{($L=R_m=1$~m)}.}
\vskip-\dp\strutbox
\label{fig:G}
\end{figure}

Figure~\ref{fig:G} shows $\tilde{G}$ for multiple choices of $k$ and $n_m$.  For $G_r$, increasing $n_m$ improves the approximation $\tilde{G}_r$ but introduces spurious oscillations that require finer discretizations for mesh-convergence studies.  Therefore, we approximate the Green's function with only the imaginary contribution by setting $G_{r_n}=0$.
%

\subsection{Incident Field Integral Evaluations} 

%

In $b\big(\mathbf{E}^\mathcal{I},\mathbf{H}^\mathcal{I},\boldsymbol{\Lambda}_{\testidx}\big)$~\eqref{eq:proj_disc}, $\mathbf{E}^\mathcal{I}$~\eqref{eq:E_i} and $\mathbf{H}^\mathcal{I}$~\eqref{eq:H_i} take the form
%
%
%
\begin{align}
\mathbf{E}^\mathcal{I} &{}= \frac{j}{\epsilon\omega}\big(k^2 \mathbf{I}_{\mathcal{E}_\mathbf{A}} - \mathbf{I}_{\mathcal{E}_\Phi}\big), \label{eq:E_i_2}
\\
\mathbf{H}^\mathcal{I} &{}= \frac{1}{2}\mathbf{J}_\text{MS}\times\mathbf{n} - \mathbf{I}_\mathcal{M},
\label{eq:H_i_2}
\end{align}
where
%
\begin{align}
\mathbf{I}_{\mathcal{E}_\mathbf{A}}(\mathbf{x}) &{}= \int_{S'} \mathbf{J}_\text{MS}(\mathbf{x}') G(\mathbf{x},\mathbf{x}') dS', 
\label{eq:i_1}\\
\mathbf{I}_{\mathcal{E}_\Phi}(\mathbf{x}) &{}= \int_{S'} \nabla'\cdot\mathbf{J}_\text{MS}(\mathbf{x}') \nabla'G(\mathbf{x},\mathbf{x}') dS', 
\label{eq:i_2}\\
\mathbf{I}_\mathcal{M}(\mathbf{x}) &{}= \int_{S'} \mathbf{J}_\text{MS}(\mathbf{x}')\times\nabla'G(\mathbf{x},\mathbf{x}') dS'
\label{eq:i_3}
\end{align}
are the integrals that contain the Green's function.    
\reviewerTwo{To evaluate these integrals, we replace $G$ with $\tilde{G}$~\eqref{eq:G_approx}}, which can be written alternatively as 
\begin{align}
\tilde{G}(R) = \sum_{n=0}^{n_m} \tilde{G}_n R^{2n}, \label{eq:G_approx2} 
\end{align}
such that
\begin{align}
\nabla'\tilde{G}(R) = -\frac{\tilde{G}'(R)}{R}\mathbf{R} = -2\mathbf{R}\sum_{n=1}^{n_m} n \tilde{G}_n R^{2(n-1)}. \label{eq:G_approx_grad} 
\end{align}
%
%
Inserting~\eqref{eq:G_approx2} into~\eqref{eq:i_1} and inserting~\eqref{eq:G_approx_grad} into~\eqref{eq:i_2} and~\eqref{eq:i_3}, \eqref{eq:i_1}--\eqref{eq:i_3} become
\begin{alignat}{9}
\mathbf{I}_{\mathcal{E}_\mathbf{A}}(\mathbf{x}) &{}= &&\sum_{n=0}^{n_m}  &&\tilde{G}_n \int_{S'} R^{2n} \mathbf{J}_\text{MS}(\mathbf{x}') dS', \label{eq:i_e_a}
\\
\mathbf{I}_{\mathcal{E}_\Phi}(\mathbf{x}) &{}= -2&&\sum_{n=1}^{n_m} n&&\tilde{G}_n \int_{S'} R^{2(n-1)} \nabla'\cdot\mathbf{J}_\text{MS}(\mathbf{x}') \mathbf{R} dS', \label{eq:i_e_phi}
\\
\mathbf{I}_\mathcal{M}(\mathbf{x}) &{}= -2&&\sum_{n=1}^{n_m} n&&\tilde{G}_n  \int_{S'}  R^{2(n-1)}\mathbf{J}_\text{MS}(\mathbf{x}')\times\mathbf{R}dS'. \label{eq:i_m}
\end{alignat}
The evaluation of~\eqref{eq:i_e_a}--\eqref{eq:i_m} is discussed in~\ref{app:integrals} for the cases presented in Section~\ref{sec:results}.

%% file: results.tex
\section{Numerical Examples} 
\label{sec:results}

\begin{figure}
\centering
\includegraphics[scale=.215,clip=true,trim=0in 0in 0in 0in]{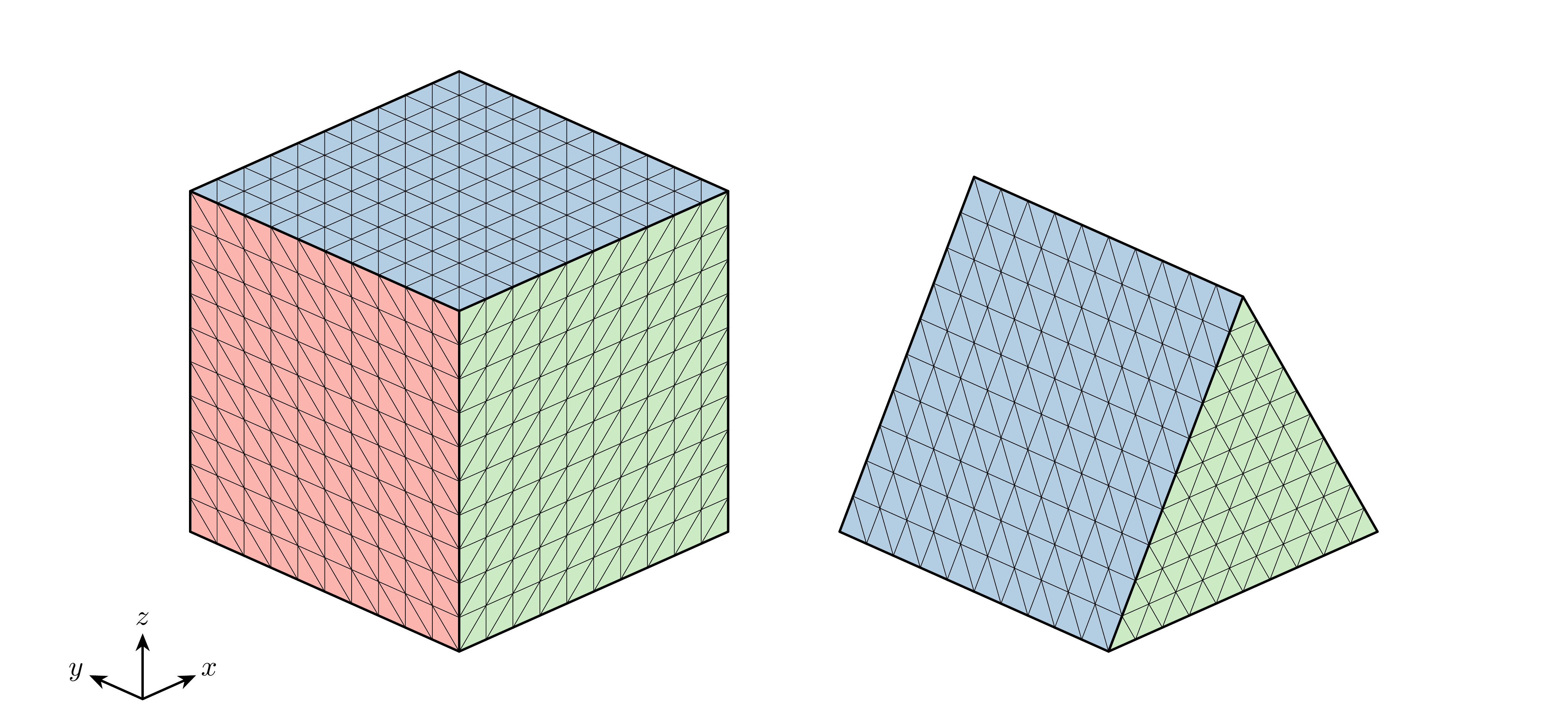}
\caption{Dimetric view of meshes for the cube with $\ntriangles=1200$ (left) and triangular prism with $\ntriangles=800$ (right).}
\vskip-\dp\strutbox
\label{fig:3d_domains}
\end{figure}

\begin{figure}
\vspace{10em}
\centering
\includegraphics[scale=.215,clip=true,trim=0.25in 0.5in 0.25in 0in]{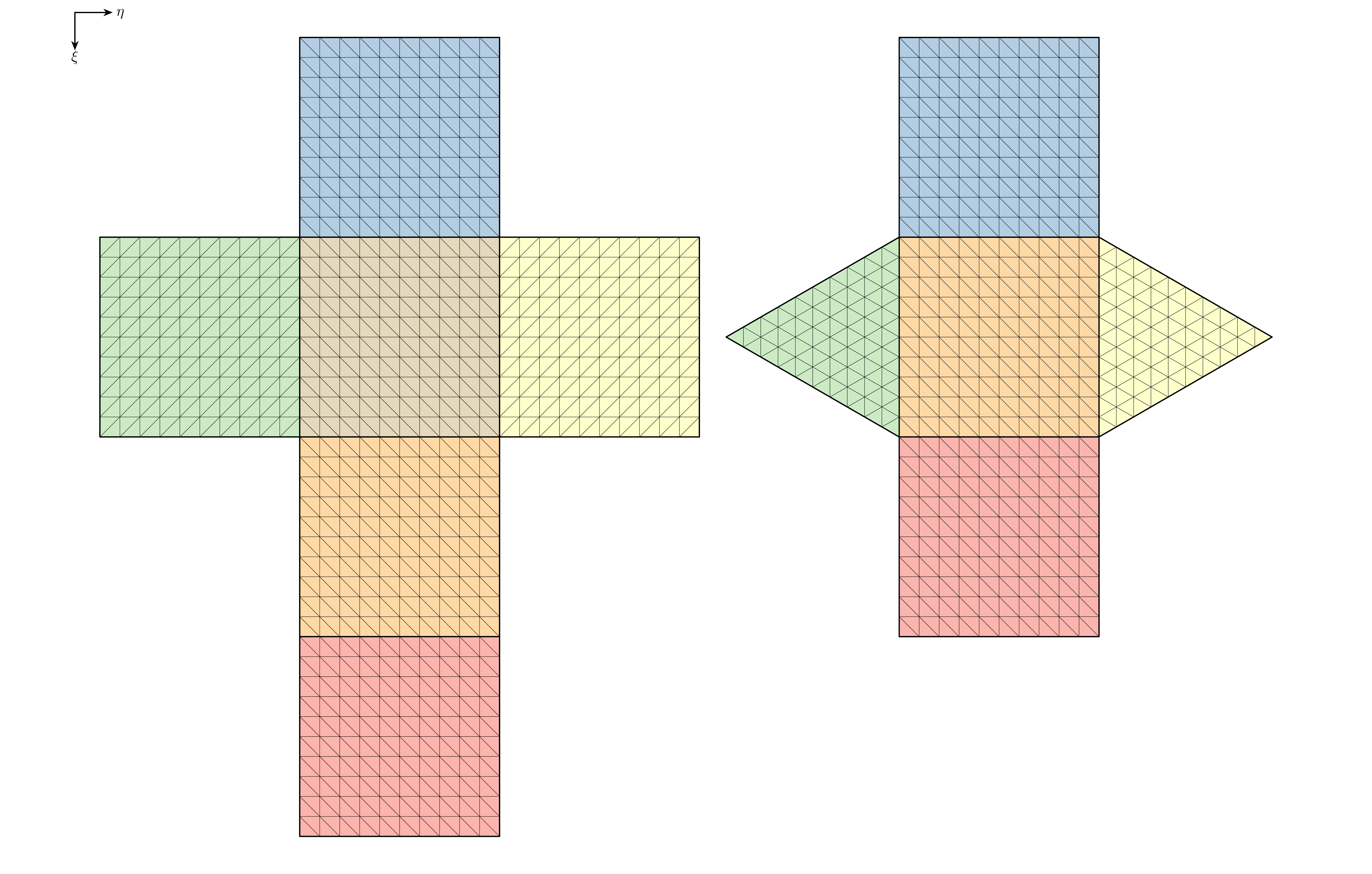}
\caption{Net view of meshes for the cube with $\ntriangles=1200$ (left) and the triangular prism with $\ntriangles=800$ (right).}
\vskip-\dp\strutbox
\label{fig:2d_domains}
\end{figure}

In this section, we demonstrate the approaches described in Section~\ref{sec:mms} by isolating and measuring the solution-discretization error (Section~\ref{sec:sde}) and the numerical-integration error (Section~\ref{sec:nie}).
%
%
%
We consider two domains: a cube and an equilateral triangular prism, each with all edges of length \reviewerOne{$L = 1$~m}.  These domains are shown in Figures~\ref{fig:3d_domains} and~\ref{fig:2d_domains} with the total number of triangles $\ntriangles=1200$ for the cube and $\ntriangles=800$ for the triangular prism.  

For both domains, we introduce a coordinate system $\boldsymbol{\xi}$ that is fixed to the $n_s$ surfaces for which $\mathbf{n}\cdot\mathbf{e}_y=0$. 
For the cube, \reviewerOne{$\xi\in[0,\,4 L]$} is perpendicular to $y$, wrapping around the surfaces for which $\mathbf{n}\cdot\mathbf{e}_y=0$, beginning at \reviewerOne{$x=0$} and \reviewerOne{$z=L$}.
For the triangular prism, \reviewerOne{$\xi\in[0,\,3L]$} is perpendicular to $y$, wrapping around the surfaces for which $\mathbf{n}\cdot\mathbf{e}_y=0$, beginning at \reviewerOne{$x=z=0$}.
For both cases, \reviewerOne{$\eta\in[0,\,L]$} is equal to $y$.
$\boldsymbol{\xi}$ is depicted in Figure~\ref{fig:2d_domains}, which shows the nets of these domains.
The transformation between $\boldsymbol{\xi}$ and $\mathbf{x}$ for the two domains, which is defined in a piecewise manner for each of the $n_s$ surfaces, is listed in Table~\ref{tab:transformations}.  Table~\ref{tab:transformations} additionally lists the $\xi$ domain $[\xi_{a_j},\,\xi_{b_j}]$ for each of the $n_s$ surfaces.

\begin{figure}
\centering
\includegraphics[scale=.215,clip=true,trim=0in 0in 0in 0in]{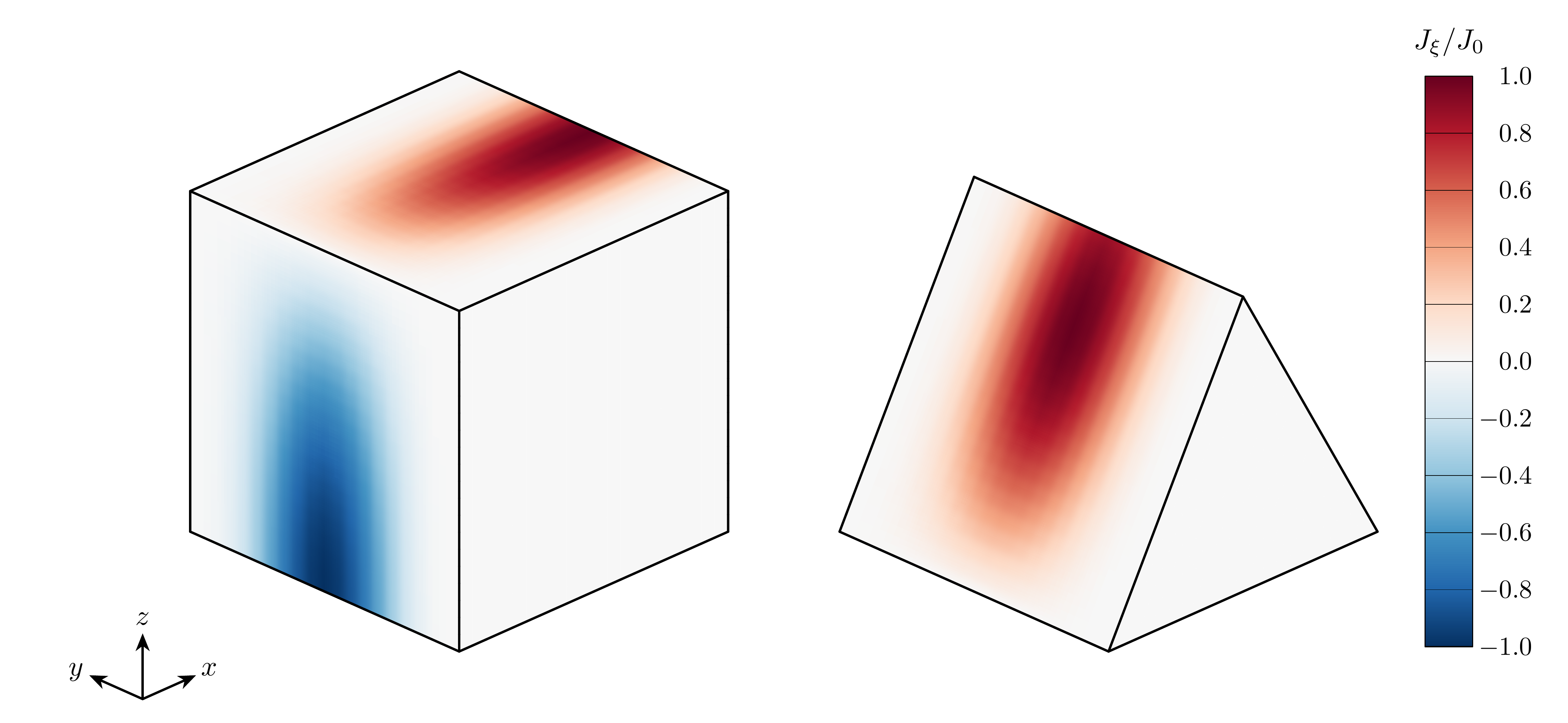}
\caption{\strut Dimetric view of manufactured surface current density $\mathbf{J}_\text{MS}$ for the cube (left) and triangular prism (right).}
\vskip-\dp\strutbox
\label{fig:3d_solutions}
\end{figure}

\begin{figure}
\centering
\includegraphics[scale=.215,clip=true,trim=0.25in 0.5in 0.25in 0in]{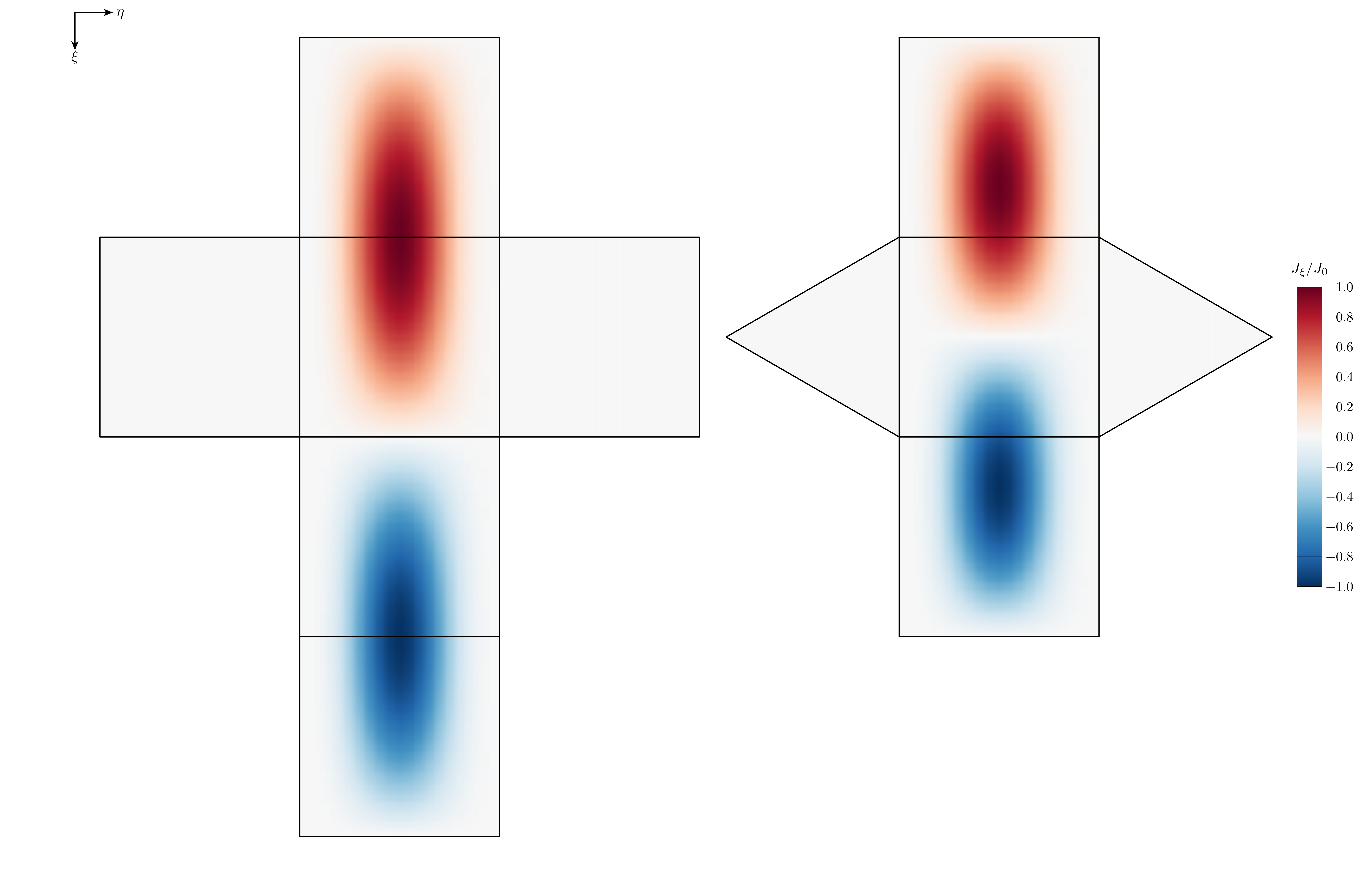}
\caption{\strut Net view of manufactured surface current density $\mathbf{J}_\text{MS}$ for the cube (left) and triangular prism (right).}
\vskip-\dp\strutbox
\label{fig:2d_solutions}
\end{figure}

We manufacture the surface current density $\mathbf{J}_\text{MS}(\mathbf{x}) = J_\xi(\boldsymbol{\xi})\mathbf{e}_\xi$, where 
\begin{align}
J_\xi(\boldsymbol{\xi})=J_0\left\{\begin{matrix}
\sin(\beta\xi\reviewerOne{/L})\sin^3(\pi\eta\reviewerOne{/L}), & \text{for }\mathbf{n}\cdot\mathbf{e}_y=0 \\[.5em]
0, & \text{for }\mathbf{n}\cdot\mathbf{e}_y\ne 0
\end{matrix}\right.,
\label{eq:jxi}
\end{align}
$J_0=1$~A/m, and \reviewerOne{$\beta=\pi/2$} for the cube and \reviewerOne{$\beta=2\pi/3$} for the triangular prism.  In the $\mathbf{x}$-coordinate system, $\mathbf{e}_\xi=(\partial\mathbf{x}/\partial\xi)_j$. 
\reviewerTwo{Equation~\eqref{eq:jxi} is chosen because it is of class $C^2$ and its oscillations are minimal, such that finer meshes are not required for mesh-convergence studies.}
Figures~\ref{fig:3d_solutions} and~\ref{fig:2d_solutions} show plots of~\eqref{eq:jxi}. Additionally we set the permeability and permittivity of the surrounding medium to those of free space: $\mu = 1.25663706212\times 10^{-6}$~N/A$^2$ and $\epsilon = 8.8541878128\times 10^{-12}$~F/m.

\begin{table}
\centering
\begin{tabular}{c c c c c c}
\toprule
    & & \multicolumn{2}{c}{Cube} & \multicolumn{2}{c}{Triangular Prism} \\
    \cmidrule(r){3-4} \cmidrule(l){5-6}
$j$ & $[\xi_{a_j},\,\xi_{b_j}]$ & $\mathbf{x}_j(\boldsymbol{\xi})$ & $\displaystyle \bigg(\frac{\partial \mathbf{x}}{\partial \boldsymbol{\xi}}\bigg)_j$ & $\mathbf{x}_j(\boldsymbol{\xi})$ & $\displaystyle \bigg(\frac{\partial \mathbf{x}}{\partial \boldsymbol{\xi}}\bigg)_j$\\[1em] \midrule
1   & $[0,\,1] \reviewerOne{L}$ &
$\displaystyle\bigg(\frac{\partial \mathbf{x}}{\partial \boldsymbol{\xi}}\bigg)_j\boldsymbol{\xi} + \left\{\begin{matrix}\pn0 \\ \pn0 \\ \pn1\end{matrix}\right\} \reviewerOne{L}$ & 
$\displaystyle \left[\begin{matrix}\pn1 & \pn0 & \pn0 \\ \pn0 & \pn1 & \pn0 \\ \pn0 & \pn0 & \pn1\end{matrix}\right]$ &
$\displaystyle\bigg(\frac{\partial \mathbf{x}}{\partial \boldsymbol{\xi}}\bigg)_j\boldsymbol{\xi} + \left\{\begin{matrix}\pn0 \\ \pn0 \\ \pn0\end{matrix}\right\} \reviewerOne{L}$ &
$\displaystyle \left[\begin{array}{@{} r @{} c @{} l c r @{} c @{}l @{}} 1 & / & 2 & \pn0 & -\sqrt{3} & / & 2 \\ & 0 & & \pn1 & & 0 \\ \pn\sqrt{3} &/ & 2 & \pn0 & 1 &/ & 2
\end{array}\right]$ 
\\[1.5em]
2   & $[1,\,2] \reviewerOne{L}$ &
$\displaystyle\bigg(\frac{\partial \mathbf{x}}{\partial \boldsymbol{\xi}}\bigg)_j\boldsymbol{\xi} + \left\{\begin{matrix}\pn1 \\ \pn0 \\ \pn2\end{matrix}\right\} \reviewerOne{L}$ & 
$\displaystyle \left[\begin{matrix} \pn0 & \pn0 & \pn1 \\ \pn0 & \pn1 & \pn0 \\ -1 & \pn0 & \pn0\end{matrix}\right]$ &
$\displaystyle\bigg(\frac{\partial \mathbf{x}}{\partial \boldsymbol{\xi}}\bigg)_j\boldsymbol{\xi} + \left\{\begin{matrix}\pn0 \\ \pn0 \\ \sqrt{3}\end{matrix}\right\} \reviewerOne{L}$ & 
$\displaystyle \left[\begin{array}{@{} r @{} c @{} l c r @{} c @{}l @{}} 1 & / & 2 & \pn0 & \pn\sqrt{3} & / & 2 \\ & 0 & & \pn1 & & 0 \\ -\sqrt{3} & / & 2 & \pn0 & 1 & / & 2\end{array}\right]$
\\[1.5em]
3   & $[2,\,3] \reviewerOne{L}$ &
$\displaystyle\bigg(\frac{\partial \mathbf{x}}{\partial \boldsymbol{\xi}}\bigg)_j\boldsymbol{\xi} + \left\{\begin{matrix}\pn3 \\ \pn0 \\ \pn0\end{matrix}\right\} \reviewerOne{L}$ & 
$\displaystyle \left[\begin{matrix} -1 & \pn0 & \pn0 \\ \pn0 & \pn1 & \pn0 \\ \pn0 & \pn0 & -1\end{matrix}\right]$ & 
$\displaystyle\bigg(\frac{\partial \mathbf{x}}{\partial \boldsymbol{\xi}}\bigg)_j\boldsymbol{\xi} + \left\{\begin{matrix}\pn3 \\ \pn0 \\ \pn0\end{matrix}\right\} \reviewerOne{L}$ & 
$\displaystyle \left[\begin{array}{@{} r @{} c @{} l c r @{} c @{}l @{}} 
                  - & 1 & & \pn0 &                     & 0 & \pz \\ 
\phantom{-\sqrt{3}} & 0 & \pz  & \pn1 & \phantom{-\sqrt{3}} & 0       \\ 
                    & 0 & & \pn0 & -                   & 1 
\end{array}\right]$
\\[1.5em]
4   & $[3,\,4] \reviewerOne{L}$ &
$\displaystyle\bigg(\frac{\partial \mathbf{x}}{\partial \boldsymbol{\xi}}\bigg)_j\boldsymbol{\xi} + \left\{\begin{matrix}\pn0 \\ \pn0 \\ -3\end{matrix}\right\} \reviewerOne{L}$ & 
$\displaystyle \left[\begin{matrix} \pn0 & \pn0 & -1 \\ \pn0 & \pn1 & \pn0 \\ \pn1 & \pn0 & \pn0\end{matrix}\right]$
\\
\bottomrule
\end{tabular}
\caption{Transformations between $\boldsymbol{\xi}$ and $\mathbf{x}$ for the cube and triangular prism.}
\label{tab:transformations}
\end{table}

\begin{table}
\centering
\begin{tabular}{c c c c c c}
\toprule
Maximum & Number & Convergence &
Maximum & Number & Convergence \\
integrand degree & of points & rate &
integrand degree & of points & rate \\
\cmidrule(r){1-3} \cmidrule(l){4-6}
1 & \pz1 & $\mathcal{O}(h^2)$ & \pz7 & 13 & $\mathcal{O}(h^{8\pz}) $\\
2 & \pz3 & $\mathcal{O}(h^4)$ & \pz8 & 16 & $\mathcal{O}(h^{10})$\\
3 & \pz4 & $\mathcal{O}(h^4)$ & \pz9 & 19 & $\mathcal{O}(h^{10})$\\
4 & \pz6 & $\mathcal{O}(h^6)$ &   10 & 25 & $\mathcal{O}(h^{12})$\\
5 & \pz7 & $\mathcal{O}(h^6)$ &   11 & 27 & $\mathcal{O}(h^{12})$\\
6 &   12 & $\mathcal{O}(h^8)$ &   12 & 33 & $\mathcal{O}(h^{14})$\\
\bottomrule
\end{tabular}
\caption{Polynomial triangle quadrature properties.}
\vskip-\dp\strutbox
\label{tab:dunavant_properties}
\end{table}

\begin{figure}
\centering
\begin{subfigure}[b]{.49\textwidth}
\centering
\input{six_a.tex}
\caption{Maximum polynomial degree: 4}
\label{fig:quad_optimal_6}
\end{subfigure}
\begin{subfigure}[b]{.49\textwidth}
\centering
\input{six_b.tex}
\caption{Maximum polynomial degree: 3}
\label{fig:quad_suboptimal_6}
\end{subfigure}
\caption{6-point quadrature rules.}
\vskip-\dp\strutbox
\label{fig:quadrature}
\end{figure}
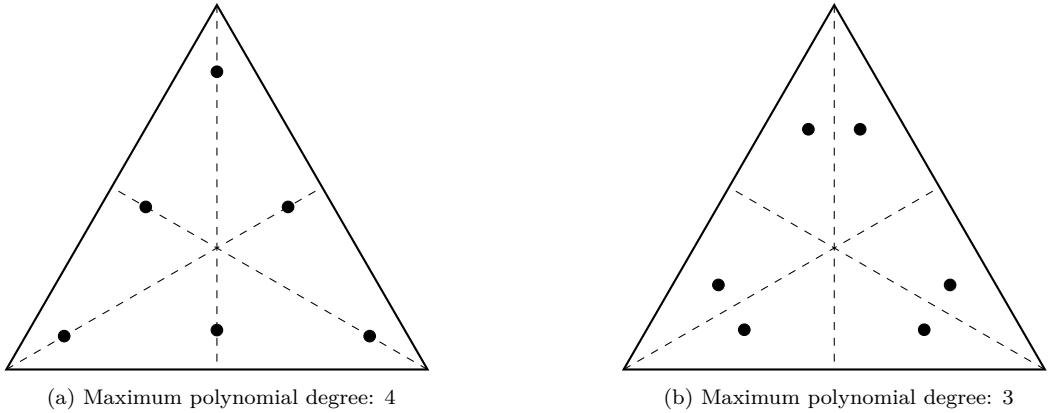

When solving~\eqref{eq:proj_disc}, numerical integration is performed using polynomial quadrature rules for triangles.  For multiple quadrature point amounts, Table~\ref{tab:dunavant_properties} lists the maximum polynomial degree of the integrand the points can integrate exactly~\cite{lyness_1975,dunavant_1985}, as well as the convergence rates of the errors for inexact integrations of nonsingular integrands.  These properties correspond to the optimal point locations and weights.  Figure~\ref{fig:quad_optimal_6} shows the optimal 6-point quadrature rule, which can exactly integrate polynomials up to degree 4.

To evaluate $b\big(\mathbf{E}^\mathcal{I},\mathbf{H}^\mathcal{I}, \boldsymbol{\Lambda}_{\testidx}\big)$ in~\eqref{eq:proj_disc}, we use $\tilde{G}$~\eqref{eq:G_approx} to compute the integrals $\mathbf{I}_{\mathcal{E}_\mathbf{A}}$~\eqref{eq:i_1}, $\mathbf{I}_{\mathcal{E}_\Phi}$~\eqref{eq:i_2}, and $\mathbf{I}_\mathcal{M}$~\eqref{eq:i_3} in $\mathbf{E}^\mathcal{I}$~\eqref{eq:E_i_2} and $\mathbf{H}^\mathcal{I}$~\eqref{eq:H_i_2} analytically, as shown in~\ref{app:integrals}.  
Setting $n_m=5$, $\mathbf{I}_{\mathcal{E}_\mathbf{A}}$, $\mathbf{I}_{\mathcal{E}_\Phi}$, and $\mathbf{I}_\mathcal{M}$ yield polynomials in $\mathbf{x}$ of degrees 10, 9, and 9, respectively.  
In $b\big(\mathbf{E}^\mathcal{I} , \mathbf{H}^\mathcal{I}, \boldsymbol{\Lambda}_{\testidx}\big)$, multiplication with $\boldsymbol{\Lambda}_{\testidx}$ increases these degrees by a power.  Therefore, a 27-point polynomial quadrature rule, which can exactly evaluate integrals up to degree 11, can be used to evaluate the integrals.

To evaluate $a(\mathbf{J}_h,\boldsymbol{\Lambda}_{\testidx})$ in~\eqref{eq:proj_disc}, we note that $\tilde{G}$~\eqref{eq:G_approx} is a polynomial in $\mathbf{x}$ and $\mathbf{x}'$ of degree 10, and $\nabla'\tilde{G}$ is a polynomial in $\mathbf{x}$ and $\mathbf{x}'$ of degree 9.  In $a(\mathbf{J}_h,\boldsymbol{\Lambda}_{\testidx})$, multiplication with $\boldsymbol{\Lambda}_{\testidx}(\mathbf{x})$ and $\boldsymbol{\Lambda}_{\srcidx}(\mathbf{x}')$ increases these degrees by a power, such that a 27-point polynomial quadrature rule can be used to evaluate the integrals with respect to $\mathbf{x}$ and $\mathbf{x}'$.

\subsection{Solution-Discretization Error} 

To isolate and measure the solution-discretization error, we proceed with the assessment described in Section~\ref{sec:sde}.  As stated previously, integrals on both sides of~\eqref{eq:proj_disc} are computed exactly with a 27-point polynomial quadrature rule.  

\begin{figure}
\centering
\begin{subfigure}[b]{.49\textwidth}
\includegraphics[scale=.64,clip=true,trim=2.3in 0in 2.8in 0in]{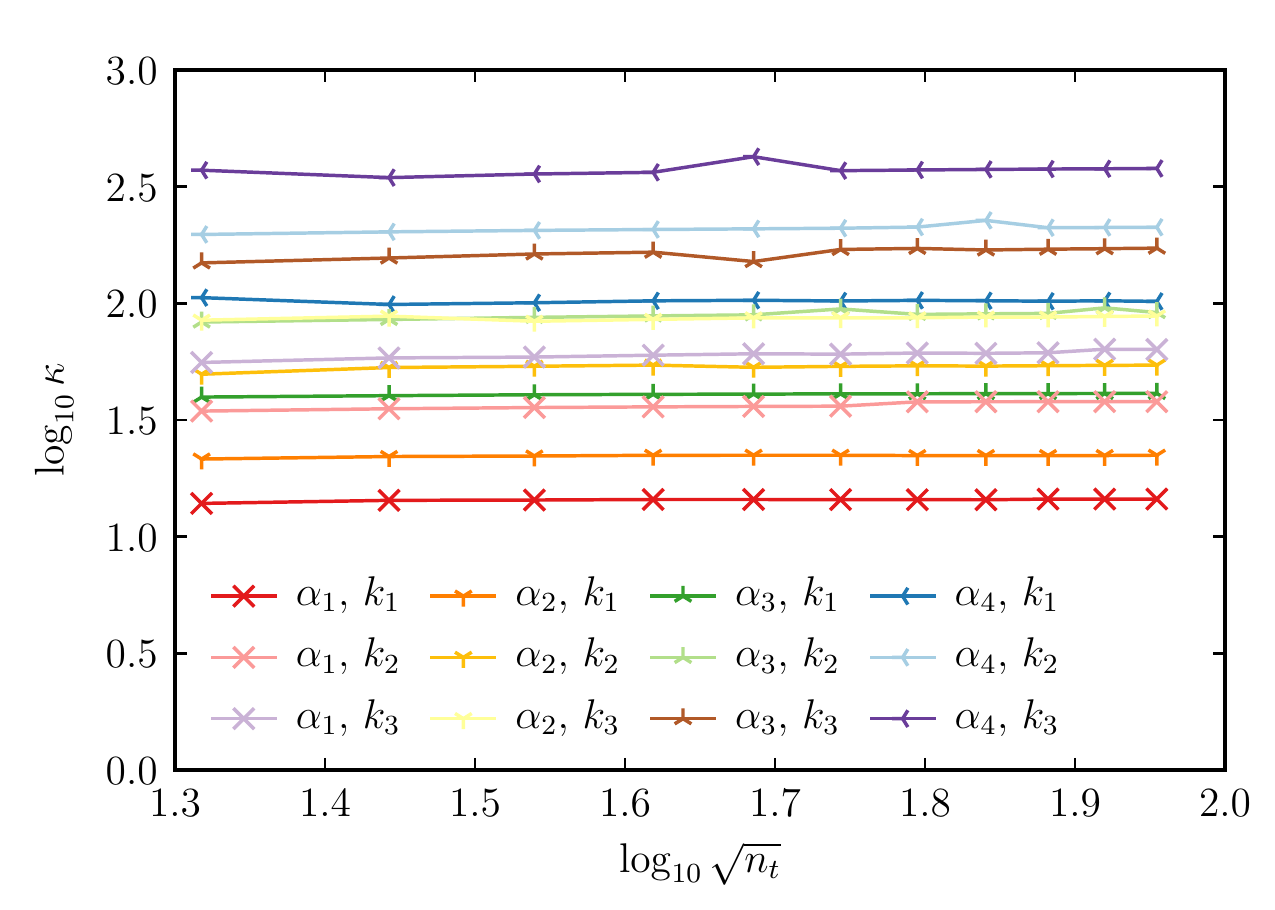}
\caption{Cube\vpad}
\end{subfigure}
\hspace{0.25em}
\begin{subfigure}[b]{.49\textwidth}
\includegraphics[scale=.64,clip=true,trim=2.3in 0in 2.8in 0in]{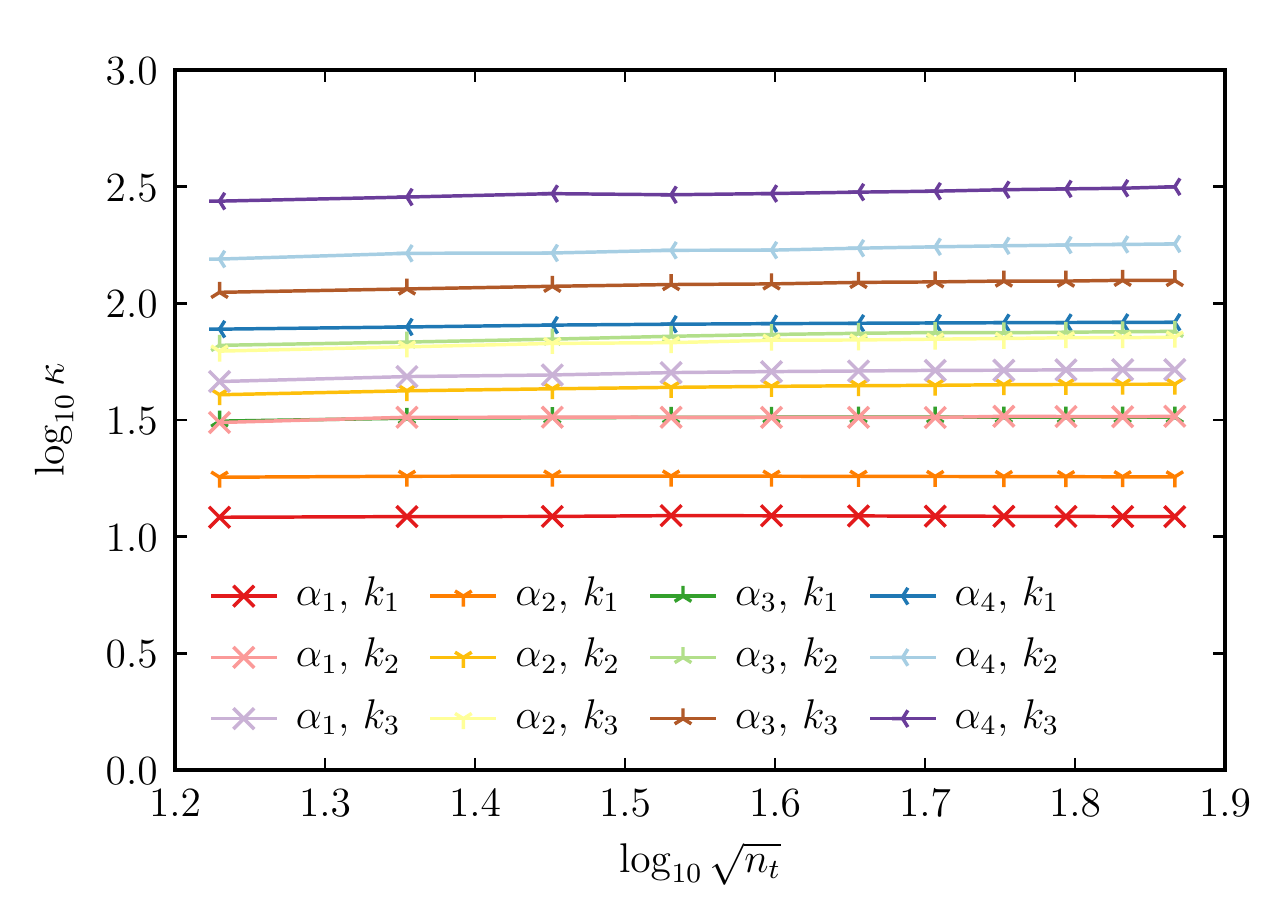}
\caption{Triangular Prism\vpad}
\end{subfigure}
\caption{Solution-discretization error: Condition numbers \reviewerTwo{of the impedance matrix $\mathbf{Z}$} \reviewerOne{for $\alpha_m=m/5$, $k_n=n\pi/L$}.}
\vskip-\dp\strutbox
\label{fig:p1_cond}
\end{figure}

To account for potential disparities in the magnitudes of the terms, we consider four values of $\alpha$: \reviewerOne{$\alpha_m=m/5$, for $m=1,\hdots,4$}, and three values of $k$: \reviewerOne{$k_n=n\pi/L$, for $n=1,\hdots,3$}.  For both domains, Figure~\ref{fig:p1_cond} shows how the condition numbers $\kappa$ \reviewerTwo{of the impedance matrix $\mathbf{Z}$} from these 12 combinations vary with respect to mesh size.  The condition numbers are low enough to conduct meaningful convergence studies. 

\begin{figure}
\centering
\begin{subfigure}[b]{.49\textwidth}
\includegraphics[scale=.64,clip=true,trim=2.3in 0in 2.8in 0in]{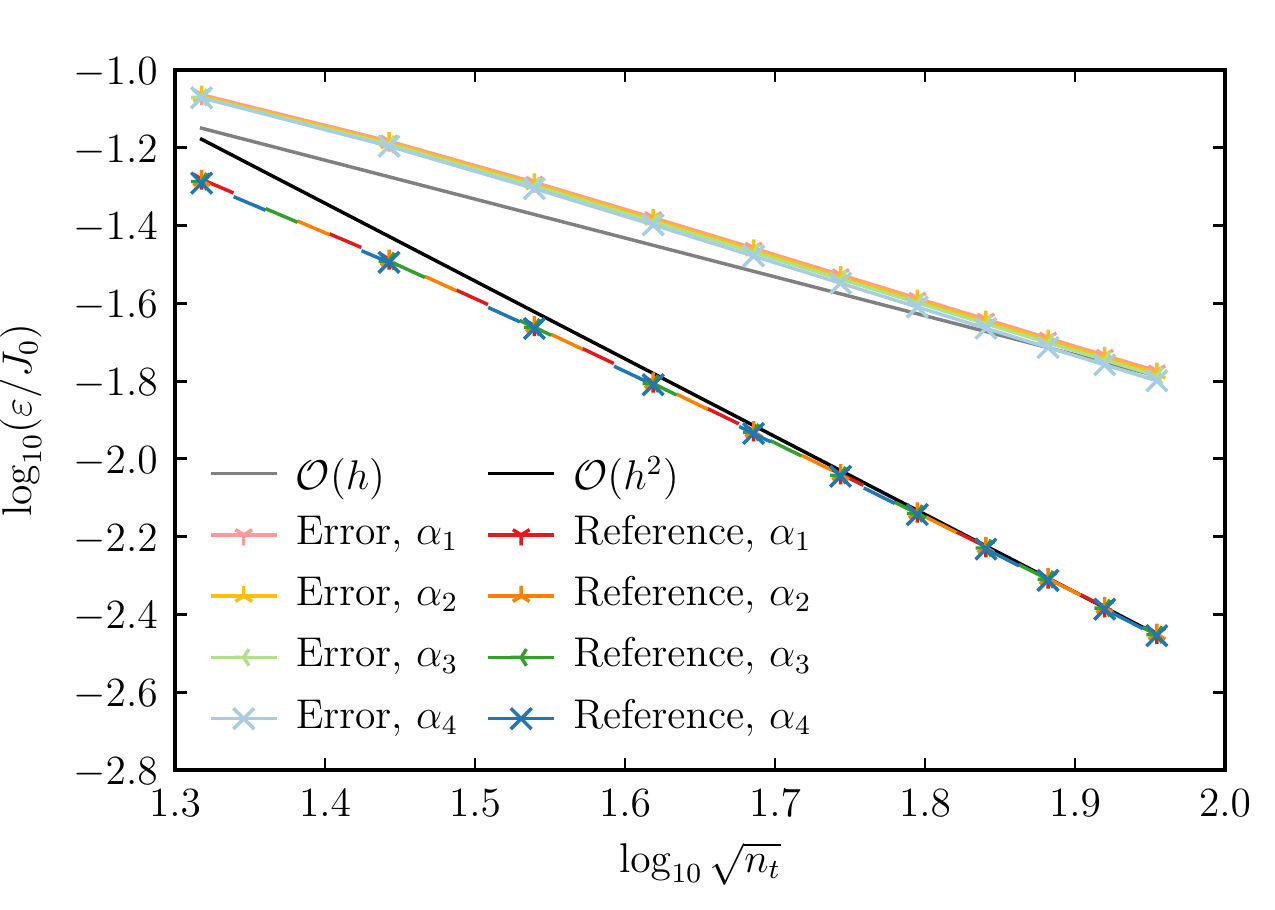}
\caption{Cube, \reviewerOne{$\alpha_m=m/5$, $k=\pi/L$}\vpad}
\end{subfigure}
\hspace{0.25em}
\begin{subfigure}[b]{.49\textwidth}
\includegraphics[scale=.64,clip=true,trim=2.3in 0in 2.8in 0in]{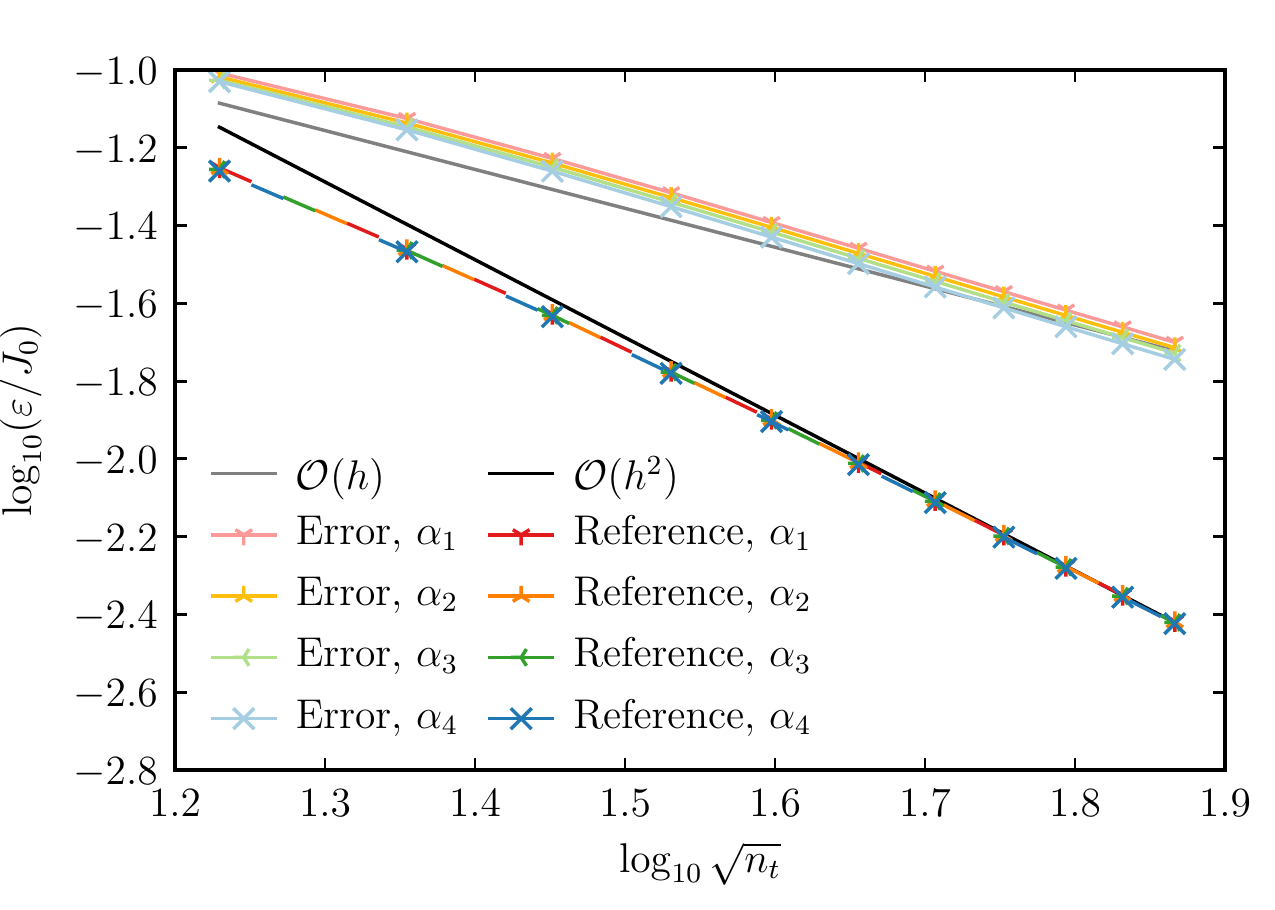}
\caption{Triangular Prism, \reviewerOne{$\alpha_m=m/5$, $k=\pi/L$}\vpad}
\end{subfigure}
\\
\begin{subfigure}[b]{.49\textwidth}
\includegraphics[scale=.64,clip=true,trim=2.3in 0in 2.8in 0in]{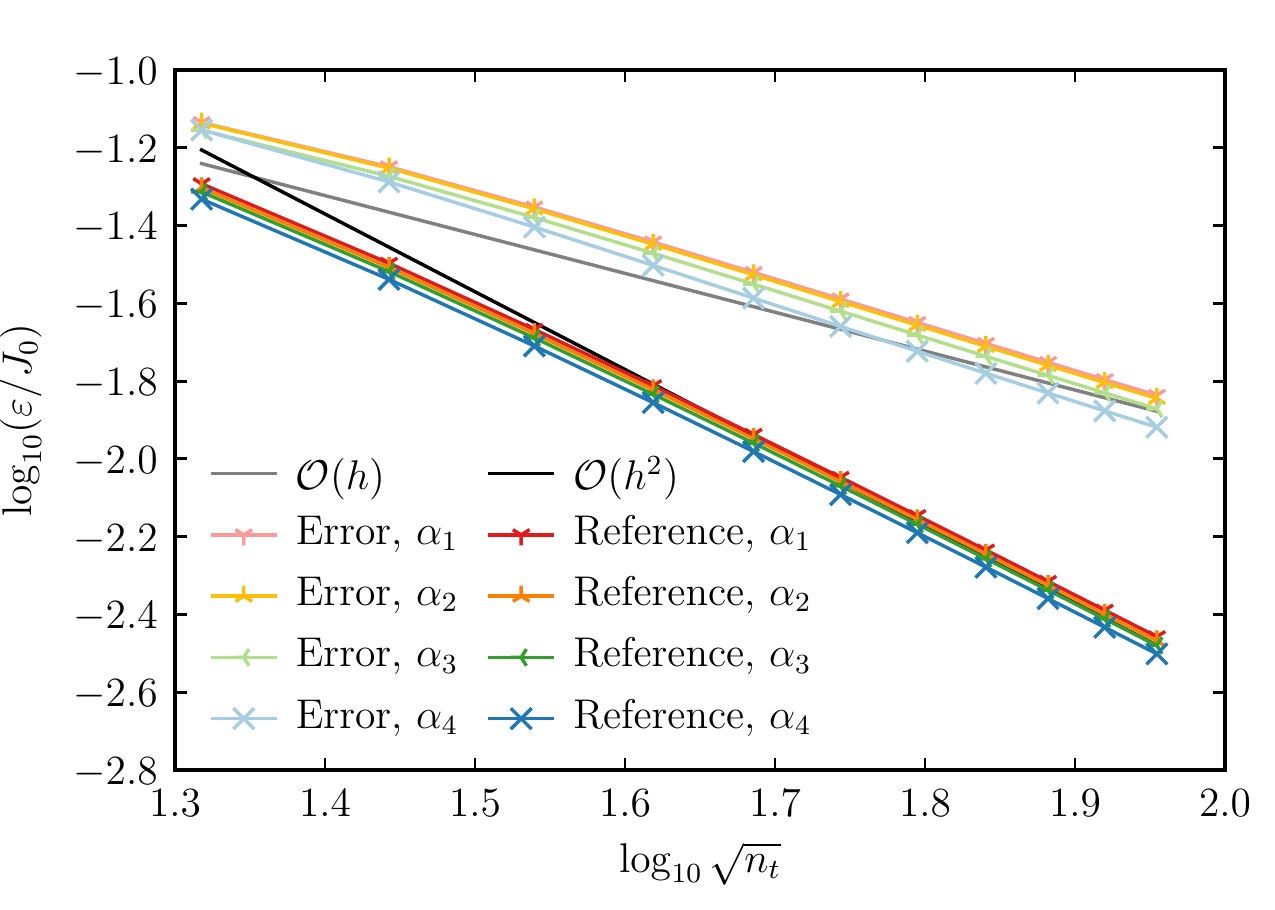}
\caption{Cube, \reviewerOne{$\alpha_m=m/5$, $k=2\pi/L$}\vpad}
\end{subfigure}
\hspace{0.25em}
\begin{subfigure}[b]{.49\textwidth}
\includegraphics[scale=.64,clip=true,trim=2.3in 0in 2.8in 0in]{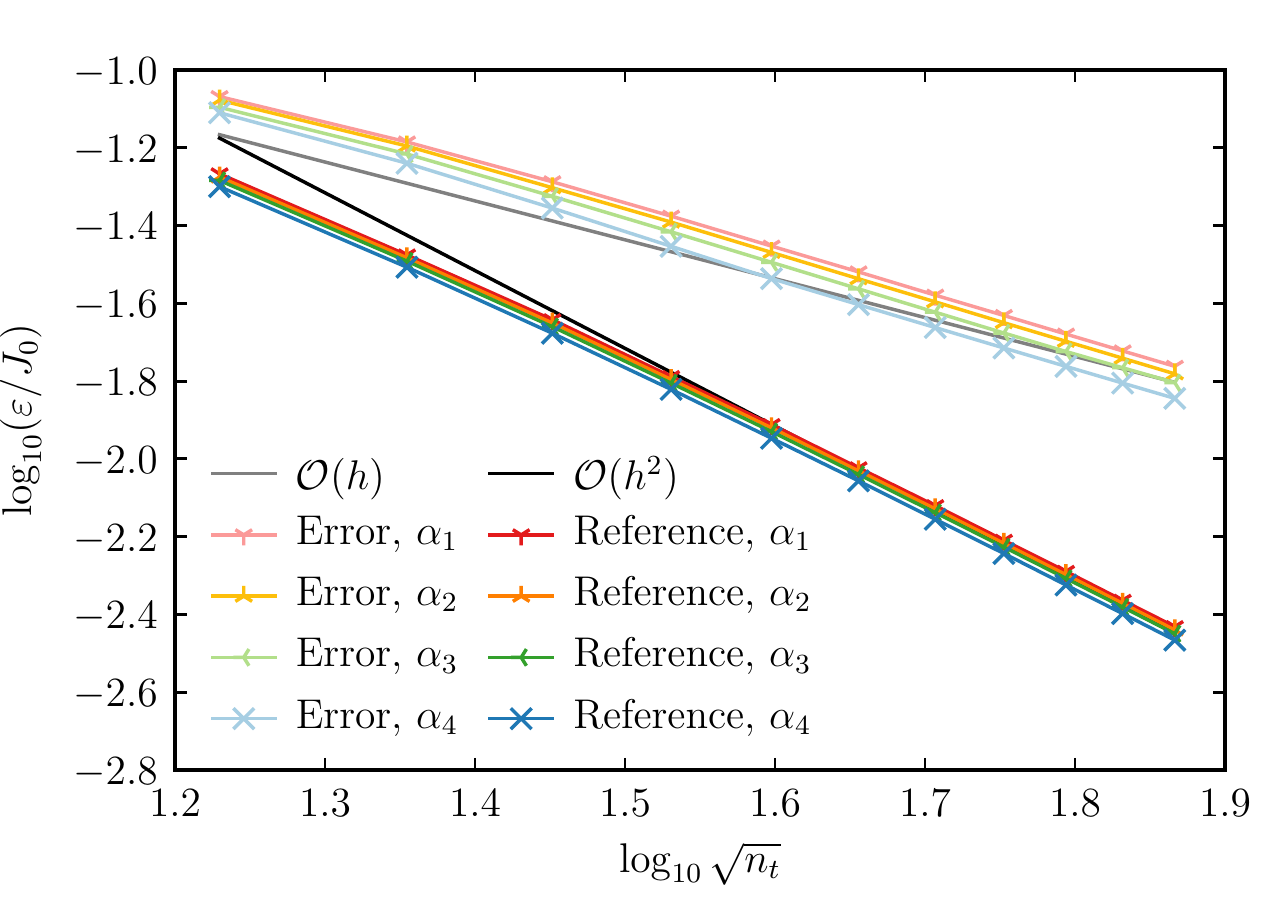}
\caption{Triangular Prism, \reviewerOne{$\alpha_m=m/5$, $k=2\pi/L$}\vpad}
\end{subfigure}
\\
\begin{subfigure}[b]{.49\textwidth}
\includegraphics[scale=.64,clip=true,trim=2.3in 0in 2.8in 0in]{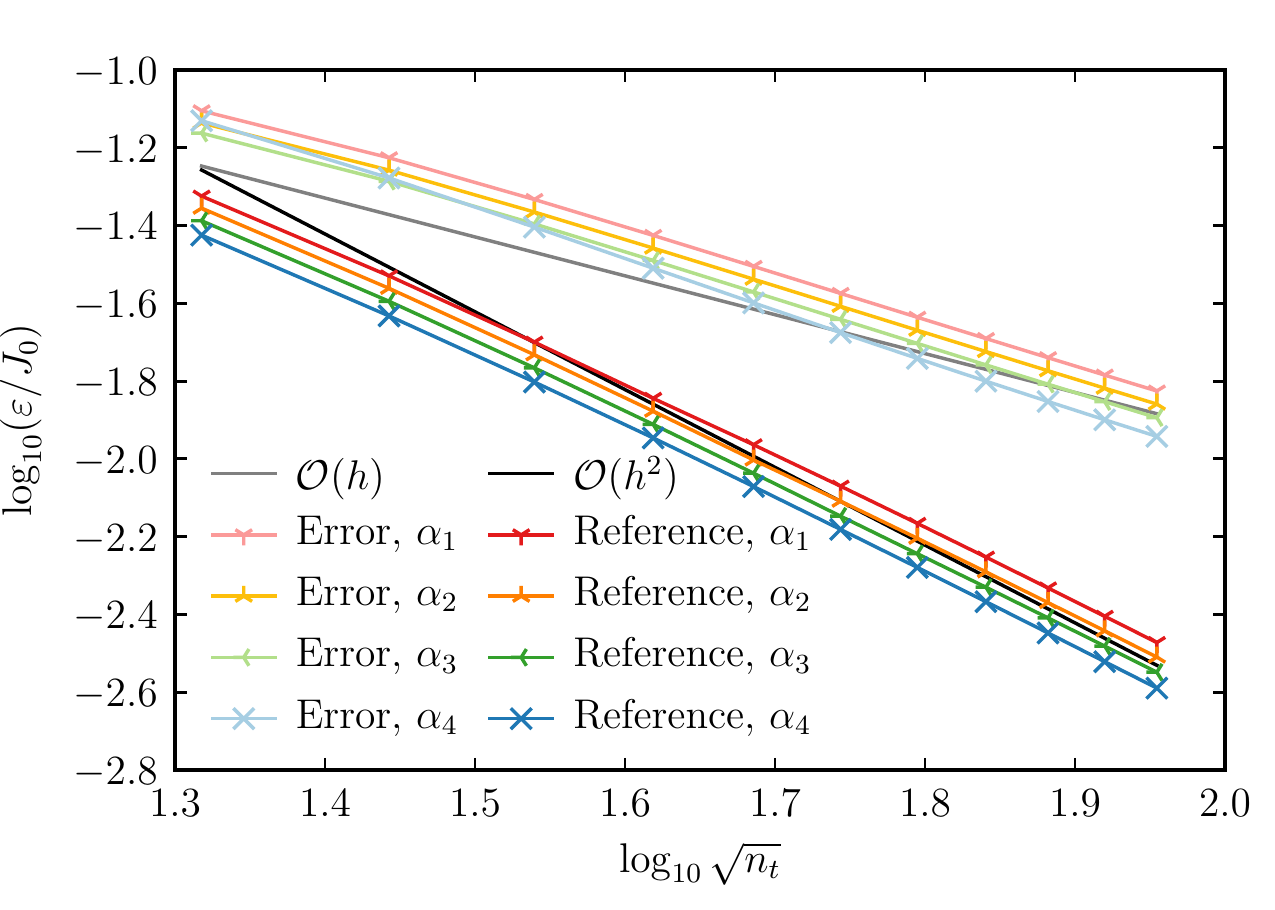}
\caption{Cube, \reviewerOne{$\alpha_m=m/5$, $k=3\pi/L$}\vpad}
\end{subfigure}
\hspace{0.25em}
\begin{subfigure}[b]{.49\textwidth}
\includegraphics[scale=.64,clip=true,trim=2.3in 0in 2.8in 0in]{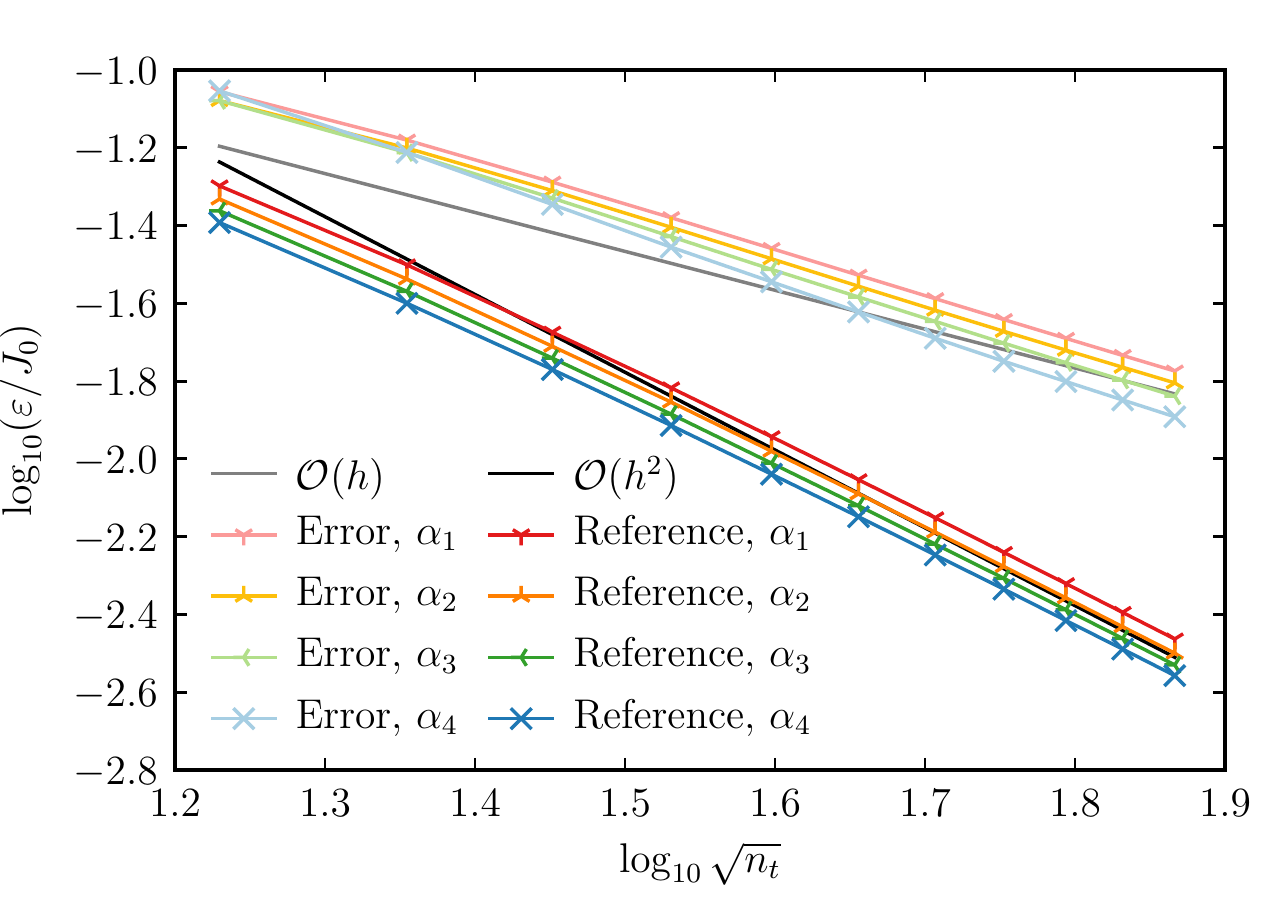}
\caption{Triangular Prism, \reviewerOne{$\alpha_m=m/5$, $k=3\pi/L$}\vpad}
\end{subfigure}
\caption{Solution-discretization error: $\varepsilon={\|\mathbf{e}_\mathbf{J}\|}_\infty$.}
\vskip-\dp\strutbox
\label{fig:p1_error}
\end{figure}

Figure~\ref{fig:p1_error} shows the $L^\infty$-norm of the discretization error ${\|\mathbf{e}_\mathbf{J}\|}_\infty$~\eqref{eq:solution_error} arising from only the solution-discretization error for both domains.  The convergence rates for these cases, which are denoted as reference cases, are all $\mathcal{O}(h^2)$ as expected.

To test the ability to detect a coding error, we multiply the diagonal elements of the \reviewerTwo{impedance} matrix by 
$\Big(1 + \ntriangles^{-1/2}\Big)$. 
In Figure~\ref{fig:p1_error}, these cases with the coding error are $\mathcal{O}(h)$, instead of $\mathcal{O}(h^2)$.  Therefore, the coding error is detected.

\subsection{Numerical-Integration Error} 

To isolate and measure the numerical-integration error, we perform the assessments described in Section~\ref{sec:nie}.

\begin{figure}
\centering
\begin{subfigure}[b]{.49\textwidth}
\includegraphics[scale=.64,clip=true,trim=2.3in 0in 2.8in 0in]{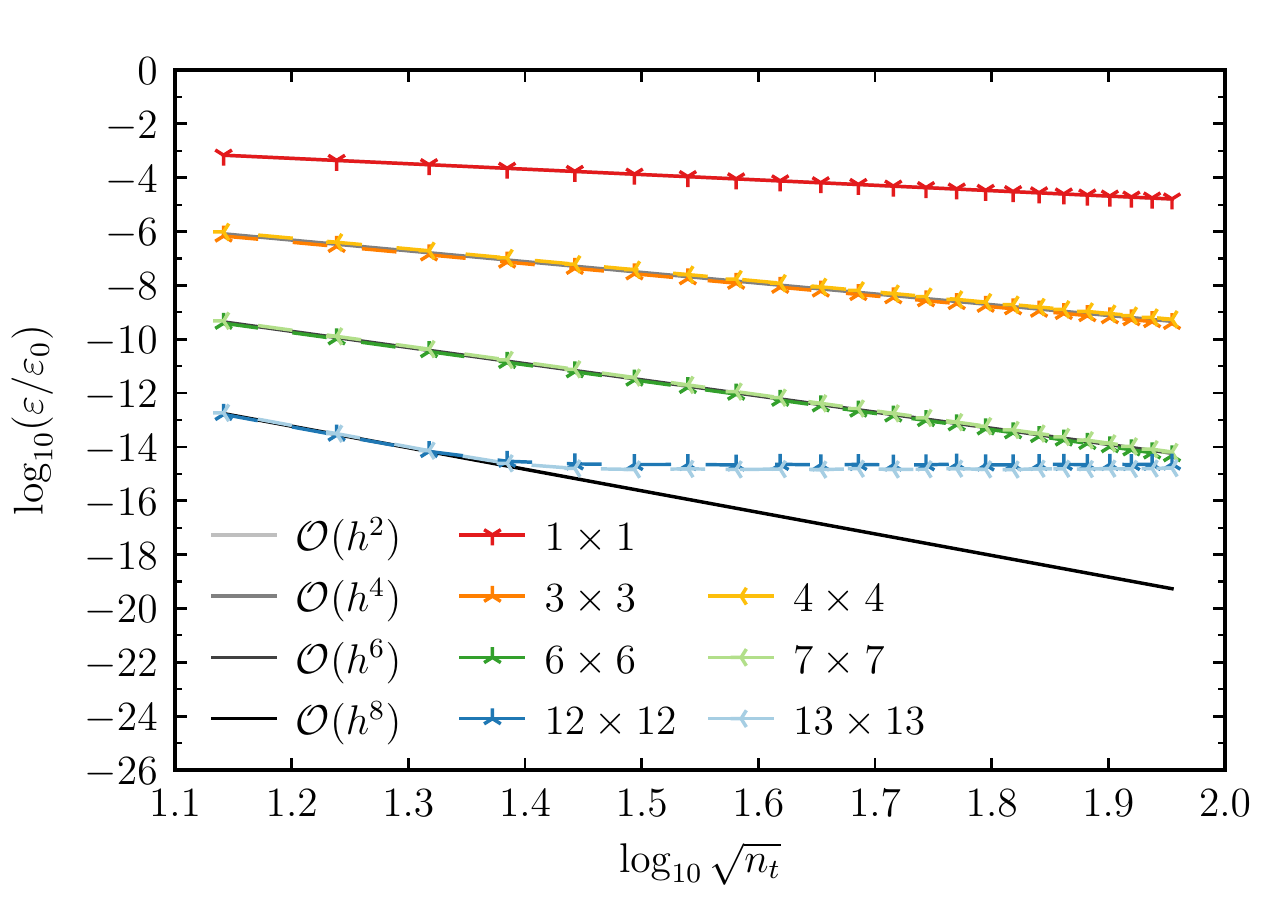}
\caption{Cube, \reviewerOne{$\alpha=1/5$, $k=\pi/L$}\vpad}
\end{subfigure}
\hspace{0.25em}
\begin{subfigure}[b]{.49\textwidth}
\includegraphics[scale=.64,clip=true,trim=2.3in 0in 2.8in 0in]{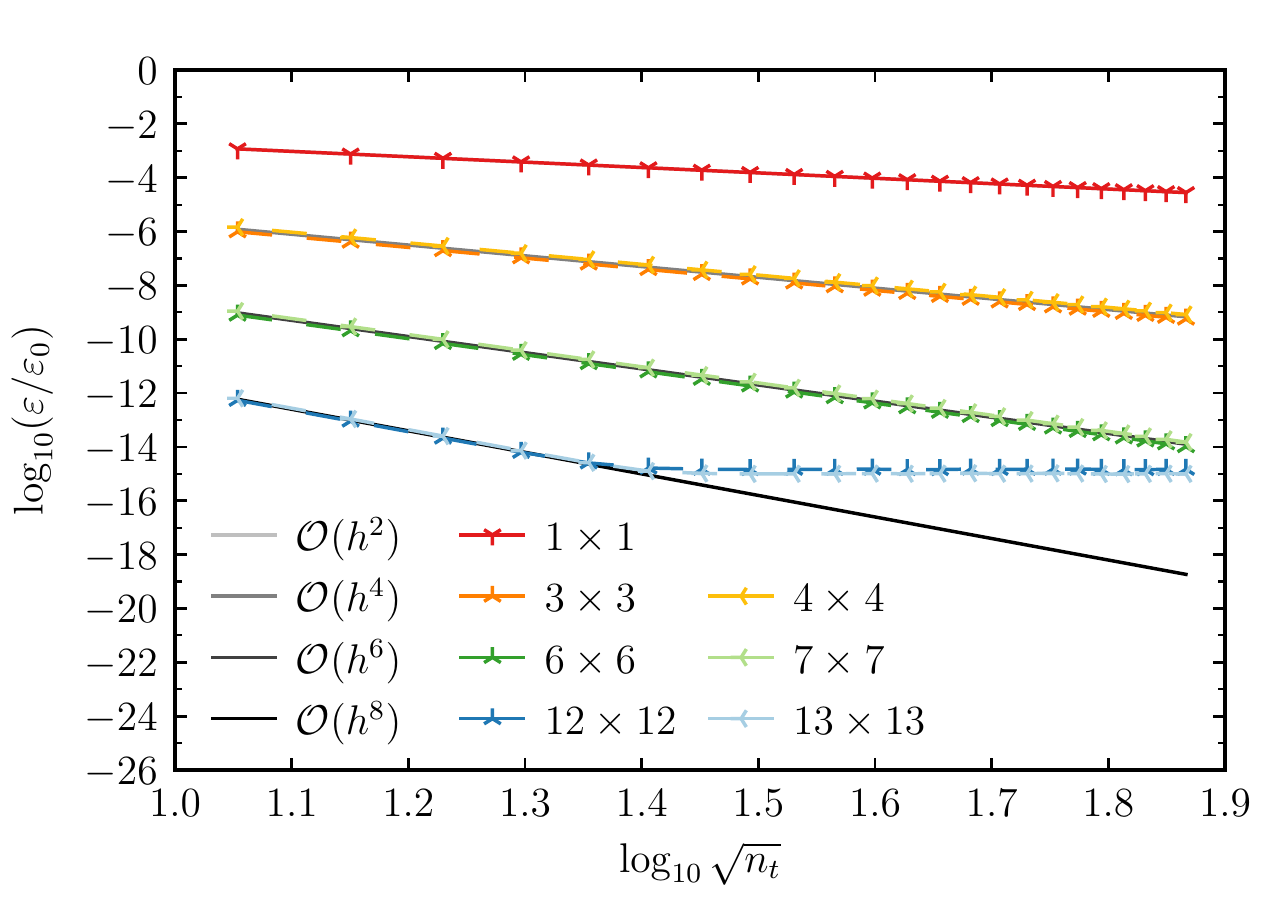}
\caption{Triangular Prism, \reviewerOne{$\alpha=1/5$, $k=\pi/L$}\vpad}
\end{subfigure}
\\
\begin{subfigure}[b]{.49\textwidth}
\includegraphics[scale=.64,clip=true,trim=2.3in 0in 2.8in 0in]{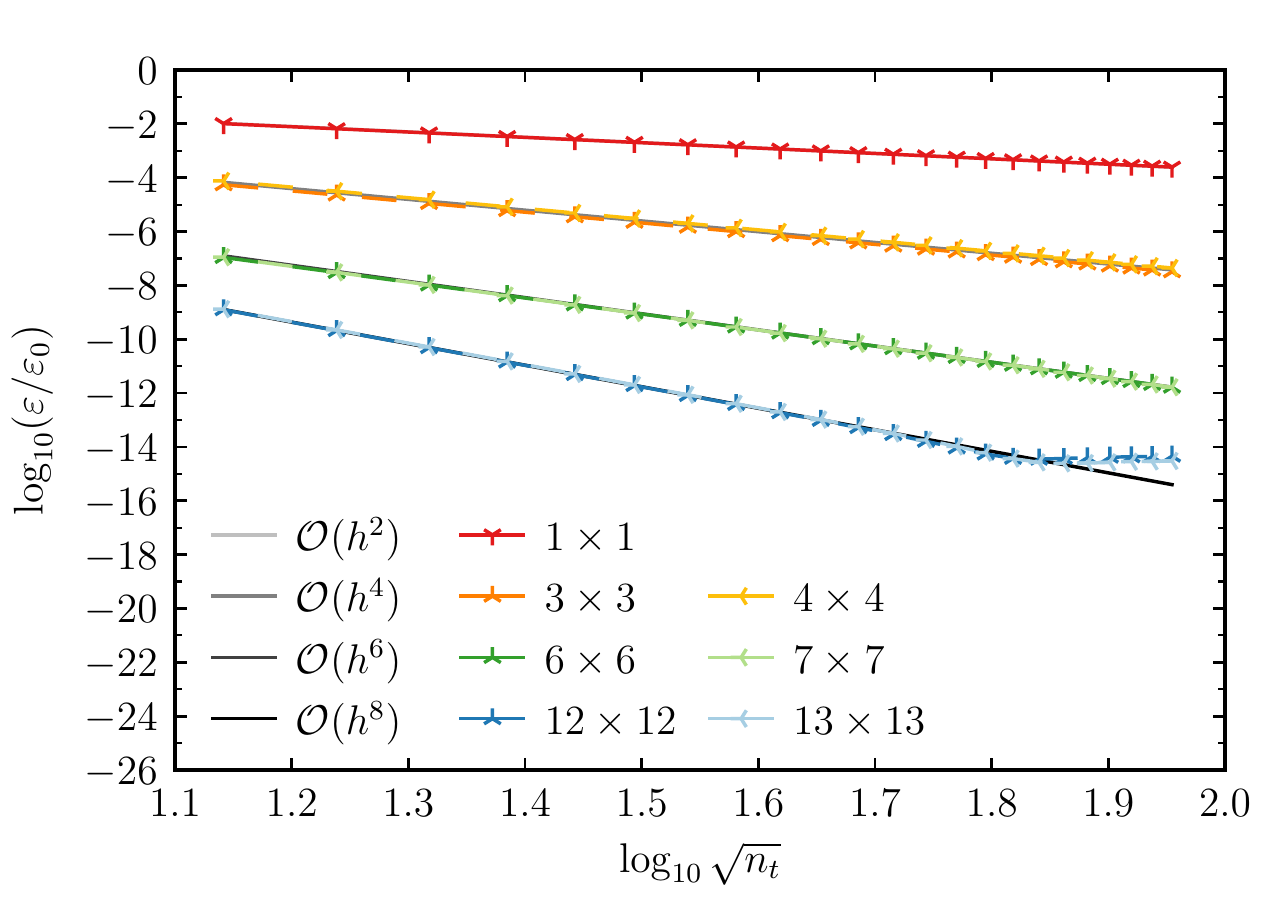}
\caption{Cube, \reviewerOne{$\alpha=1/5$, $k=3\pi/L$}\vpad}
\end{subfigure}
\hspace{0.25em}
\begin{subfigure}[b]{.49\textwidth}
\includegraphics[scale=.64,clip=true,trim=2.3in 0in 2.8in 0in]{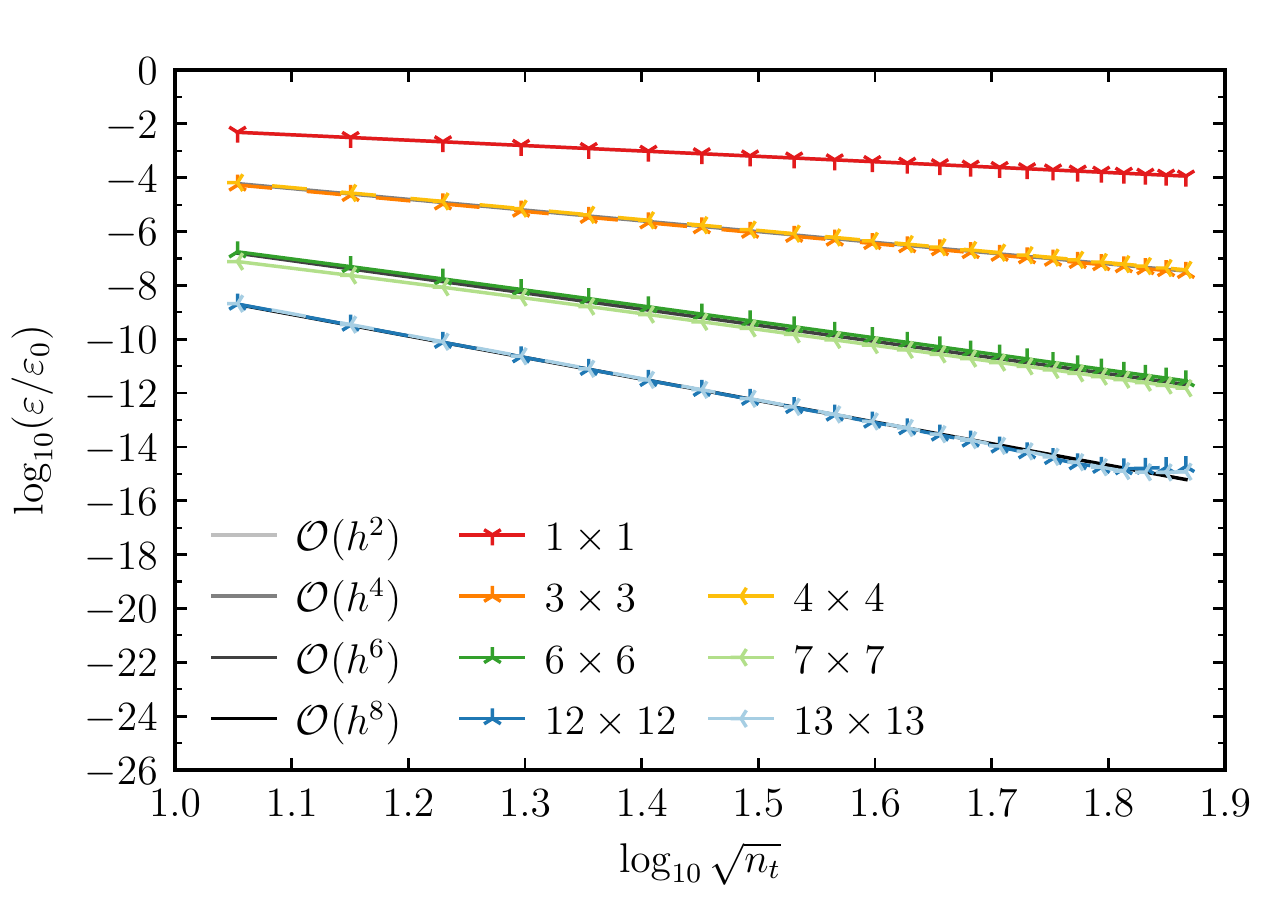}
\caption{Triangular Prism, \reviewerOne{$\alpha=1/5$, $k=3\pi/L$}\vpad}
\end{subfigure}
\caption{Numerical-integration error: $\varepsilon=|e_a(\mathbf{J}_{h_\text{MS}})|$~\eqref{eq:a_error_cancel} for different amounts of quadrature points.}
\vskip-\dp\strutbox
\label{fig:part5a}
\end{figure}

\begin{figure}
\centering
\begin{subfigure}[b]{.49\textwidth}
\includegraphics[scale=.64,clip=true,trim=2.3in 0in 2.8in 0in]{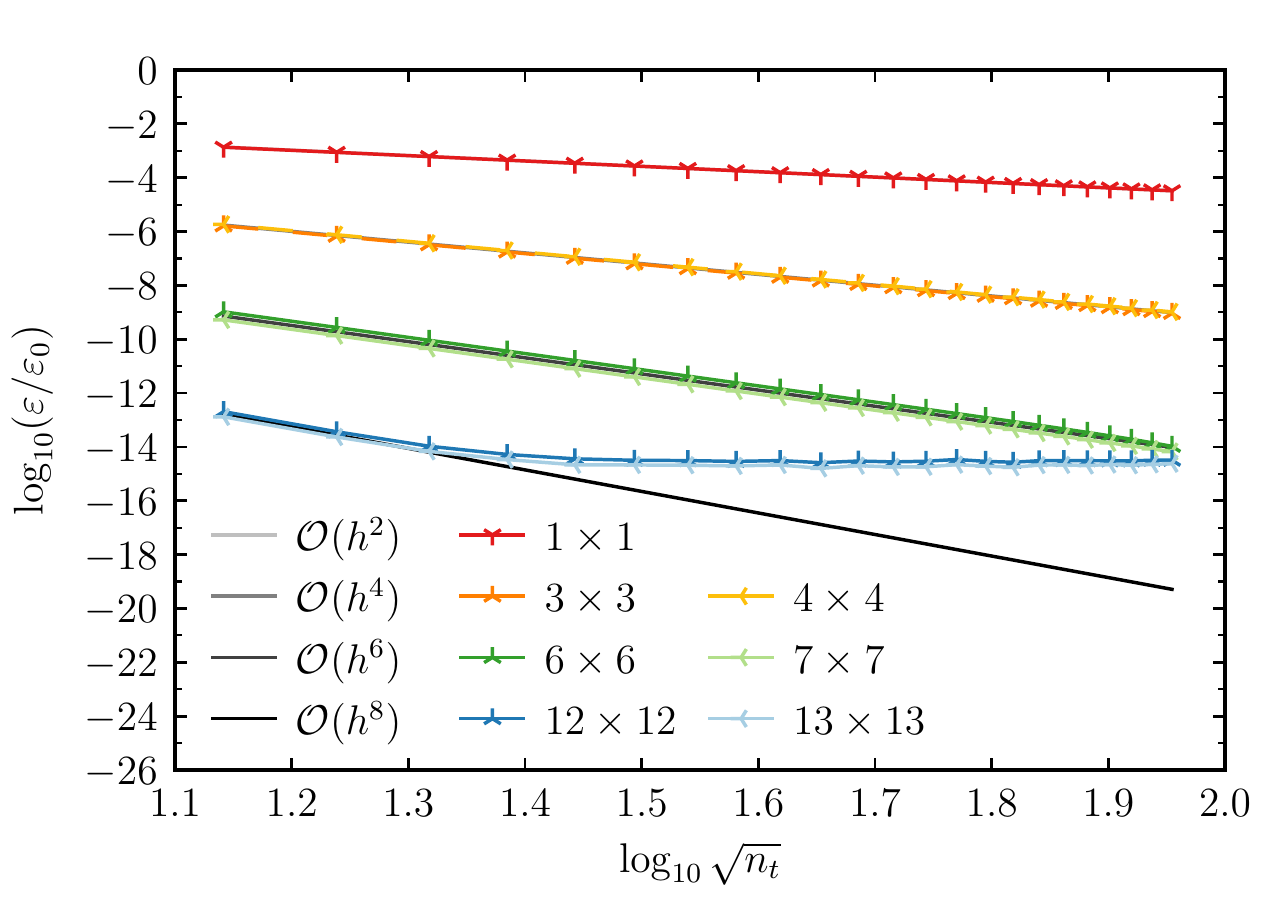}
\caption{Cube, \reviewerOne{$\alpha=4/5$, $k=\pi/L$}\vpad}
\end{subfigure}
\hspace{0.25em}
\begin{subfigure}[b]{.49\textwidth}
\includegraphics[scale=.64,clip=true,trim=2.3in 0in 2.8in 0in]{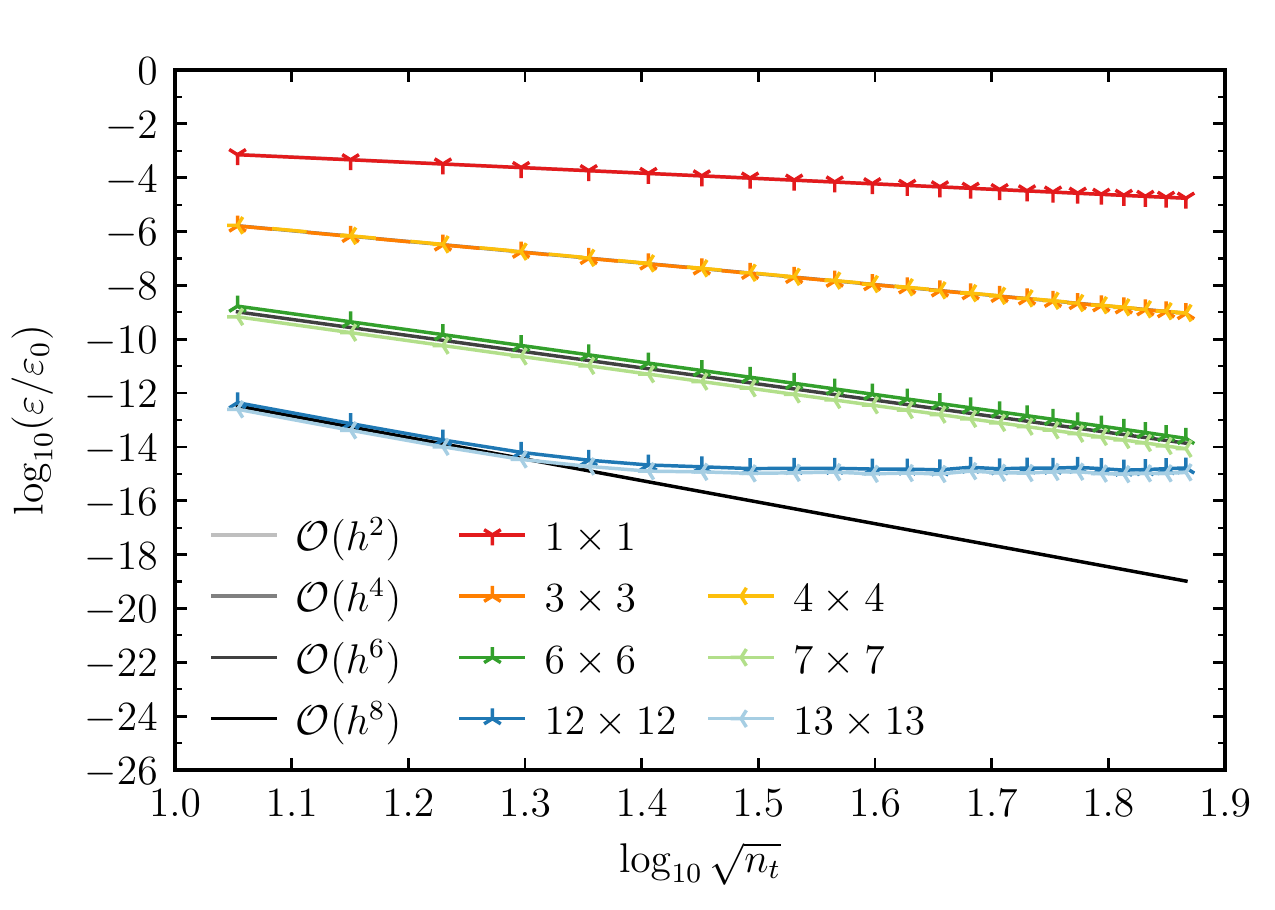}
\caption{Triangular Prism, \reviewerOne{$\alpha=4/5$, $k=\pi/L$}\vpad}
\end{subfigure}
\\
\begin{subfigure}[b]{.49\textwidth}
\includegraphics[scale=.64,clip=true,trim=2.3in 0in 2.8in 0in]{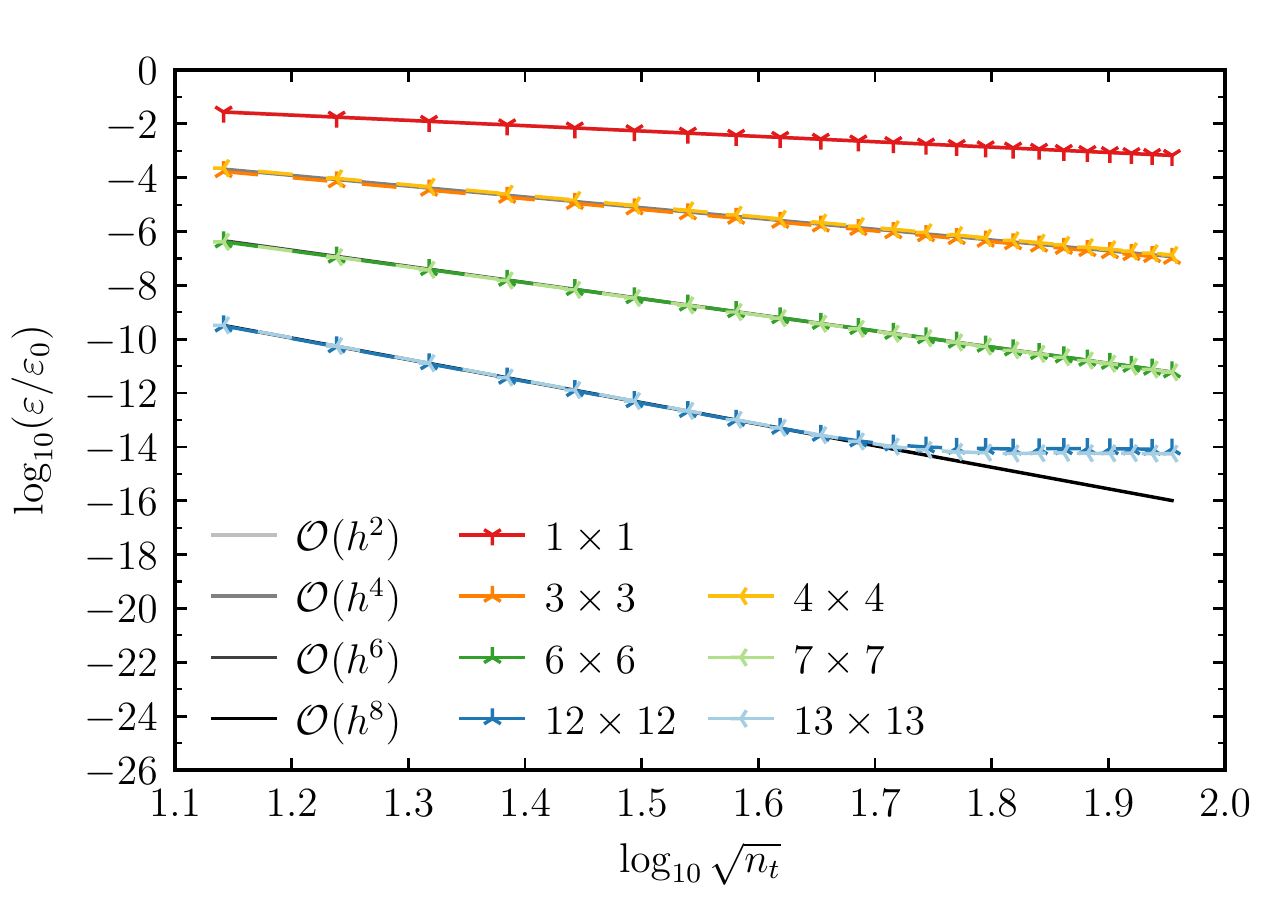}
\caption{Cube, \reviewerOne{$\alpha=4/5$, $k=3\pi/L$}\vpad}
\end{subfigure}
\hspace{0.25em}
\begin{subfigure}[b]{.49\textwidth}
\includegraphics[scale=.64,clip=true,trim=2.3in 0in 2.8in 0in]{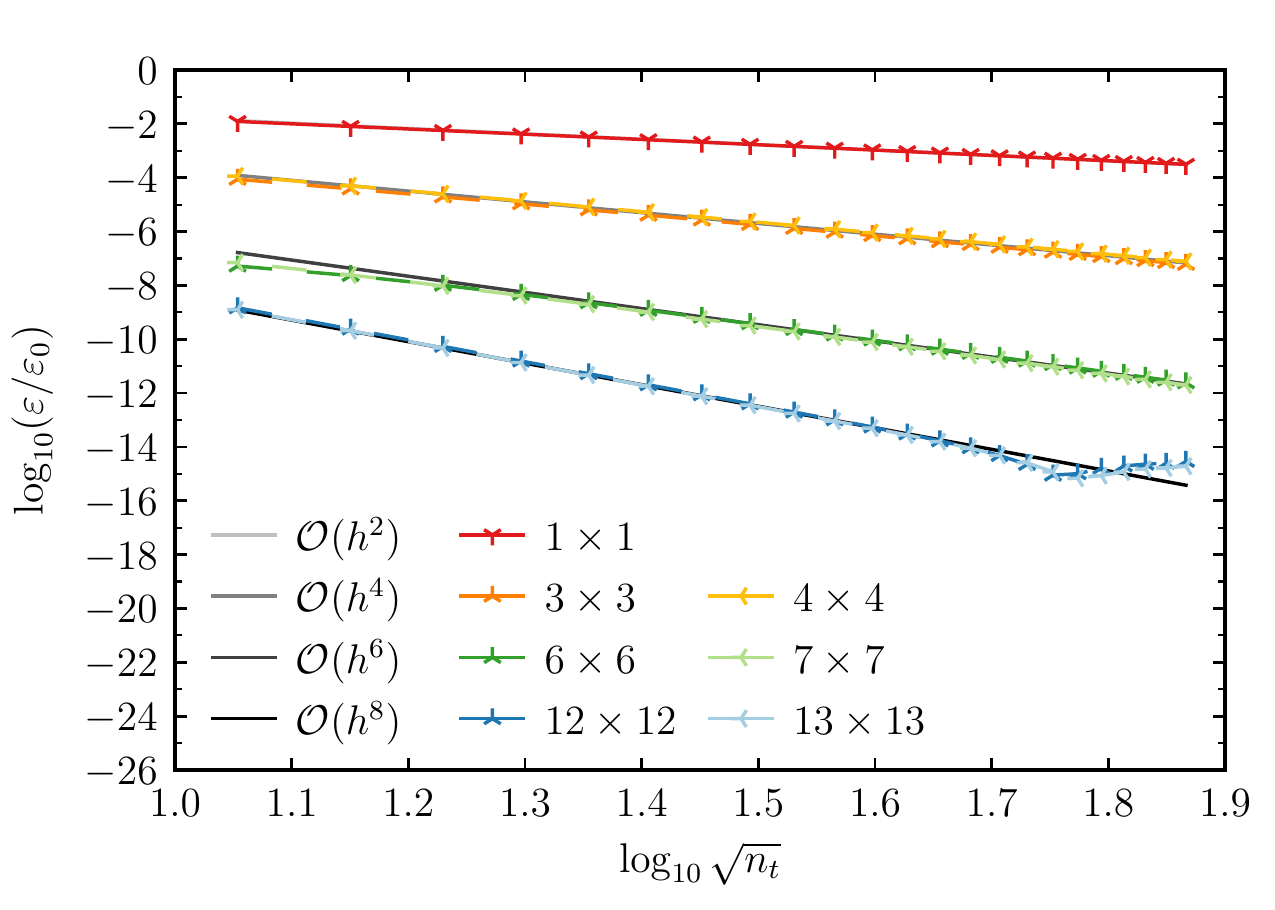}
\caption{Triangular Prism, \reviewerOne{$\alpha=4/5$, $k=3\pi/L$}\vpad}
\end{subfigure}
\caption{Numerical-integration error: $\varepsilon=|e_a(\mathbf{J}_{h_\text{MS}})|$~\eqref{eq:a_error_cancel} for different amounts of quadrature points.}
\vskip-\dp\strutbox
\label{fig:part5b}
\end{figure}

For both domains and the most extreme combinations of $\alpha$ and $k$, Figures~\ref{fig:part5a} and~\ref{fig:part5b} show the numerical-integration error $e_a(\mathbf{J}_{h_\text{MS}})$~\eqref{eq:a_error_cancel} when the solution-discretization error is canceled.  In the legend entries, the first number is the amount of quadrature points used to compute the integral over $S$, whereas the second is the amount used to compute the integral over $S'$.  The numerical-integration error is nondimensionalized by the constant $\varepsilon_0=1$ A$^2$.  Each of the solutions converges at the expected rate listed in Table~\ref{tab:dunavant_properties}.  \reviewerOne{For the finest meshes considered, the round-off error arising from the double-precision calculations exceeds the numerical-integration error.}

\begin{figure}
\centering
\begin{subfigure}[b]{.49\textwidth}
\includegraphics[scale=.64,clip=true,trim=2.3in 0in 2.8in 0in]{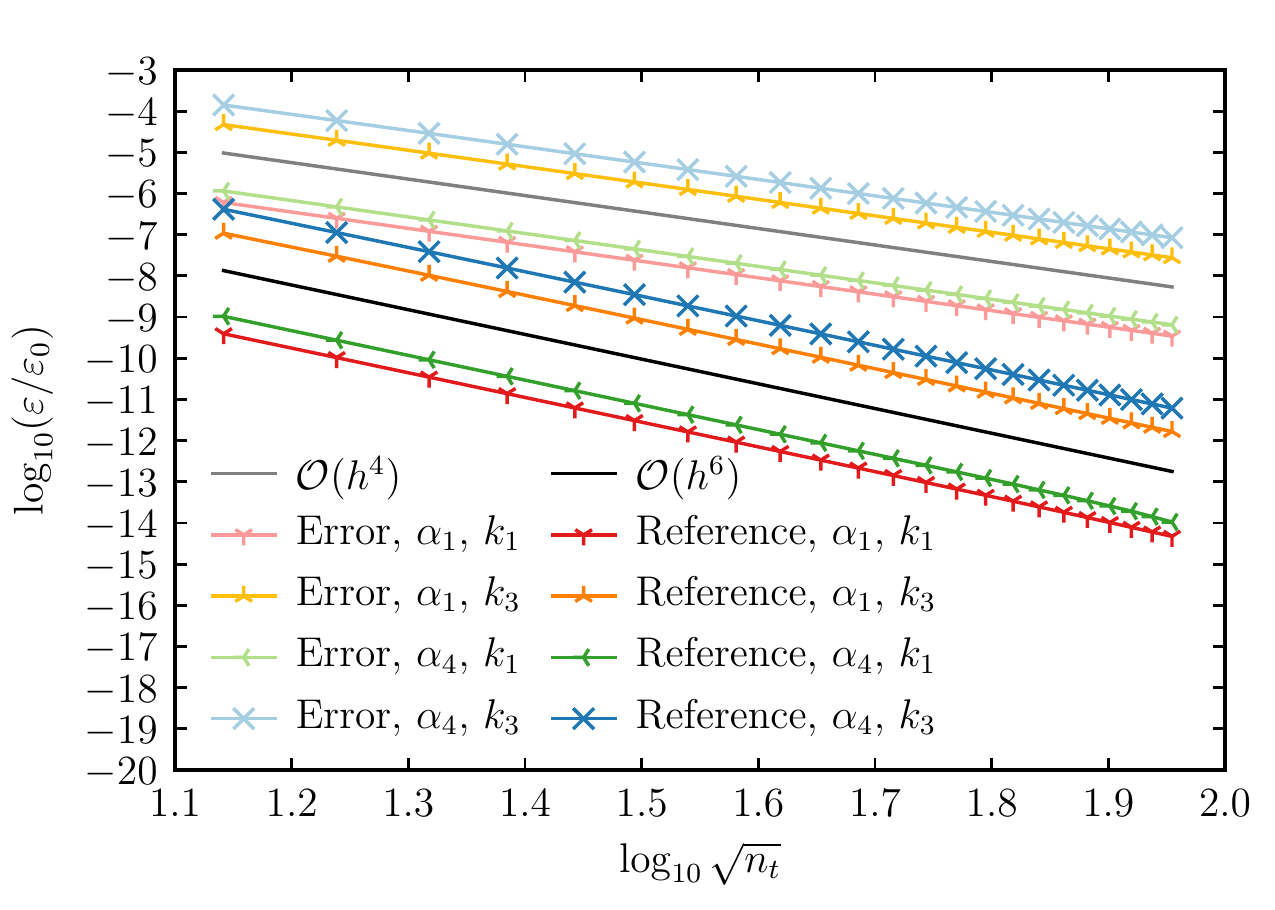}
\caption{Cube, \reviewerOne{$\alpha_m=m/5$, $k_n=n\pi/L$}\vpad}
\end{subfigure}
\hspace{0.25em}
\begin{subfigure}[b]{.49\textwidth}
\includegraphics[scale=.64,clip=true,trim=2.3in 0in 2.8in 0in]{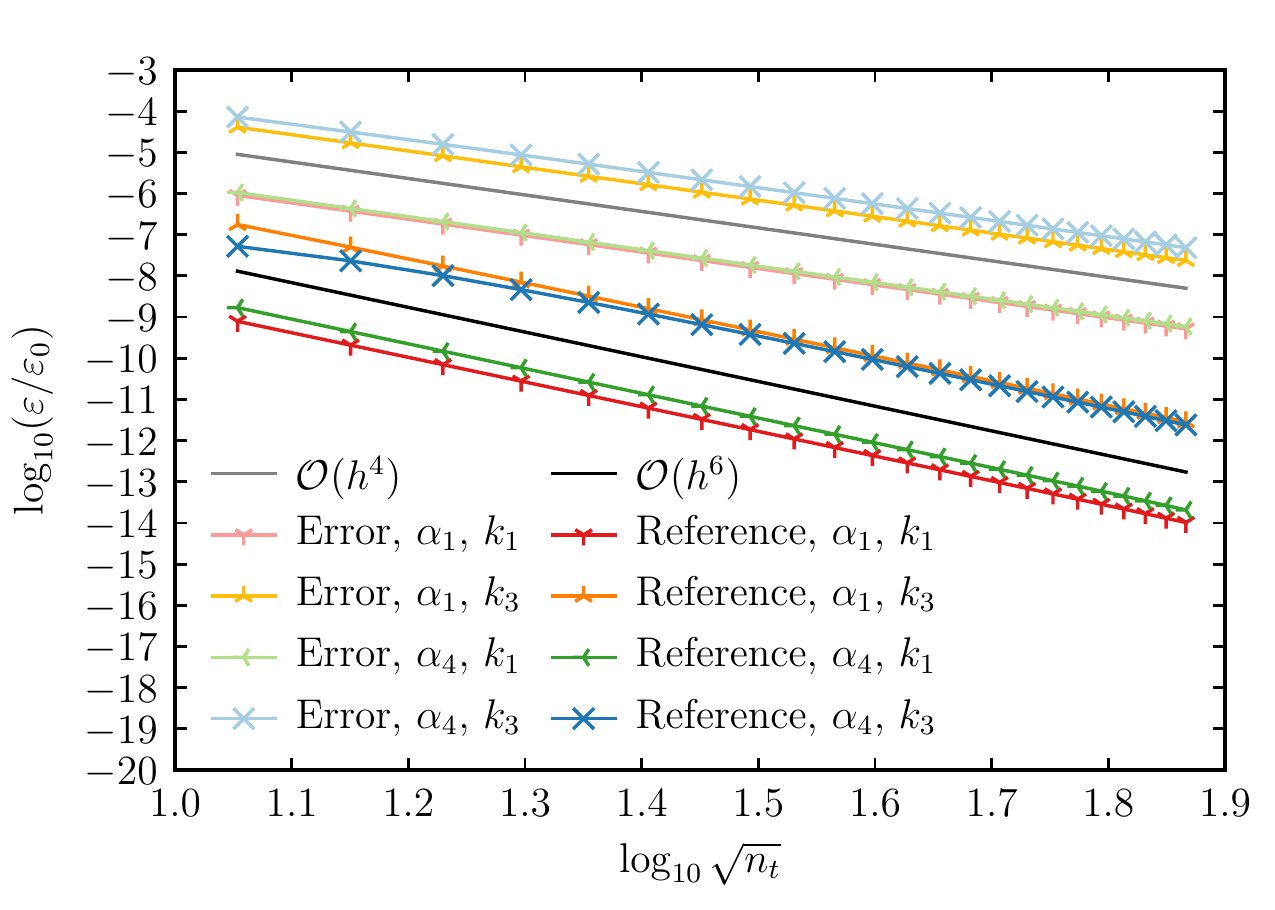}
\caption{Triangular Prism, \reviewerOne{$\alpha_m=m/5$, $k_n=n\pi/L$}\vpad}
\end{subfigure}
\caption{Numerical-integration error: $\varepsilon=|e_a(\mathbf{J}_{h_\text{MS}})|$~\eqref{eq:a_error_cancel} in the presence of a coding error.}
\vskip-\dp\strutbox
\label{fig:part5_bug}
\end{figure}

To test the ability to detect a coding error, we replace the optimal 6-point quadrature rule that can exactly integrate polynomials up to degree 4 (Figure~\ref{fig:quad_optimal_6}) with a suboptimal rule~\cite{papanicolopulos_2015} (Figure~\ref{fig:quad_suboptimal_6}) that can integrate polynomials up to degree 3.  
Figure~\ref{fig:part5_bug} shows how the cases with this coding error compare with the cases presented in Figures~\ref{fig:part5a} and~\ref{fig:part5b}.  In Figure~\ref{fig:part5_bug}, the convergence rates are $\mathcal{O}(h^4)$ for the cases with the coding error, compared to the expected $\mathcal{O}(h^6)$ rates without.  Therefore, $e_a(\mathbf{J}_{h_\text{MS}})$ detects the coding error.

\begin{figure}
\centering
\begin{subfigure}[b]{.49\textwidth}
\includegraphics[scale=.64,clip=true,trim=2.3in 0in 2.8in 0in]{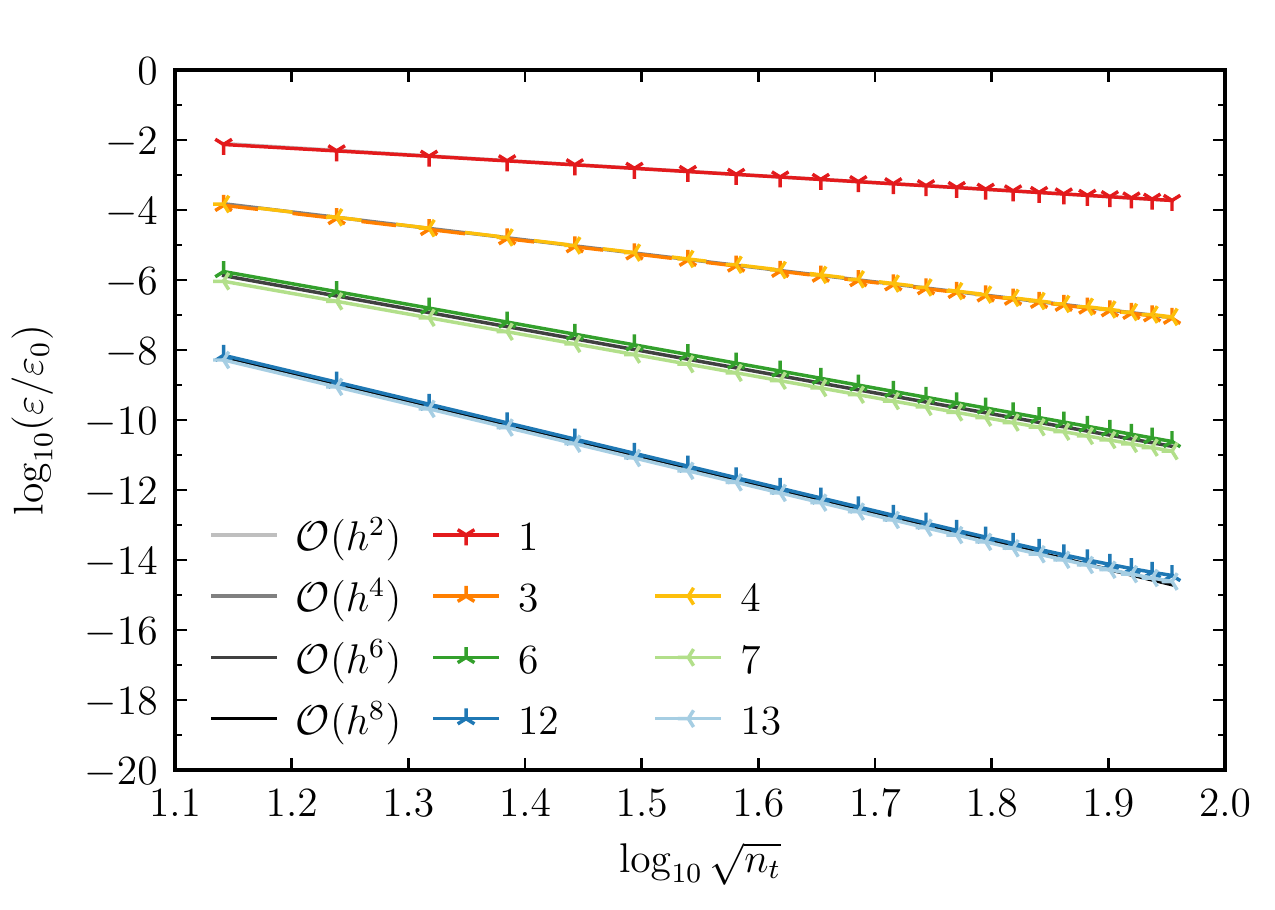}
\caption{Cube, \reviewerOne{$\alpha=1/5$, $k=\pi/L$}\vpad}
\end{subfigure}
\hspace{0.25em}
\begin{subfigure}[b]{.49\textwidth}
\includegraphics[scale=.64,clip=true,trim=2.3in 0in 2.8in 0in]{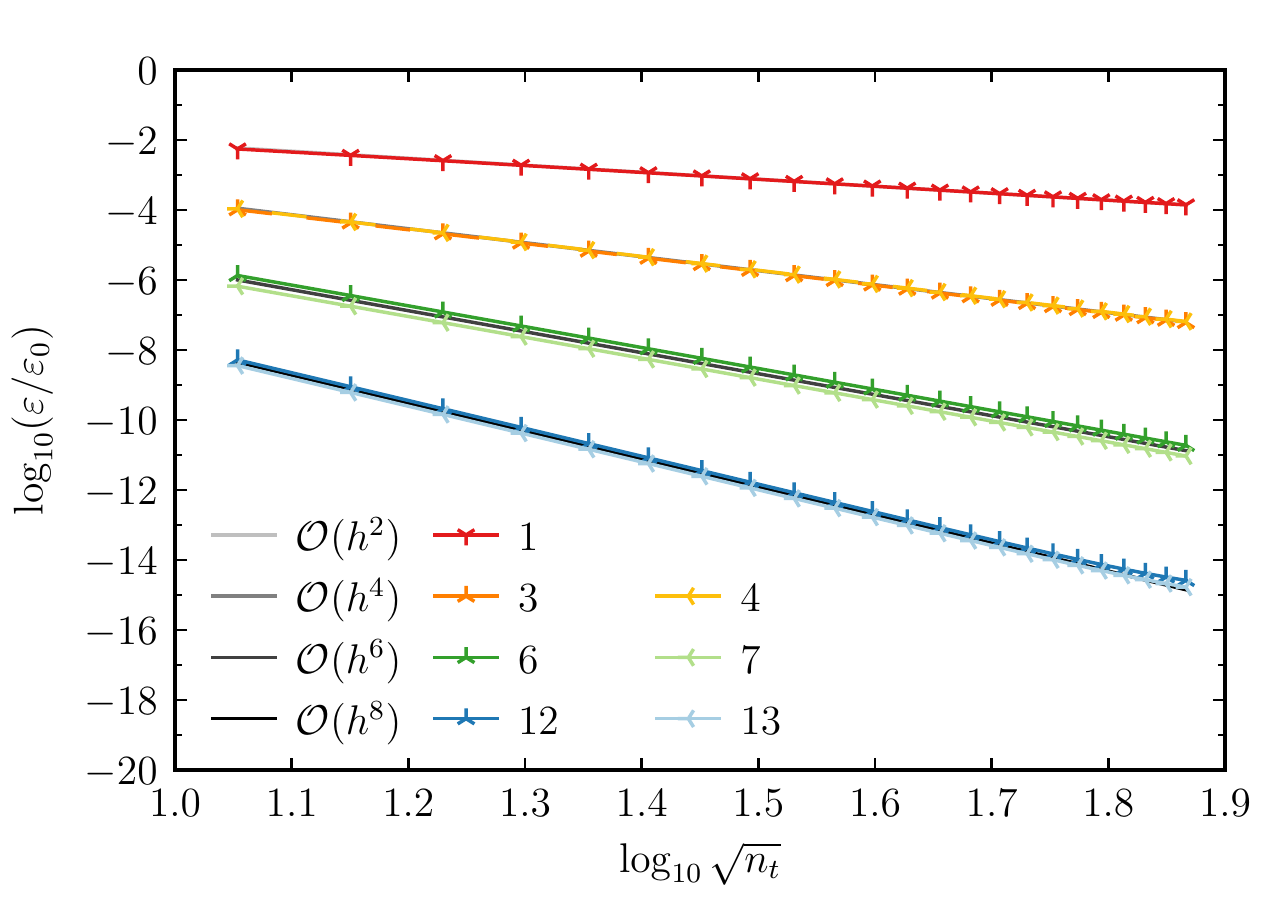}
\caption{Triangular Prism, \reviewerOne{$\alpha=1/5$, $k=\pi/L$}\vpad}
\end{subfigure}
\\
\begin{subfigure}[b]{.49\textwidth}
\includegraphics[scale=.64,clip=true,trim=2.3in 0in 2.8in 0in]{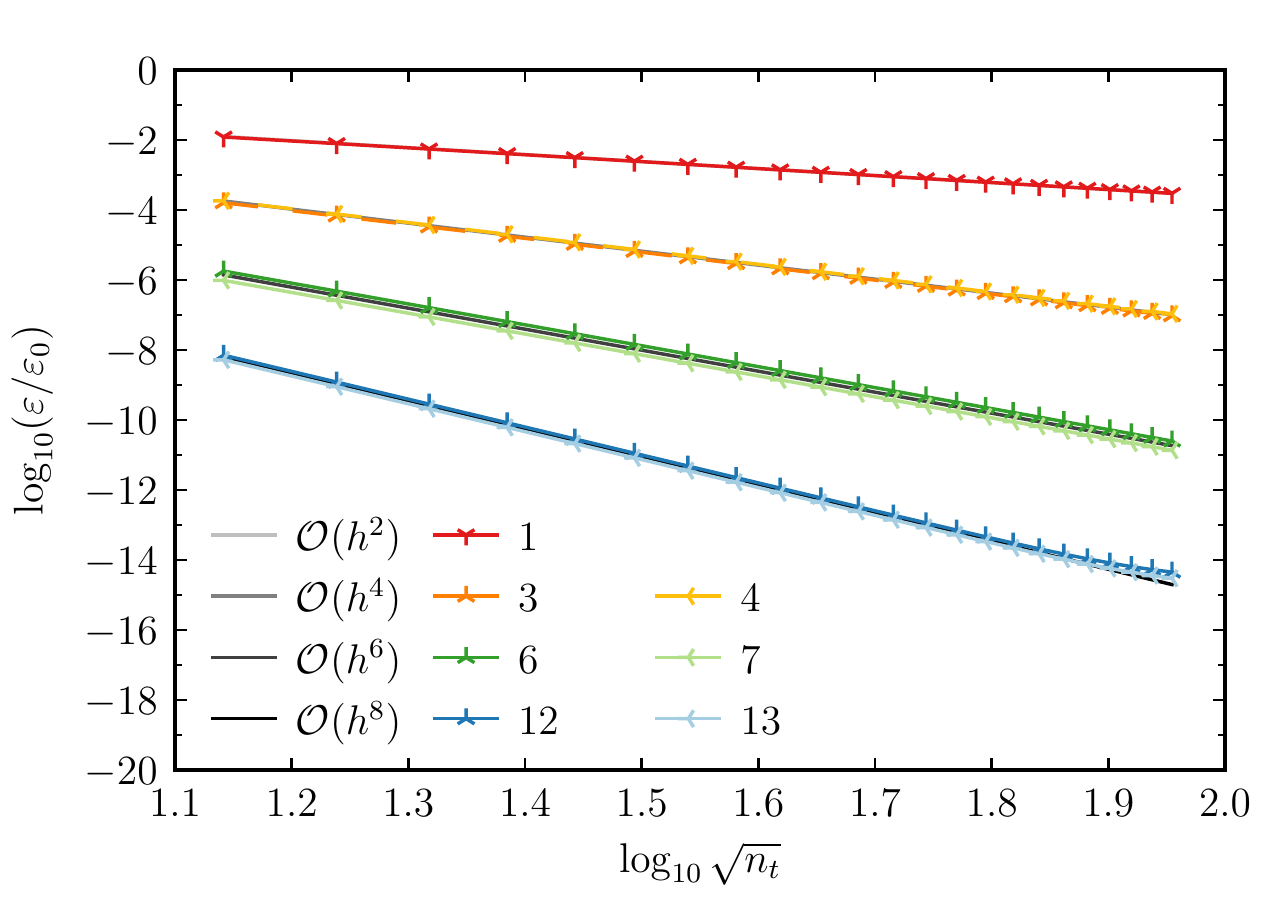}
\caption{Cube, \reviewerOne{$\alpha=1/5$, $k=3\pi/L$}\vpad}
\end{subfigure}
\hspace{0.25em}
\begin{subfigure}[b]{.49\textwidth}
\includegraphics[scale=.64,clip=true,trim=2.3in 0in 2.8in 0in]{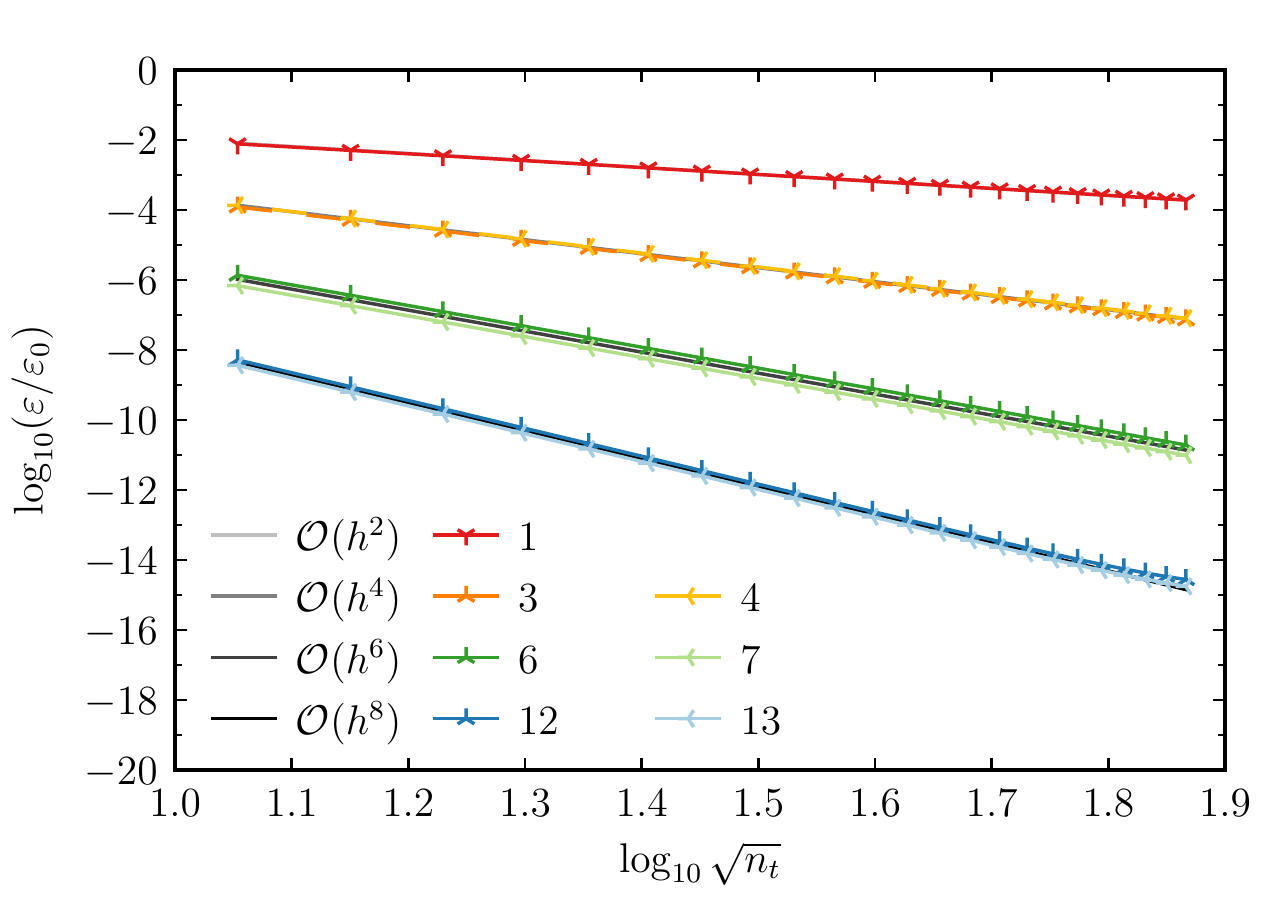}
\caption{Triangular Prism, \reviewerOne{$\alpha=1/5$, $k=3\pi/L$}\vpad}
\end{subfigure}
\caption{Numerical-integration error: $\varepsilon=|e_b(\mathbf{J}_{h_\text{MS}})|$~\eqref{eq:b_error_cancel} for different amounts of quadrature points.}
\vskip-\dp\strutbox
\label{fig:part6a}
\end{figure}

\begin{figure}
\centering
\begin{subfigure}[b]{.49\textwidth}
\includegraphics[scale=.64,clip=true,trim=2.3in 0in 2.8in 0in]{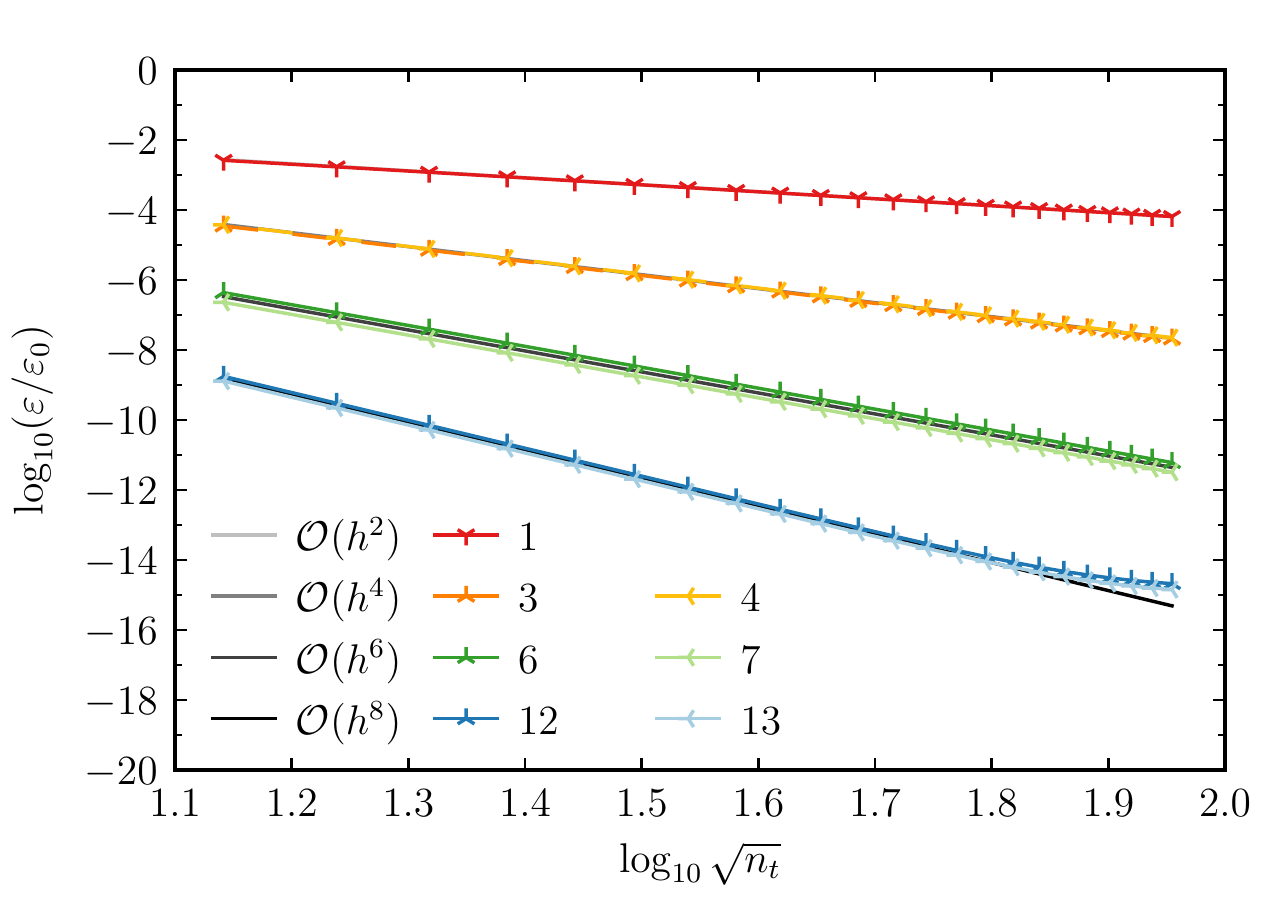}
\caption{Cube, \reviewerOne{$\alpha=4/5$, $k=\pi/L$}\vpad}
\end{subfigure}
\hspace{0.25em}
\begin{subfigure}[b]{.49\textwidth}
\includegraphics[scale=.64,clip=true,trim=2.3in 0in 2.8in 0in]{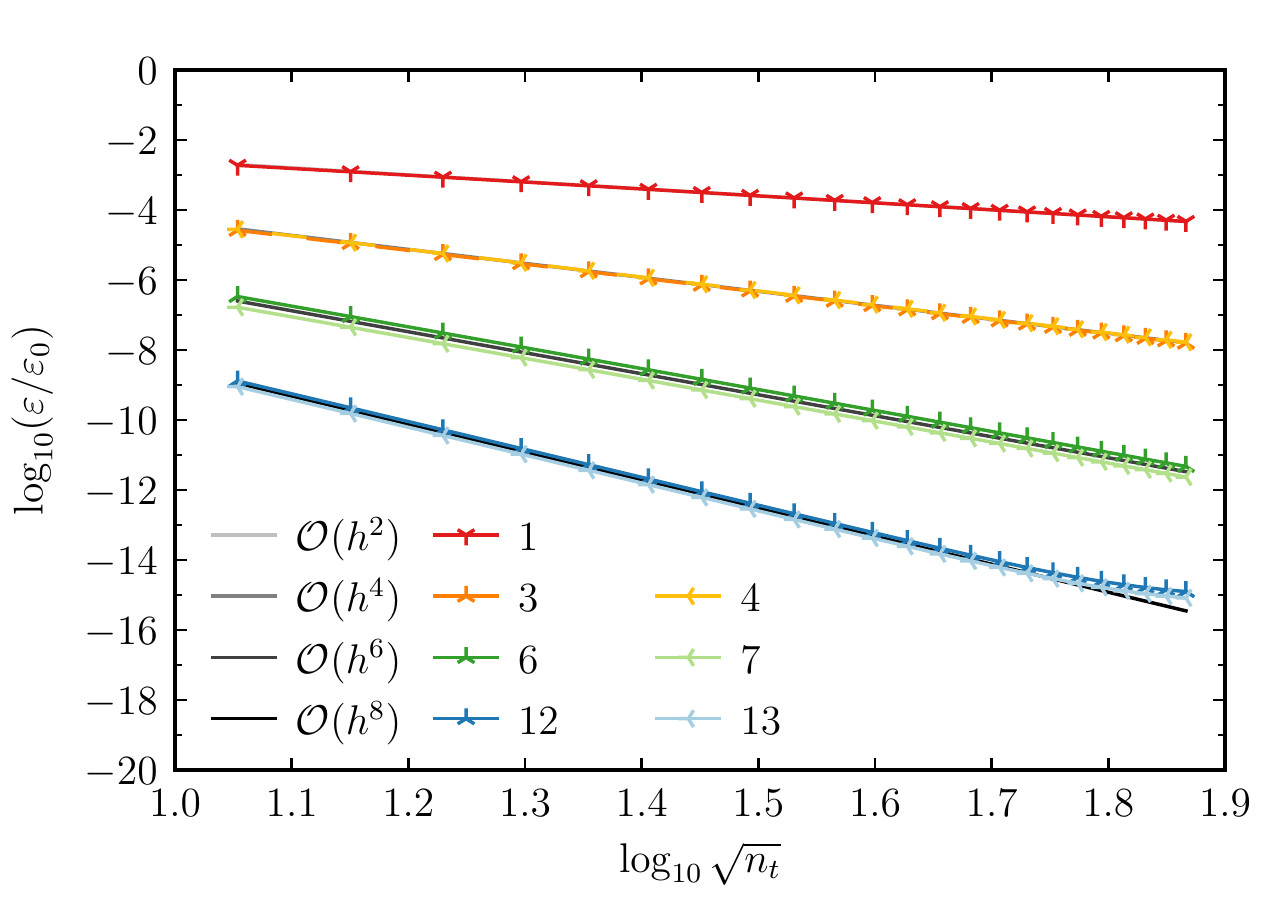}
\caption{Triangular Prism, \reviewerOne{$\alpha=4/5$, $k=\pi/L$}\vpad}
\end{subfigure}
\\
\begin{subfigure}[b]{.49\textwidth}
\includegraphics[scale=.64,clip=true,trim=2.3in 0in 2.8in 0in]{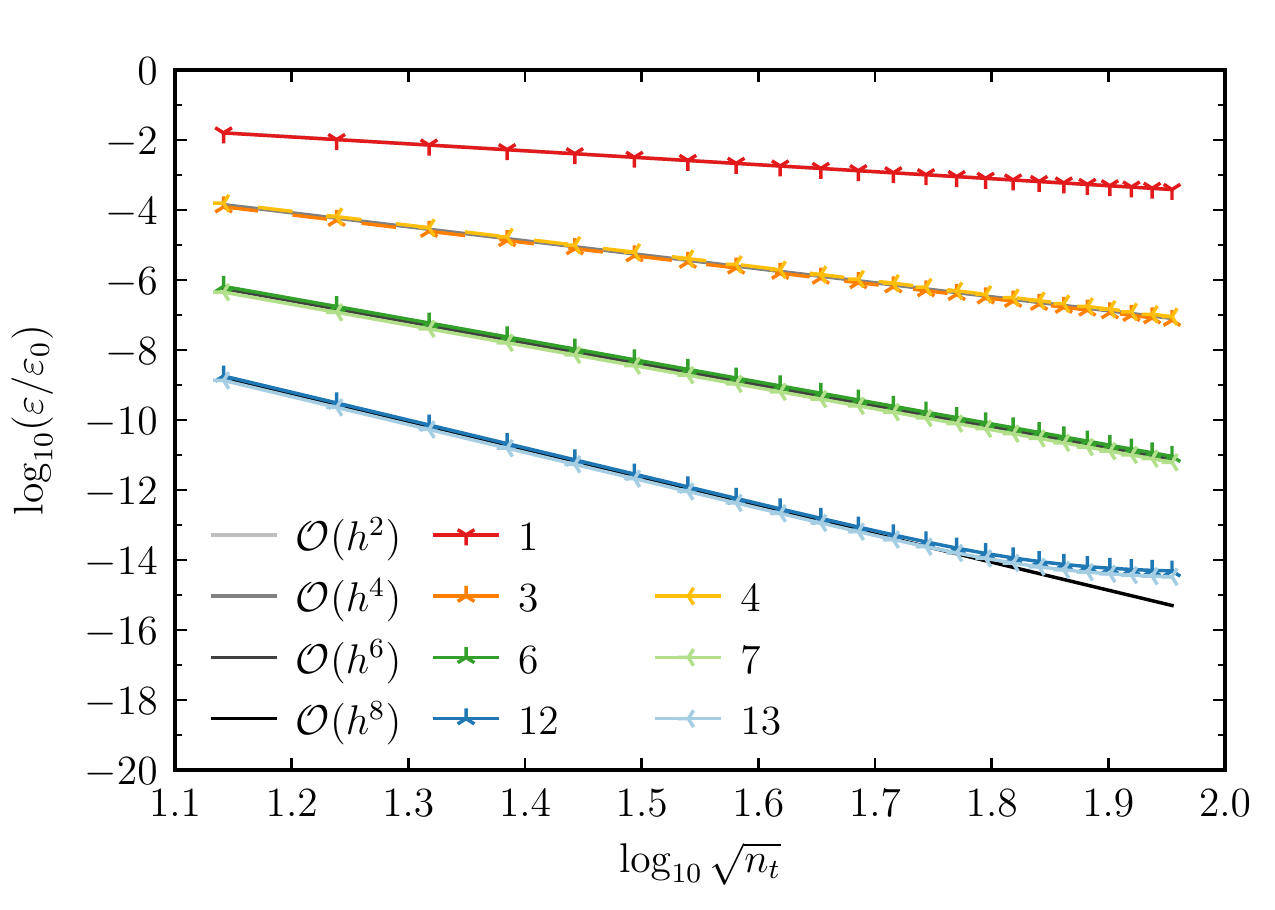}
\caption{Cube, \reviewerOne{$\alpha=4/5$, $k=3\pi/L$}\vpad}
\end{subfigure}
\hspace{0.25em}
\begin{subfigure}[b]{.49\textwidth}
\includegraphics[scale=.64,clip=true,trim=2.3in 0in 2.8in 0in]{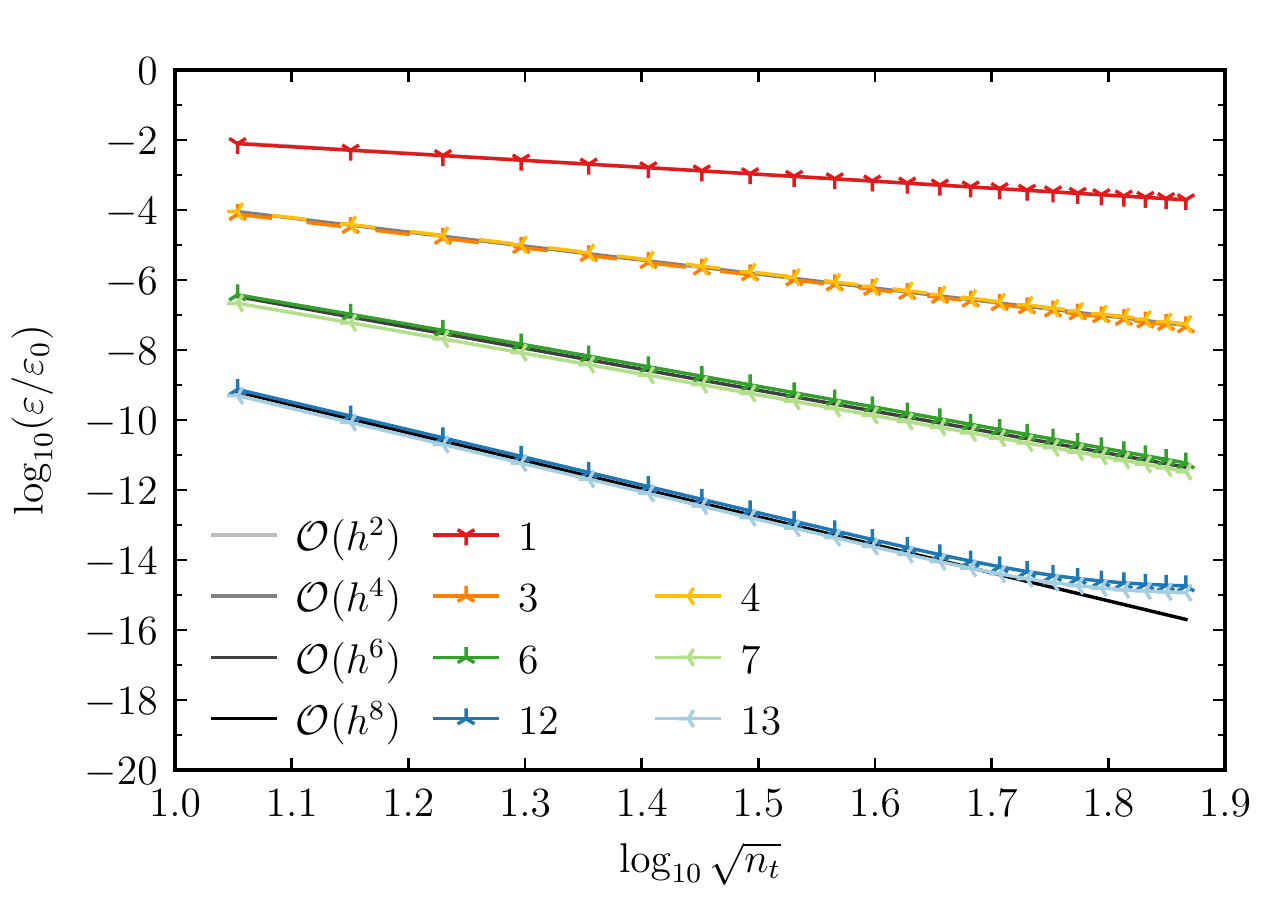}
\caption{Triangular Prism, \reviewerOne{$\alpha=4/5$, $k=3\pi/L$}\vpad}
\end{subfigure}
\caption{Numerical-integration error: $\varepsilon=|e_b(\mathbf{J}_{h_\text{MS}})|$~\eqref{eq:b_error_cancel} for different amounts of quadrature points.}
\vskip-\dp\strutbox
\label{fig:part6b}
\end{figure}

Figures~\ref{fig:part6a} and~\ref{fig:part6b} show the numerical-integration error $e_b(\mathbf{J}_{h_\text{MS}})$~\eqref{eq:b_error_cancel} when the solution-discretization error is canceled.  In the legend entries, the number is the amount of quadrature points used to compute the integral.  Each of the solutions converges at the expected rate.  \reviewerOne{For the finest meshes considered, the round-off error arising from the double-precision calculations exceeds the numerical-integration error.}

\begin{figure}
\centering
\begin{subfigure}[b]{.49\textwidth}
\includegraphics[scale=.64,clip=true,trim=2.3in 0in 2.8in 0in]{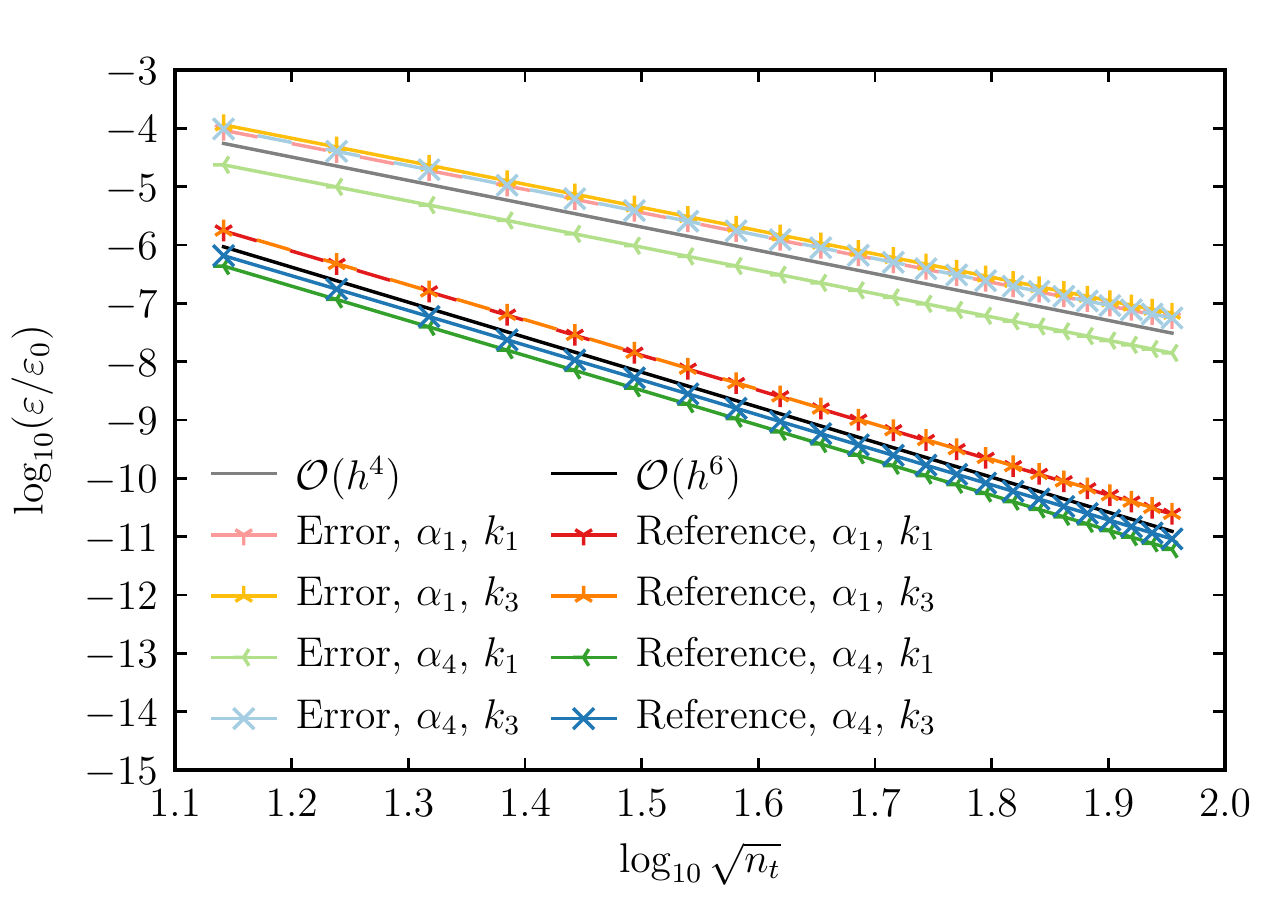}
\caption{Cube, \reviewerOne{$\alpha_m=m/5$, $k_n=n\pi/L$}\vpad}
\end{subfigure}
\hspace{0.25em}
\begin{subfigure}[b]{.49\textwidth}
\includegraphics[scale=.64,clip=true,trim=2.3in 0in 2.8in 0in]{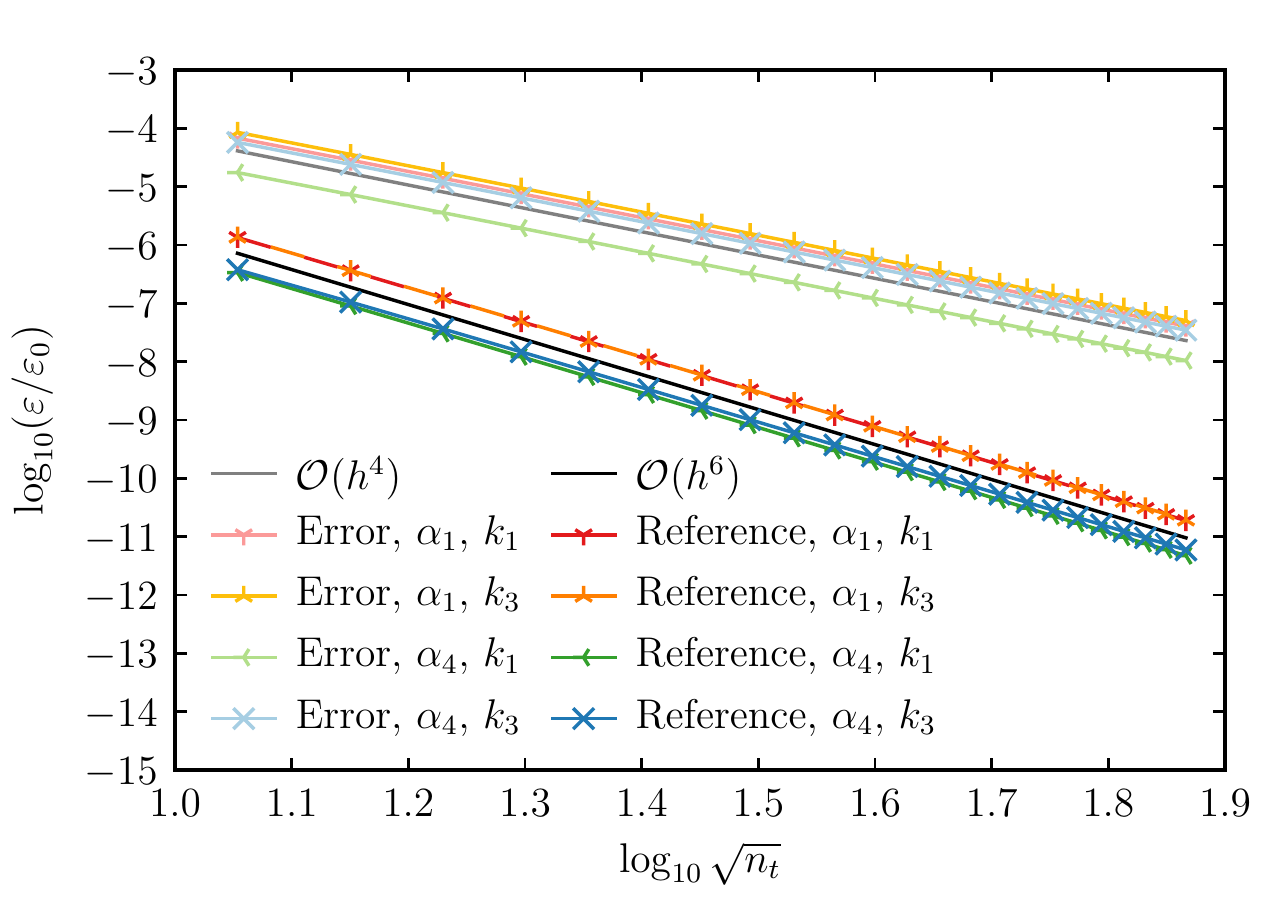}
\caption{Triangular Prism, \reviewerOne{$\alpha_m=m/5$, $k_n=n\pi/L$}\vpad}
\end{subfigure}
\caption{Numerical-integration error: $\varepsilon=|e_b(\mathbf{J}_{h_\text{MS}})|$~\eqref{eq:b_error_cancel} in the presence of a coding error.}
\vskip-\dp\strutbox
\label{fig:part6_bug}
\end{figure}

Figure~\ref{fig:part6_bug} shows how the cases with this coding error compare with the cases presented in Figures~\ref{fig:part6a} and~\ref{fig:part6b}.  In Figure~\ref{fig:part6_bug}, the convergence rates are $\mathcal{O}(h^4)$ for the cases with the coding error, compared to the expected $\mathcal{O}(h^6)$ rates without.  Therefore, $e_b(\mathbf{J}_{h_\text{MS}})$ detects the coding error.

\input{conclusions.tex}

\input{acknowledgments.tex}

%% file: six_a.tex
\begin{tikzpicture}[scale=0.7, every node/.style={scale=0.7}]

\def\ts{8}; 

\coordinate (A) at (-0.5,{-sqrt(3)/6});
\coordinate (B) at ( 0.5,{-sqrt(3)/6});
\coordinate (C) at ( 0  , {sqrt(1/3)});
\coordinate (D) at ($0.5*(A)+0.5*(B)$);
\coordinate (E) at ($0.5*(B)+0.5*(C)$);
\coordinate (F) at ($0.5*(C)+0.5*(A)$);
\coordinate (O) at (0,0);

\def\tx{0}
\def\ty{{\ts*sqrt(3)/6}}

\coordinate (T1) at ($\ts*(A)+(\tx,\ty)$);
\coordinate (T2) at ($\ts*(B)+(\tx,\ty)$);
\coordinate (T3) at ($\ts*(C)+(\tx,\ty)$);
\coordinate (T4) at ($\ts*(D)+(\tx,\ty)$);
\draw[thick] (T1) -- (T2) -- (T3) -- cycle;

\draw[dashed] ($\ts*(C)+(\tx,\ty)$) -- ($\ts*(D)+(\tx,\ty)$);
\draw[dashed] ($\ts*(A)+(\tx,\ty)$) -- ($\ts*(E)+(\tx,\ty)$);
\draw[dashed] ($\ts*(B)+(\tx,\ty)$) -- ($\ts*(F)+(\tx,\ty)$);

\draw[fill=black] ($\ts*(-0.168922736373948, 0.097527587317747)+(\tx,\ty)$) circle (.11);
\draw[fill=black] ($\ts*( 0.168922736373948, 0.097527587317747)+(\tx,\ty)$) circle (.11);
\draw[fill=black] ($\ts*( 0.000000000000000,-0.195055174635494)+(\tx,\ty)$) circle (.11);
\draw[fill=black] ($\ts*( 0.362635679735344,-0.209367807312963)+(\tx,\ty)$) circle (.11);
\draw[fill=black] ($\ts*(-0.362635679735344,-0.209367807312963)+(\tx,\ty)$) circle (.11);
\draw[fill=black] ($\ts*( 0.000000000000000, 0.418735614625928)+(\tx,\ty)$) circle (.11);

\end{tikzpicture}

%% file: six_b.tex
\begin{tikzpicture}[scale=0.7, every node/.style={scale=0.7}]

\def\ts{8}; 

\coordinate (A) at (-0.5,{-sqrt(3)/6});
\coordinate (B) at ( 0.5,{-sqrt(3)/6});
\coordinate (C) at ( 0  , {sqrt(1/3)});
\coordinate (D) at ($0.5*(A)+0.5*(B)$);
\coordinate (E) at ($0.5*(B)+0.5*(C)$);
\coordinate (F) at ($0.5*(C)+0.5*(A)$);
\coordinate (O) at (0,0);

\def\tx{0}
\def\ty{{\ts*sqrt(3)/6}}

\coordinate (T1) at ($\ts*(A)+(\tx,\ty)$);
\coordinate (T2) at ($\ts*(B)+(\tx,\ty)$);
\coordinate (T3) at ($\ts*(C)+(\tx,\ty)$);
\coordinate (T4) at ($\ts*(D)+(\tx,\ty)$);
\draw[thick] (T1) -- (T2) -- (T3) -- cycle;

\draw[dashed] ($\ts*(C)+(\tx,\ty)$) -- ($\ts*(D)+(\tx,\ty)$);
\draw[dashed] ($\ts*(A)+(\tx,\ty)$) -- ($\ts*(E)+(\tx,\ty)$);
\draw[dashed] ($\ts*(B)+(\tx,\ty)$) -- ($\ts*(F)+(\tx,\ty)$);

\draw[fill=black] ($\ts*( 6.1447179740076699e-2, 2.8205952817680885e-1)+(\tx,\ty)$) circle (.11);
\draw[fill=black] ($\ts*( 2.1354712691053079e-1,-1.9424458273421930e-1)+(\tx,\ty)$) circle (.11);
\draw[fill=black] ($\ts*(-2.7499430665060748e-1,-8.7814945442589498e-2)+(\tx,\ty)$) circle (.11);
\draw[fill=black] ($\ts*(-6.1447179740076699e-2, 2.8205952817680885e-1)+(\tx,\ty)$) circle (.11);
\draw[fill=black] ($\ts*(-2.1354712691053079e-1,-1.9424458273421930e-1)+(\tx,\ty)$) circle (.11);
\draw[fill=black] ($\ts*( 2.7499430665060748e-1,-8.7814945442589498e-2)+(\tx,\ty)$) circle (.11);

\end{tikzpicture}

%% file: conclusions.tex
\section{Conclusions} 
\label{sec:conclusions}

In this paper, we presented code-verification approaches for the method-of-moments implementation of the combined-field integral equation to isolate and measure the solution-discretization error and numerical-integration error.  To isolate the solution-discretization error, we approximated the Green's function using a basis that can be integrated exactly, eliminating the numerical-integration error.  To isolate the numerical-integration error, we removed the solution-discretization error by canceling the basis-function contribution.

For both approaches, we considered different wavenumbers and combination parameters to vary the relative weights between the terms in the combined-field integral equation. For these different cases, we achieved the expected orders of accuracy for cases without coding errors, and we were able to detect cases with coding errors.

\reviewerTwo{This work can be complemented with unit tests that assess the evaluation of the (nearly) singular integrals that arise from the actual Green's function.}

%% file: acknowledgments.tex
\section*{Acknowledgments} 
\label{sec:acknowledgments}

The authors thank Timothy Smith for his insightful feedback. 
This article has been authored by employees of National Technology \& Engineering Solutions of Sandia, LLC under Contract No.~DE-NA0003525 with the U.S.~Department of Energy (DOE). The employees own all right, title, and interest in and to the article and are solely responsible for its contents. The United States Government retains and the publisher, by accepting the article for publication, acknowledges that the United States Government retains a non-exclusive, paid-up, irrevocable, world-wide license to publish or reproduce the published form of this article or allow others to do so, for United States Government purposes. The DOE will provide public access to these results of federally sponsored research in accordance with the DOE Public Access Plan \url{https://www.energy.gov/downloads/doe-public-access-plan}.

%% file: appendix.tex
\section{Incident Field Integral Evaluations for Section~\ref{sec:results}} 
\label{app:integrals}

The integration with respect to $\mathbf{x}'$ in $\mathbf{I}_{\mathcal{E}_\mathbf{A}}$~\eqref{eq:i_e_a}, $\mathbf{I}_{\mathcal{E}_\Phi}$~\eqref{eq:i_e_phi}, and $\mathbf{I}_\mathcal{M}$~\eqref{eq:i_m} can be evaluated in terms of $\boldsymbol{\xi}'$ on the $n_s$ surfaces where $\mathbf{n}\cdot\mathbf{e}_y\ne 0$.  From Table~\ref{tab:transformations}, $(\partial \mathbf{x}/\partial \boldsymbol{\xi})_j$ is constant, and $|\partial \mathbf{x}/\partial \boldsymbol{\xi}|_j=1$. 
Therefore,
\begin{alignat}{9}
\mathbf{I}_{\mathcal{E}_\mathbf{A}}(\mathbf{x}) 
&{}={}& &\sum_{n=0}^{n_m}  &&\tilde{G}_n \int_{S'} R(\mathbf{x},\mathbf{x}')^{2n} \mathbf{J}_\text{MS}(\mathbf{x}') dS' \nonumber
\\
&{}={}& &\sum_{n=0}^{n_m}  &&\tilde{G}_n \sum_{j=1}^{n_s}\int_0^\reviewerOne{L} \int_{\xi_{a_j}'}^{\xi_{b_j}'} R(\mathbf{x},\mathbf{x}_j(\boldsymbol{\xi}'))^{2n} J_\xi(\boldsymbol{\xi}')\bigg|\frac{\partial \mathbf{x}}{\partial \boldsymbol{\xi}}\bigg|_j\bigg(\frac{\partial \mathbf{x}}{\partial \boldsymbol{\xi}}\bigg)_j\mathbf{e}_\xi d\xi' d\eta'\nonumber
\\
&{}={}& &\sum_{n=0}^{n_m}  &&\tilde{G}_n \sum_{j=1}^{n_s} \bigg(\frac{\partial \mathbf{x}}{\partial \xi}\bigg)_j\int_0^\reviewerOne{L} \int_{\xi_{a_j}'}^{\xi_{b_j}'}R(\mathbf{x},\mathbf{x}_j(\boldsymbol{\xi}'))^{2n} J_\xi(\boldsymbol{\xi}') d\xi'd\eta' ,
\label{eq:i_e_a2}
\\
\mathbf{I}_{\mathcal{E}_\Phi}(\mathbf{x}) 
&{}={}& -2&\sum_{n=1}^{n_m} n&&\tilde{G}_n \int_{S'} R^{2(n-1)} \nabla'\cdot\mathbf{J}_\text{MS}(\mathbf{x}') \mathbf{R} dS' \nonumber
\\
&{}={}& -2&\sum_{n=1}^{n_m} n&&\tilde{G}_n \sum_{j=1}^{n_s} \int_0^\reviewerOne{L} \int_{\xi_{a_j}'}^{\xi_{b_j}'}R(\mathbf{x},\mathbf{x}_j(\boldsymbol{\xi}'))^{2(n-1)} \frac{\partial J_\xi}{\partial \xi}(\boldsymbol{\xi}')\mathbf{R}(\mathbf{x},\mathbf{x}_j(\boldsymbol{\xi}'))\bigg|\frac{\partial \mathbf{x}}{\partial \boldsymbol{\xi}}\bigg|_j d\xi'd\eta' \nonumber
\\
&{}={}& -2&\sum_{n=1}^{n_m} n&&\tilde{G}_n \sum_{j=1}^{n_s} \int_0^\reviewerOne{L} \int_{\xi_{a_j}'}^{\xi_{b_j}'}R(\mathbf{x},\mathbf{x}_j(\boldsymbol{\xi}'))^{2(n-1)} \frac{\partial J_\xi}{\partial \xi}(\boldsymbol{\xi}')\mathbf{R}(\mathbf{x},\mathbf{x}_j(\boldsymbol{\xi}')) d\xi'd\eta', 
\label{eq:i_e_phi2}
\\
\mathbf{I}_\mathcal{M}(\mathbf{x}) 
&{}={}& -2&\sum_{n=1}^{n_m} n&&\tilde{G}_n \int_{S'}  R^{2(n-1)}\mathbf{J}_\text{MS}(\mathbf{x}')\times\mathbf{R}dS'\nonumber
\\
&{}={}& -2&\sum_{n=1}^{n_m} n&&\tilde{G}_n \sum_{j=1}^{n_s} \int_0^\reviewerOne{L} \int_{\xi_{a_j}'}^{\xi_{b_j}'}R(\mathbf{x},\mathbf{x}_j(\boldsymbol{\xi}'))^{2(n-1)} J_\xi(\boldsymbol{\xi}') \mathbf{e}_\xi \times \mathbf{R}(\mathbf{x},\mathbf{x}_j(\boldsymbol{\xi}'))\bigg|\frac{\partial \mathbf{x}}{\partial \boldsymbol{\xi}}\bigg|_j d\xi'd\eta'  \nonumber
\\
&{}={}& -2&\sum_{n=1}^{n_m} n&&\tilde{G}_n \sum_{j=1}^{n_s} \int_0^\reviewerOne{L} \int_{\xi_{a_j}'}^{\xi_{b_j}'}R(\mathbf{x},\mathbf{x}_j(\boldsymbol{\xi}'))^{2(n-1)} J_\xi(\boldsymbol{\xi}') \mathbf{e}_\xi \times \mathbf{R}(\mathbf{x},\mathbf{x}_j(\boldsymbol{\xi}')) d\xi'd\eta',  
\label{eq:i_m2}
\end{alignat}
where $\zeta'=0$ in $\boldsymbol{\xi}'$, and the directions of $\mathbf{I}_{\mathcal{E}_\mathbf{A}}^n$, $\mathbf{I}_{\mathcal{E}_\Phi}^n$, and $\mathbf{I}_\mathcal{M}^n$ remain expressed in the $\mathbf{x}$-coordinate system.

In~\eqref{eq:i_e_a2}--\eqref{eq:i_m2}, we can replace $\mathbf{x}$ with $\mathbf{x}_j(\boldsymbol{\xi})$, such that
\begin{align}
\mathbf{R}(\mathbf{x},\mathbf{x}_j(\boldsymbol{\xi}')) = \mathbf{x}_j(\boldsymbol{\xi})-\mathbf{x}_j(\boldsymbol{\xi}') = \bigg(\frac{\partial \mathbf{x}}{\partial \boldsymbol{\xi}}\bigg)_j(\boldsymbol{\xi}_j-\boldsymbol{\xi}'),
\label{eq:R}
\end{align}
and
\begin{align}
R(\mathbf{x}_j(\boldsymbol{\xi}),\mathbf{x}_j(\boldsymbol{\xi}'))^2 
= \|\mathbf{R}(\mathbf{x},\mathbf{x}_j(\boldsymbol{\xi}'))\|_2^2
= \bigg\|\bigg(\frac{\partial \mathbf{x}}{\partial \boldsymbol{\xi}}\bigg)_j(\boldsymbol{\xi}_j-\boldsymbol{\xi}')\bigg\|_2^2
= \|\boldsymbol{\xi}_j-\boldsymbol{\xi}'\|_2^2.
\label{eq:R2}
\end{align}
Accounting for~\eqref{eq:R2}, $R^{2n}$ in~\eqref{eq:i_e_a2} can be written as
\begin{align}
R^{2n} = \sum_{m=0}^n \sum_{k=0}^m \binom{n}{m}\binom{m}{k} (\xi_j-\xi')^{2(n-m)}(\eta_j-\eta')^{2(m-k)}(\zeta_j-\zeta')^{2k},
\label{eq:R2n}
\end{align}
and $R^{2(n-1)}$ in~\eqref{eq:i_e_phi2} and~\eqref{eq:i_m2} can be written as
\begin{align}
R^{2(n-1)} = \sum_{m=0}^{n-1} \sum_{k=0}^m \binom{n-1}{m}\binom{m}{k} (\xi_j-\xi')^{2(n-m-1)}(\eta_j-\eta')^{2(m-k)}(\zeta_j-\zeta')^{2k},
\label{eq:R2nm1}
\end{align}
where $\boldsymbol{\xi}_j=\boldsymbol{\xi}_j(\mathbf{x})$.

\clearpage
Inserting~\eqref{eq:R2n} into~\eqref{eq:i_e_a2} and inserting~\eqref{eq:R2nm1} and~\eqref{eq:R} into~\eqref{eq:i_e_phi2} and~\eqref{eq:i_m2}, \eqref{eq:i_e_a2}--\eqref{eq:i_m2} become
\begin{alignat*}{9}
\mathbf{I}_{\mathcal{E}_\mathbf{A}}(\mathbf{x}) 
&{}={}& &\sum_{j=1}^{n_s} \sum_{n=0}^{n_m}  && \sum_{m=0}^n \sum_{k=0}^m \tilde{G}_n\binom{n}{m}\binom{m}{k} \bigg(\frac{\partial \mathbf{x}}{\partial \xi}\bigg)_j
I_\mathbf{J}^j(\boldsymbol{\xi}_j;2(n-m),2(m-k),2k) ,
\\
\mathbf{I}_{\mathcal{E}_\Phi}(\mathbf{x}) 
&{}={}& -2&\sum_{j=1}^{n_s} \sum_{n=1}^{n_m} && \sum_{m=0}^{n-1} \sum_{k=0}^m n \tilde{G}_n\binom{n-1}{m}\binom{m}{k} \bigg(\frac{\partial \mathbf{x}}{\partial \boldsymbol{\xi}}\bigg)_j\mathbf{I}_{\nabla'\cdot \mathbf{J}}^j(\boldsymbol{\xi}_j;2(n-m-1),2(m-k),2k), 
\\
\mathbf{I}_\mathcal{M}(\mathbf{x}) 
&{}={}& -2&\sum_{j=1}^{n_s} \sum_{n=1}^{n_m} && \sum_{m=0}^{n-1} \sum_{k=0}^m n \tilde{G}_n\binom{n-1}{m}\binom{m}{k} \bigg(\frac{\partial \mathbf{x}}{\partial \boldsymbol{\xi}}\bigg)_j\big(
\mathbf{e}_\xi
\times 
\mathbf{I}_\mathbf{J}^j(\boldsymbol{\xi}_j;2(n-m-1),2(m-k),2k)\big),  
\end{alignat*}
where 
\begin{align*}
\mathbf{I}_{\nabla'\cdot \mathbf{J}}^j(\boldsymbol{\xi};p,q,r) = \left\{\begin{matrix}
I_{\nabla'\cdot \mathbf{J}}^j(\boldsymbol{\xi};p+1            ,q\phantom{{}+1},r\phantom{{}+1}) \\ 
I_{\nabla'\cdot \mathbf{J}}^j(\boldsymbol{\xi};p\phantom{{}+1},q+1            ,r\phantom{{}+1}) \\ 
I_{\nabla'\cdot \mathbf{J}}^j(\boldsymbol{\xi};p\phantom{{}+1},q\phantom{{}+1},r+1            )
\end{matrix}\right\}, 
\qquad
\mathbf{I}_\mathbf{J}^j(\boldsymbol{\xi};p,q,r) = \left\{\begin{matrix}
I_\mathbf{J}^j(\boldsymbol{\xi};p+1            ,q\phantom{{}+1},r\phantom{{}+1}) \\ 
I_\mathbf{J}^j(\boldsymbol{\xi};p\phantom{{}+1},q+1            ,r\phantom{{}+1}) \\ 
I_\mathbf{J}^j(\boldsymbol{\xi};p\phantom{{}+1},q\phantom{{}+1},r+1            )
\end{matrix}\right\}.
\end{align*}
The directions of $\mathbf{I}_{\nabla'\cdot \mathbf{J}}^j$, $\mathbf{e}_\xi$ and $\mathbf{I}_\mathbf{J}^j$ are expressed in the $\boldsymbol{\xi}$-coordinate system, and
\begin{align}
I_\mathbf{J}^j(\boldsymbol{\xi};p,q,r) &{}= \int_0^\reviewerOne{L} \int_{\xi_{a_j}'}^{\xi_{b_j}'}(\xi-\xi')^{p}(\eta-\eta')^{q}(\zeta-\zeta')^{r} J_\xi(\boldsymbol{\xi}') d\xi'd\eta', 
\label{eq:i_j_j}
\\
I_{\nabla'\cdot \mathbf{J}}^j(\boldsymbol{\xi};p,q,r) &{}= \int_0^\reviewerOne{L} \int_{\xi_{a_j}'}^{\xi_{b_j}'}(\xi-\xi')^{p}(\eta-\eta')^{q}(\zeta-\zeta')^{r} \frac{\partial J_\xi}{\partial \xi}(\boldsymbol{\xi}') d\xi'd\eta'. 
\label{eq:i_divj_j}
\end{align}

Recalling that $\zeta'=0$ in $\boldsymbol{\xi}'$, for $\displaystyle J(\boldsymbol{\xi})=J_0\sin (\beta\xi\reviewerOne{/L}) \sin^3 (\pi\eta\reviewerOne{/L})$~\eqref{eq:jxi}, \eqref{eq:i_j_j} and~\eqref{eq:i_divj_j} become
\begin{alignat*}{9}
I_\mathbf{J}^j(\boldsymbol{\xi};p,q,r) &{}={}& J_0\zeta^r\bigg(\int_{\xi'_{a_j}}^{\xi'_{b_j}}  (\xi-\xi')^p \sin \bigg(\reviewerOne{\frac{\beta\xi'}{L}}\bigg) d\xi'\bigg)\bigg(\int_0^\reviewerOne{L} (\eta-\eta')^q\sin^3 \bigg(\reviewerOne{\frac{\pi\eta'}{L}}\bigg)  d\eta'\bigg) &{}={}& J_0 I_p^j(\xi)  I_q(\eta) I_r(\zeta),
\\
I_\mathbf{\nabla'\cdot J}^j(\boldsymbol{\xi};p,q,r) &{}={}&\frac{\beta J_0\zeta^r}{\reviewerOne{L}}\bigg(\int_{\xi'_{a_j}}^{\xi'_{b_j}}  (\xi-\xi')^p \cos \bigg(\reviewerOne{\frac{\beta\xi'}{L}}\bigg) d\xi'\bigg) \bigg(\int_0^\reviewerOne{L} (\eta-\eta')^q\sin^3 \bigg(\reviewerOne{\frac{\pi\eta'}{L}}\bigg)  d\eta'\bigg) &{}={}&  \frac{\beta J_0 I_{\bar{p}}^j(\xi)  I_q(\eta) I_r(\zeta)}{\reviewerOne{L}}.
\end{alignat*}
where $\xi'_{a_j}=\xi_{a_j}$ and $\xi'_{b_j}=\xi_{b_j}$ in Table~\ref{tab:transformations}, and
\begin{alignat*}{9}
I_p^j(\xi) 
&{}={}&&
\int_{\xi'_{a_j}}^{\xi'_{b_j}}  (\xi-\xi')^p \sin \bigg(\reviewerOne{\frac{\beta\xi'}{L}}\bigg) d\xi' 
&{}={}&
\sum_{p'=0}^p \binom{p}{p'} (-1)^{p'} \xi^{p-p'} \int_{\xi'_{a_j}}^{\xi'_{b_j}} {\xi'}^{p'} \sin\bigg(\reviewerOne{\frac{\beta\xi'}{L}}\bigg)d\xi' 
&{}={}&
\sum_{p'=0}^p \binom{p}{p'} (-1)^{p'} \xi^{p-p'} I_{p'}^j,
\\[1em]
I_{\bar{p}}^j(\xi) 
&{}={}&& 
\int_{\xi'_{a_j}}^{\xi'_{b_j}}  (\xi-\xi')^p \cos \bigg(\reviewerOne{\frac{\beta\xi'}{L}}\bigg) d\xi' 
&{}={}&
\sum_{p'=0}^p \binom{p}{p'} (-1)^{p'} \xi^{p-p'} \int_{\xi'_{a_j}}^{\xi'_{b_j}} {\xi'}^{p'} \cos\bigg(\reviewerOne{\frac{\beta\xi'}{L}}\bigg)d\xi' 
&{}={}&
\sum_{p'=0}^p \binom{p}{p'} (-1)^{p'} \xi^{p-p'} I_{\bar{p}'}^j,
\\[1em]
I_q(\eta) &{}={}&& \int_0^\reviewerOne{L} (\eta-\eta')^q\sin^3 \bigg(\reviewerOne{\frac{\pi\eta'}{L}}\bigg)  d\eta' &{}={}& \sum_{q'=0}^q \binom{q}{q'} (-1)^{q'} \eta^{q-q'} \int_0^\reviewerOne{L} {\eta'}^{q'} \sin^3\bigg(\reviewerOne{\frac{\pi\eta'}{L}}\bigg)d\eta' &{}={}& \sum_{q'=0}^q \binom{q}{q'} (-1)^{q'} \eta^{q-q'} I_{q'},
\\[1em]
I_r(\zeta) &{}={}&& \zeta^r,
\end{alignat*}
and
\begin{align*}
I_{p'}^j
={}&
\int_{\xi'_{a_j}}^{\xi'_{b_j}} {\xi'}^{p'} \sin\bigg(\reviewerOne{\frac{\beta\xi'}{L}}\bigg)d\xi'\\
={}&
-\sum_{p''=0}^{p'} \frac{p'!}{(p'-p'')!}\reviewerOne{\bigg(\frac{L}{\beta}\bigg)^{1+p''}}\bigg[{\xi'_{b_j}}^{p'-p''}\cos\bigg(\reviewerOne{\frac{\beta \xi'_{b_j}}{L}}+\frac{p''\pi}{2}\bigg)-{\xi'_{a_j}}^{p'-p''}\cos\bigg(\reviewerOne{\frac{\beta \xi'_{a_j}}{L}}+\frac{p''\pi}{2}\bigg)\bigg],
%
\\[1em]
I_{\bar{p}'}^j
={}&
\int_{\xi'_{a_j}}^{\xi'_{b_j}} {\xi'}^{p'} \cos\bigg(\reviewerOne{\frac{\beta\xi'}{L}}\bigg)d\xi'\\
={}&
\phantom{+}\sum_{p''=0}^{p'} \frac{p'!}{(p'-p'')!}\reviewerOne{\bigg(\frac{L}{\beta}\bigg)^{1+p''}}\bigg[{\xi'_{b_j}}^{p'-p''}\sin\bigg(\reviewerOne{\frac{\beta \xi'_{b_j}}{L}}+\frac{p''\pi}{2}\bigg)-{\xi'_{a_j}}^{p'-p''}\sin\bigg(\reviewerOne{\frac{\beta \xi'_{a_j}}{L}}+\frac{p''\pi}{2}\bigg)\bigg],
%
\\[1em]
I_{q'}
={}&
\int_0^\reviewerOne{L} {\eta'}^{q'} \sin^3\bigg(\reviewerOne{\frac{\pi\eta'}{L}}\bigg)d\eta' \\
={}&
\frac{q'!}{8}\reviewerOne{\bigg(\frac{L}{3\pi}\bigg)^{1+q'}}\bigg[
2\Big(-1+3^{2+q'}\Big)\cos\bigg(\frac{q'\pi}{2}\bigg) \nonumber\\
&+\sum_{q''=0}^{q'} \frac{\pi^{q''}}{q''!}\big(-3^{2+q'}+3^{q''}\big)\left\{%
\begin{array}{c l}
(-1)^{(q''-q'-2)/2}(1-(-1)^{q''-q'-1}), & \text{for } q'+q'' \text{ even}\\ 0, & \text{for } q'+q'' \text{ odd}
\end{array}\right.
\bigg].
\end{align*}

%
%
%
%
%
%
%
%
%
%
%
%
%
%
%
%
%
%
%
%